\batchmode
\makeatletter
\def\input@path{{/Users/nicolaborri/Dropbox/Research/INDEX-EVERYTHING/lyx/}}
\makeatother
\documentclass[12pt,english]{article}
\usepackage{amsmath}
\usepackage{amsthm}
\usepackage{mathptmx}
\usepackage{newtxmath}
\usepackage[T1]{fontenc}
\usepackage[latin9]{inputenc}
\usepackage{geometry}
\usepackage{color}
\usepackage{babel}
\usepackage{float}
\usepackage{graphicx}
\usepackage{rotfloat}
\usepackage{setspace}
\usepackage[authoryear]{natbib}
\usepackage[pdftex,unicode=true,pdfusetitle,
 bookmarks=true,bookmarksnumbered=false,bookmarksopen=false,
 breaklinks=false,pdfborder={0 0 0},pdfborderstyle={},backref=false,colorlinks=true]
 {hyperref}
\hypersetup{
 linkcolor=red,urlcolor=red,citecolor=blue}
\usepackage[dot]{bibtopic}

\makeatletter

\providecommand{\tabularnewline}{\\}

\usepackage[bottom]{footmisc}
\usepackage{indentfirst}
\usepackage{lscape}
\usepackage{xcolor}
\usepackage{titlesec}

\newtheorem{theorem}{Theorem}[section]              \newtheorem{lemma}{Lemma}[section]       \newtheorem{assumption}{Assumption}[section]
  \def\R{\mathbb{R}}     

\date{\today}

\@ifundefined{showcaptionsetup}{}{%
 \PassOptionsToPackage{caption=false}{subfig}}
\usepackage{subfig}
\makeatother

\begin{document}
\title{One Factor to Bind the Cross-Section of Returns\thanks{Nicola Borri is with Luiss University, Rome. Denis Chetverikov is
with the University of California, Los Angeles. Yukun Liu is with
the University of Rochester, Simon Business School. Aleh Tsyvinski
is with Yale University. We thank Yacine Ait-Sahalia, Bryan Kelly,
Leonid Kogan, Ralph Koijen, Toby Moskowitz and Semyon Malamud for
comments. We thank Andrei Voronin for excellent research assistance. }}
\author{Nicola Borri, Denis Chetverikov, Yukun Liu, and Aleh Tsyvinski}
\maketitle
\begin{abstract}
\noindent We propose a new non-linear single-factor asset pricing
model $r_{it}=h(f_{t}\lambda_{i})+\epsilon_{it}$. Despite its parsimony,
this model represents exactly any non-linear model with an arbitrary
number of factors and loadings -- a consequence of the Kolmogorov-Arnold
representation theorem. It features only one pricing component $h(f_{t}\lambda_{i})$,
comprising a nonparametric link function of the time-dependent factor
and factor loading that we jointly estimate with sieve-based estimators.
Using 171 assets across major classes, our model delivers superior
cross-sectional performance with a low-dimensional approximation of
the link function. Most known finance and macro factors become insignificant
controlling for our single-factor.
\end{abstract}
\ 



\thispagestyle{empty}

\pagebreak{}

\setcounter{page}{1}

\section{Introduction}

Factor models play an important role in finance and economics. The
existing literature, however, focuses primarily on linear factor models.
In this paper, we propose a new non-linear single factor asset pricing
model 
\begin{equation}
r_{it}=h(f_{t}\lambda_{i})+\varepsilon_{it},\quad i=1,\dots,N,\ t=1,\dots,T,\label{eq: model}
\end{equation}
where $r_{it}\in\mathbb{R}$ is the return of asset $i$ at time period
$t$, $f_{t}>0$ is the single factor, $\lambda_{i}>0$ is the factor
loading, $h$ is a continuous function, and $\varepsilon_{it}\in\mathbb{R}$
is an idiosyncratic mean-zero component. Here, the function $h$,
the factor values $f_{t}$, and the factor loading values $\lambda_{i}$
are all assumed to be unknown, and both $N$ and $T$ are assumed
to be large. We refer to our model as the HFL model.

Our model is different from the typical linear factor model. On the
one hand, our model is seemingly more restrictive because it allows
for only one time-dependent factor and requires both the time factor
and the factor loading to be strictly positive, whereas the linear
factor model allows for an arbitrary number of factors and allows
all factors and factor loadings to take negative values. On the other
hand, our model is more flexible because it includes a nonparametric
link function $h$, whereas the linear factor model sets the link
function to be identity.

Even though our model has only one factor, the presence of the nonparametric
link function $h$ makes our model surprisingly flexible. In particular,
using the Kolmogorov-Arnold representation theorem (\citet{Kolmogorov1956on,Arnold1957on}),
we show that any nonparametric factor model with an arbitrary number
of factors and with arbitrary interactions between factors and factor
loadings can be reduced to model (\ref{eq: model}). That is, any
arbitrary nonparametric model can be exactly represented by our model.
It is thus without loss of generality to focus on our model with just
one factor.

We apply our model to characterize risk premia in a large cross-section
of assets across major asset classes, and propose a sieve-based least
squares estimator of the model. This estimator is defined as a solution
to an optimization problem minimizing the sum of squared residuals
for the model not only over $T$ potential factor values $\{f_{t}\}_{t=1}^{T}$
and $N$ potential factor loading values $\{\lambda_{i}\}_{i=1}^{N}$
but also over potential values of the function $h$ in a sieve space,
which we choose to consist of all polynomials of a given order that
is slowly increasing in the sample dimensions $(N,T)$. 

The Kolmogorov-Arnold representation ensures that the function $h$
must be continuous under very mild regularity conditions, and so can
be arbitrarily well approximated by a suitable sieve space for estimation
purposes. At the same time, readers familiar with the Kolmogorov-Arnold
representation theorem might argue that this theorem implies that
even though the function $h$ must be continuous, it may nonetheless
be rather irregular, and one may need a high-dimensional sieve space
in order to approximate it sufficiently well, making estimation of
this model difficult. However, whether the function $h$ can be well
approximated by a relatively low-dimensional sieve space for the purposes
of a particular application, making estimation feasible, is ultimately
an empirical question. Our main empirical result is that, for the
cross-section of asset returns across major asset classes, the estimated
version of our model based on a relatively low-dimensional sieve space
outperforms linear factor models with multiple factors in a number
of dimensions.

We test the cross-sectional asset pricing predictions of our model
using as test assets U.S. equity portfolios, U.S. and international
government bonds, commodities, and foreign currency portfolios, for
a total of 171 assets. Specifically, we estimate a cross-sectional
regression of the average asset realized excess returns on the average
asset excess returns predicted by the model and test several implications
of our one factor model, such as that the intercept is zero and the
significance of the slope coefficient. Furthermore, we evaluate the
fit of the model. Our empirical strategy follows the classic two-step
\citet{fama1973risk}'s procedure. The first step of the \citet{fama1973risk}'s
procedure, which requires the estimation of the asset specific betas
in time-series regressions, is implicit in the algorithm to estimate
our single factor model. The second step of the procedure, which requires
a cross-sectional regression of average asset excess returns on factor
betas, corresponds to our cross-sectional regression.

Our single factor model explains a large fraction of cross-sectional
differences in asset returns. We find that a relatively low-dimensional
sieve space, with the polynomials of degree as low as four, suffices
for the majority of our empirical results. When using all asset classes
as test assets, the slope coefficient on the pricing component is
significantly different from zero. Furthermore, we cannot reject the
null that the slope coefficient is equal to 1, as implied by our model.
Moreover, the pricing error, measured by the intercept in the cross-sectional
regression, is not significantly different from zero. The HFL model
exhibits a cross-sectional regression adjusted R-squared value of
89\%.

Next, we compare the HFL model with a number of prominent cross-sectional
asset pricing models that have been proposed in the finance and economics
literature. First, we consider benchmark equity factor models, such
as the CAPM, the \citet{fama1993common}'s three factor model, the
\citet{fama2015five}'s five factor model and the five factor model
augmented with the momentum factor of \citet{jegadeesh1993returns}.
Second, we consider models using factors based on principal component
analysis (PCA). Specifically, we consider three estimators of latent
factors based on principal components: the standard estimator based
on the covariance matrix of asset returns; the RP-PCA estimator, which
is constructed to better account for the cross-section of average
asset returns; and the kernel PCA estimator, which is a nonlinear
form of PCA. Third, we consider higher-order factor models, which
account for non-linearities by including higher order factors, such
as the squared equity market return of \citet{harvey2000conditional}.
Specifically, we consider models including the square and cube of
equity and PCA factors. Fourth, we consider recent macro-factor models,
such as the intermediary capital risk factor model of \citet*{he2017intermediary},
the downside equity market risk factor model of \citet*{lettau2014conditional},
and the liquidity factor model of \citet{pastor2003liquidity}. Fifth,
we consider the macro factors of \citet{ludvigson2009macro}. Our
main result is that, once we include our single factor in cross-sectional
regressions, the risk prices associated with the alternative factors
are all indistinguishable from zero, with the exception of the size
factor in \citet{fama1993common}.

Furthermore, we compare the HFL model with a large set of factors
proposed in the last decades to account for asset risk premia, which
is commonly referred to as the factor zoo (see, e.g., \citet*{feng2020taming}).
We consider 153 established factors, collected by \citet*{jensen2023there},
and for each of these factors, we first estimate a standard cross-sectional
regression of average asset excess returns on the asset factor betas.
The betas are measured as slope coefficients in time-series regressions
of each asset excess return on the factor. We find that, for a large
number of these factors, the risk price, i.e., the slope coefficient
in the cross-sectional regression, is statistically significant, although
also the estimates for the intercept, a measure of the pricing error
of the model, are mostly significantly different from zero. We then
repeat the estimation additionally including the predicted values
from the HFL model. The results are striking. We find a significant
factor risk price estimate for only 3 out of the 153 factors in the
factor zoo, while the slope estimates associated with the HFL factor
are all statistically significant. Moreover, the pricing error of
these models are mostly indistinguishable from zero. Furthermore,
we conduct cross-sectional regressions employing both the HFL factor
and a diverse set of factors from the factor zoo, employing the double-selection
Lasso method. This approach enables simultaneous control over all
factors and provides a robust model selection technique for assessing
factor contributions in cross-sectional asset pricing regressions
(\citet*{feng2020taming}). We find that the HFL component is highly
significant, after controlling for the factor zoo, while the majority
of the factors in the factor zoo is not.

We then investigate the predictive ability of the HFL model by building
portfolios using as signals the predicted asset returns. Specifically,
we first estimate the model on all assets and the first 120 months
of data. Next, we use the model predicted values to sort assets in
five portfolios, and compute the equally-weighted average portfolio
returns in the following six months. Finally, we repeat the procedure
expanding the estimation window by 6 months each time until the end
of the sample. This empirical methodology is designed to capture the
strategy of an investor who uses the HFL model to predict future returns
rebalancing at a semi-annual frequency. In each period $t$, we sort
assets only using information available up to period $t$ and compute
excess returns between $t$ and $t+1,$where each period contains
6 months. The first portfolio groups the assets with, each period,
the lowest predicted excess returns by the HFL model. The last portfolio
groups the assets with, each period, the highest predicted excess
returns by the HFL model. Across the portfolios, we document a sizable,
and statistically significant, cross-section of monthly average returns,
ranging from 0.08\% for the first portfolio to 0.77\% for the last
one. A zero-cost long-short strategy, wherein assets with the highest
predicted returns are bought while those with the lowest predicted
returns are sold short, yields a monthly average excess return of
0.69\%, with a monthly Sharpe ratio of 0.15. We investigate the risk-adjusted
performance of the HFL strategy, by regressing the portfolio returns
on the \citet*{fama1993common}'s three factors, on the \citet*{fama2015five}'s
five factors, and on the five factors augmented by the momentum factor
of \citet*{jegadeesh1993returns}, respectively. The estimates of
the intercept from these regressions, a measure of risk-adjusted performance,
increase from the first to the last portfolio in all three specifications,
and are significant and quantitatively large for portfolio 3 to 5,
as well as for the long-short portfolio.

Our empirical results extend to alternative samples. First, we show
that the HFL component is statistically significant for each individual
asset class using the HFL component estimated with all asset classes.
Second, we repeat the cross-sectional asset pricing tests using the
first and second half of the sample and find results consistent with
those obtained using the full sample. Third, we consider a higher
frequency time-variation in the asset pricing test using an estimation
based on an expanding window. In this case, we find that the slope
coefficient associated with our single factor is significantly different
from zero for all months, and we can never reject the null hypothesis
that the slope coefficient equals to one as implied by our model.
Fourth, we augment the set of test assets with 10 equity momentum
portfolios and the momentum factor and report that the slope coefficient
associated with the HFL component is statistically different from
zero.

Our paper relates to a vast literature on factor models in asset pricing.
The majority of the literature focuses on linear factor models, such
as characteristic-based models (e.g., \citet*{fama1993common,fama1996multifactor,Carhart1997,koijen2017cross}),
statistical-based models (e.g., \citet{connor1986performance}), and
macro models (e.g., \citet*{adrian2014financial,he2017intermediary}).
There is a literature that studies conditional asset pricing models,
where the model is conditionally linear (e.g., \citet*{jagannathan1996conditional,lettau2001resurrecting,kelly2019characteristics}).
There is a literature that shows that the same characteristics-based
models are present across a large set of asset classes (e.g., \citet*{asness2013value,koijen2018carry,bollerslev2018risk}).
A small subset of papers emphasizes non-linearity in explaining asset
returns, such as \citet{bansal1993no}. A growing literature aims
at pricing assets with machine learning methods (e.g., \citet*{hutchinson1994nonparametric,gu2020empirical,kelly2023principal,kelly2024virtue}).
Apart for using factor models in understanding asset prices, a recent
literature highlights the significance of demand forces (e.g., \citet*{koijen2019demand,koijen2020investors}).
We contribute to this literature by proposing and testing a parsimonious
model that represents exactly any non-linear model with an arbitrary
number of factors and factor loadings.

The Kolmogorov-Arnold representation is typically used as a foundation
of deep learning and as a benchmark for deriving bounds on neural
network approximations; see \citet{SH21}, \citet{hecht1987kolmogorov},
\citet{K91}, and \citet{MP99}. \citet{C04} also used this representation
for nonparametric estimation in the regression context. We contribute
to this literature by applying this representation to factor models.

Finally, our approach complements the literature on non-linear factor
extraction. \citet*{gu2021autoencoder} used an auto-encoder approach.
\citet*{SSM98} developed a kernel PCA procedure, which estimates
a linear factor model for a non-linear transformation of the vectors
$(r_{1t},\dots,r_{Nt})$. In contrast to these alternative methods,
our approach is based on the assumption that the factor affects returns
via model (\ref{eq: model}).

The rest of the paper is organized as follows. Section \ref{sec:Motivation-and-Estimation}
discusses the motivation and estimation for our one factor model.
Section \ref{sec:rate of convergence} derives the rate of convergence
of our sieve-based least squares estimator. Section \ref{sec:Data-Summary}
discusses the data and presents summary statistics. Section \ref{sec:Cross-Section}
presents the cross-sectional asset pricing results and compares the
HFL model with alternative models. Section \ref{sec:Additional-results}
contains additional results and robustness checks. The Online Appendix
contains technical derivations and further empirical results.

\section{Motivation and Estimation\label{sec:Motivation-and-Estimation}}

We start our analysis with a very general factor-type model 
\begin{equation}
r_{it}=g(x_{t1},\dots,x_{tk},z_{i1},\dots,z_{im})+\varepsilon_{it},\quad i=1,\dots,N,\ t=1,\dots,T,\label{eq: general model}
\end{equation}
where $r_{it}$ is the excess return of asset $i$ at time period
$t$, $x_{t1},\dots,x_{tk}\in[0,1]$ are time-specific effects/factors,
$z_{i1},\dots,z_{im}\in[0,1]$ are asset-specific effects/factor loadings,
$g$ is a continuous function, and $\varepsilon_{it}$ is an idiosyncratic
mean-zero component. This model is very general as it nests many special
cases. For example, it nests any linear factor model 
\[
r_{it}=\sum_{j=1}^{k}x_{tj}z_{ij}+\varepsilon_{it},
\]
any single-index factor model 
\[
r_{it}=h\left(\sum_{j=1}^{k}x_{tj}z_{ij}\right)+\varepsilon_{it},
\]
and any non-linear factor model 
\[
r_{it}=h(x_{t1}z_{i1},\dots,x_{tk}z_{ik})+\varepsilon_{it}.
\]
In fact, model (\ref{eq: general model}) does not even require the
number of time-specific effects $k$ to coincide with the number of
asset-specific effects $m$.

Our first result shows that under very mild regularity conditions,
namely continuity of the function $g$, model (\ref{eq: general model})
can be reduced to model (\ref{eq: model}) with continuous $h$ as
a consequence of  the Kolmogorov-Arnold theorem (\citet{Kolmogorov1956on,Arnold1957on}).
Specifically, consider the version of Kolmogorov-Arnold theorem  in
\citet{SH21}, Theorem 2(ii). That representation implies that there
exists a continuous function $q$ and monotone functions $\phi_{1},\dots,\phi_{k},\psi_{1},\dots,\psi_{m}$
such that 
\[
g(x_{1},\dots,x_{k},z_{1},\dots,z_{m})=q\left(\sum_{j=1}^{k}\phi_{j}(x_{j})+\sum_{j=1}^{m}\psi_{j}(z_{j})\right)
\]
for all $x_{1},\dots,x_{k},z_{1},\dots,z_{m}\in[0,1]$. Therefore,
denoting $\widetilde{f}_{t}=\sum_{j=1}^{k}\phi_{j}(x_{tj})$ and $\widetilde{\lambda}_{i}=\sum_{j=1}^{m}\psi_{j}(z_{ij})$,
it follows that 
\[
g(x_{t1},\dots,x_{tk},z_{i1},\dots,z_{im})=q(\widetilde{f}_{t}+\widetilde{\lambda}_{i})
\]
for all $i=1,\dots,N$ and $t=1,\dots,T$. Moreover, by using the
identity 
\[
x+z=\log(\exp(x+z))=\log(\exp(x)\exp(z)),\quad x,z\in\R
\]
and denoting $h(x)=q(\log(x))$, $f_{t}=\exp(\widetilde{f}_{t})$,
and $\lambda_{i}=\exp(\widetilde{\lambda}_{i})$, we have 
\[
g(x_{t1},\dots,x_{tk},z_{i1},\dots,z_{im})=q(\log(f_{t}\lambda_{i}))=h(f_{t}\lambda_{i})
\]
for all $t=1,\dots,T$ and $i=1,\dots,N$. We summarize this result
in the following lemma.

\begin{lemma} If random variables $\{y_{it}\}_{i,t}^{N,T}$ satisfy
model (\ref{eq: general model}) with continuous $g$, there exist
a continuous function $h$, and strictly positive sequences $\{f_{t}\}_{t=1}^{T}$
and $\{\lambda_{i}\}_{i=1}^{N}$ such that 
\[
r_{it}=h(f_{t}\lambda_{i})+\varepsilon_{it},\quad i=1,\dots,N,\ t=1,\dots,T.
\]
\end{lemma} This lemma shows that any factor-type model (\ref{eq: general model})
can be reduced to model (\ref{eq: model}). It is because of this
generality, we study the non-linear factor model (\ref{eq: model})
with just one factor in this paper.

\medskip

\noindent \textbf{Remark} (Alternative Representations). We could
consider a more flexible reduction of model (\ref{eq: general model}).
Indeed, by applying the Kolmogorov-Arnold representation theorem conditional
on $z_{1},\dots,z_{m}$, we can find a continuous function $q$ and
monotone functions $\phi_{1},\dots,\phi_{k}$ such that 
\[
g(x_{1},\dots,x_{k},z_{1},\dots,z_{m})=q\left(\sum_{j=1}^{k}\phi_{j}(x_{j}),z_{1},\dots,z_{m}\right)
\]
for all $x_{1},\dots,x_{k},z_{1},\dots,z_{m}\in[0,1]$. Further, by
applying the Kolmogorov-Arnold representation theorem conditional
on $x_{1},\dots,x_{k}$, we can find a continuous function $u$ and
monotone functions $\psi_{1},\dots,\psi_{k}$ such that 
\[
q\left(\sum_{j=1}^{k}\phi_{j}(x_{j}),z_{1},\dots,z_{m}\right)=u\left(\sum_{j=1}^{k}\phi_{j}(x_{j}),\sum_{j=1}^{m}\psi_{j}(z_{j})\right)
\]
for all $x_{1},\dots,x_{k},z_{1},\dots,z_{m}\in[0,1]$. Thus, denoting
$f_{t}=\sum_{j=1}^{k}\phi_{j}(x_{tj})$ and $\lambda_{i}=\sum_{j=1}^{m}\psi_{j}(z_{ij})$,
it follows that model (\ref{eq: general model}) implies that 
\begin{equation}
r_{it}=u(f_{t},\lambda_{i})+\varepsilon_{it},\quad i=1,\dots,N,\ t=1,\dots,T.\label{eq: more flexible model}
\end{equation}
\qed

In order to estimate model (\ref{eq: model}), we assume that both
the factor $f_{t}$ and the factor loading $\lambda_{i}$ are bounded
random variables. By rescaling the function $h$, it is then without
loss of generality to assume that both $f_{t}$ and $\lambda_{i}$
are taking values in the $(0,1]$ interval. Further, since the function
$h$ is continuous, it follows from the Weierstrass theorem that it
can be approximated arbitrarily well by a finite-degree polynomial:
\[
h(x)=\sum_{j=0}^{K}h_{j}x^{j}+e_{K}(x),\quad x>0,
\]
where the residual $e_{K}$ converges to zero as the degree of the
polynomial $K$ gets large. We therefore propose a sieve-based least
squares estimator 
\begin{equation}
\left(\{\widehat{h}_{j}\}_{j=0}^{K},\{\widehat{f}_{t}\}_{t=1}^{T},\{\widehat{\lambda}_{i}\}_{i=1}^{N}\right)=\arg\min\sum_{i=1}^{N}\sum_{t=1}^{T}\left(r_{it}-\sum_{j=0}^{K}c_{j}(\phi_{t}l_{i})^{j}\right)^{2},\label{eq: estimator}
\end{equation}
where the minimum is taken over all $\{c_{j}\}_{j=0}^{K}\in\R^{K+1}$,
$\{\phi_{t}\}_{t=1}^{T}\in(0,1]^{T}$, and $\{l_{i}\}_{i=1}^{N}\in(0,1]^{N}$.
The estimators of the factor values $\{f_{t}\}_{t=1}^{T}$ and factor
loadings $\{\lambda_{i}\}_{i=1}^{N}$ are then $\{\widehat{f}_{t}\}_{t=1}^{T}$
and $\{\widehat{\lambda}_{i}\}_{i=1}^{N}$, respectively, and the
estimator of the function $h$ is the $K$th degree polynomial
\begin{equation}
\widehat{h}(x)=\sum_{j=0}^{K}\widehat{h}_{j}x^{j},\quad x>0.\label{eq: estimator 2}
\end{equation}
Note here that the optimization problem (\ref{eq: estimator}) is
highly non-convex and may be difficult to solve exactly. To make progress
in this direction, we take advantage of the fact that for given values
of $\{\phi_{t}\}_{t=1}^{T}$ and $\{l_{i}\}_{i=1}^{N}$, the optimization
over $\{c_{j}\}_{j=0}^{K}$ is easy as it can be implemented via OLS.
We therefore only need to take care of optimization over $\{\phi_{t}\}_{t=1}^{T}\in(0,1]^{T}$
and $\{l_{i}\}_{i=1}^{N}\in(0,1]^{N}$. In practice, we proceed with
optimization over these two sequences as follows. First, we calculate
the initial values of $\{\phi_{t}\}_{t=1}^{T}$ and $\{l_{i}\}_{i=1}^{N}$
as the first left and right singular vectors of the $T\times N$ matrix
$\{r_{it}\}_{t,i=1}^{T,N}$ shifted and scaled in a way to make sure
that their components are taking values in the $(0,1]$ interval.
Second, we perform gradient descent starting from the initial values
to find a local minimum of the optimization problem. Third, we reoptimize
each component of the sequences $\{\phi_{t}\}_{t=1}^{T}$ and $\{l_{i}\}_{i=1}^{N}$
in turn multiple times until the change in the criterion function
from reoptimization becomes negligible.

\noindent \textbf{Remark} (Other Sieve Spaces). We emphasize that
although we focus on the polynomial sieves, there exist many other
sieve spaces that could be used to approximate the function $h$ arbitrarily
well. For example, one could consider splines, wavelets, or neural
networks. Our theory in the next section could be extended to cover
these alternative sieve spaces but we have decided to focus on the
polynomial sieves for conciseness of the derivations.\qed

\section{Rate of Convergence\label{sec:rate of convergence}}

In this section, we derive the rate of convergence of our sieve-based
least squares estimator. To do so, we need the concept of the sub-Gaussian
norm, also known as the $\psi_{2}$ norm, which is defined as follows.
Let $\psi_{2}$ be the function defined by $\psi_{2}(x)=\exp(x^{2})$
for all $x\geq0$. Then for any random variable $X$, its sub-Gaussian
norm $\|X\|_{\psi_{2}}$ is the smallest constant $C>0$ such that
$E[\psi(|X|/C)]\leq1$. In addition, we impose the following assumptions.

\begin{assumption}[Idiosyncratic Noise]\label{as: idiosyncratic noise}
(i) Random variables $\varepsilon_{i,t}$ are mutually indepenent
across $i$ and $t$. (ii) $\max_{1\leq i\leq N}\max_{1\leq t\leq T}\|\varepsilon_{it}\|_{\psi_{2}}=O(1).$\end{assumption}

Assumption \ref{as: idiosyncratic noise}(i) is fairly standard and
can be relaxed to allow weak dependence in exchange for a more complicated
analysis. Assumption \ref{as: idiosyncratic noise}(ii) means that
the tails of the random variables $\varepsilon_{it}$ are not heavier
than those of the Gaussian random variables. This assumption is also
rather standard in the literature.

\begin{assumption}[Approximation Error]\label{as: approximation error}
There exist constants $\alpha>0$ and $L>0$ such that for each $K$,
there exist constants $h_{0}^{K},h_{1}^{K}\dots,h_{K}^{K}\in[-L,L]$
such that the polynomial $h_{K}(x)=\sum_{j=0}^{K}h_{j}^{K}x^{j}$,
$x\in[0,1]$, satisfies $\sup_{x\in[0,1]}|h_{K}(x)-h(x)|=O(K^{-\alpha})$.\end{assumption} 

This assumption means that the function $h$ can be approximated by
finite-degree polynomials and specifies how fast the approximation
error in the uniform metric converges to zero as we increase the degree
of the polynomials. 

In light of Assumption \ref{as: approximation error}, for the theoretical
analysis in this section, we assume that optimization in (\ref{eq: estimator})
over slope parameters $\{c_{j}\}_{j=0}^{K}$ is performed so that
each of these parameters remains bounded: $|c_{j}|\leq L$ for all
$j=0,\dots,K$ and the same constant $L$ as that specified in Assumption
\ref{as: approximation error}. In practice, we find that these constraints
are not binding if the constant $L$ is chosen large enough. 

The following theorem establishes the main result of this section.

\begin{theorem}[Rate of Convergence]\label{thm: consistency} Suppose
that Assumptions \ref{as: idiosyncratic noise} and \ref{as: approximation error}
are satisfied. Then
\begin{equation}
\frac{1}{NT}\sum_{i=1}^{N}\sum_{t=1}^{T}(\widehat{h}(\widehat{f}_{t}\widehat{\lambda}_{i})-h(f_{t}\lambda_{i}))^{2}=O_{P}\left(\frac{(T+N+K)\log(TNK)}{NT}+K^{-2\alpha}\right).\label{eq: convergence rate}
\end{equation}
\end{theorem}

The proof of this theorem is in Appendix \ref{sec:proof rate of convergence}.
The theorem shows that our sieve-based least-squares estimator is
consistent as long $K\to\infty$ together with $N$ and $T$ in such
a way that $K\log(TN)=o(NT).$ Following standard terminology of nonparametric
estimation, the first and the second terms on the right-hand side
of (\ref{eq: convergence rate}) represent the variance and the bias
terms, respectively. The variance term is increasing in the degree
of the polynomial sieve $K$ and the bias term is decreasing in $K$.
Moreover, if $K=o(\max(N,T))$, then it follows from the theorem that
\[
\frac{1}{NT}\sum_{i=1}^{N}\sum_{t=1}^{T}(\widehat{h}(\widehat{f}_{t}\widehat{\lambda}_{i})-h(f_{t}\lambda_{i}))^{2}=O_{P}\left(\frac{\log(NT)}{\min(N,T)}+K^{-2\alpha}\right).
\]
Thus, the theorem shows that the variance term converges to zero with
a fast rate even if we set $K$ to be rather large.\footnote{Note also that Theorem \ref{thm: consistency} derives the convergence
rate of the HFL estimator $\widehat{h}(\widehat{f}_{t}\widehat{\lambda}_{i})$
which we use for all of our empirical results. Another interesting
question is about convergence rates of the estimators $\widehat{f}_{t}$
and $\widehat{\lambda}_{i}$. It turns out that answering those questions
is more difficult. In fact, even deriving conditions under which $f_{t}$
and $\lambda_{i}$ are identified, modulo certain normalizations,
is highly non-trivial. In a separate paper (\citet*{Borri2024factor}),
we establish identification and consistency of $f_{t}$ and $\lambda_{i}$.}

\section{Data and Summary Statistics\label{sec:Data-Summary}}

We present a description of the data used for the estimation of the
HFL model. In order to estimate the HFL model and perform cross-sectional
asset pricing tests, we use data for a range of primary asset classes:
equities, bonds, commodities, and currencies. 

For equities, we use the \citet{fama1993common} 25 size and value
sorted portfolios and the \citet{fama2015five} 25 size and operating
profitability sorted portfolios. We also include the size, value and
operating profitability equity factors. These equity portfolios and
factors are from Ken French's data library. For bonds, we consider
U.S. bonds and international bonds. For U.S. bonds, we use eleven
maturity-sorted government bond portfolios from CRSP's ``Fama Bond
Portfolios'' file with maturities in six month intervals up to ten
years. For international bonds, we use the Refinitiv Datastream Government
Bond Indices for 8 advanced economies (Australia, Austria, Canada,
France, Germany, Japan, Netherlands and the United Kingdom). Specifically,
we consider bond indices with maturity 1-3 years, 2 years, 5 years,
7 years and 10 years. These bond indices include coupon payments and
bonds denominated in each country local currency. For commodities,
we use returns of 21 commodity spot price indices from Datastream.
The commodities are aluminum, Brent oil, cocoa, copper, corn, cotton,
crude oil, eggs, gasoline, gold, natural gas, oats, platinum, pork
bellies, propane, silver, soybean, sugar, tin, and wheat. For foreign
exchange, we use 46 portfolios from \citet*{nucera2024currency}.
These portfolios are based on popular investment strategies: carry,
short-term and long-term momentum, currency value, net foreign assets
and liabilities in domestic currencies, term spread, long-term yields,
and output gap.

The selection of advanced economies, for the government bond indices,
and commodities is guided by the objective of constructing a balanced
long sample, which in our case starts in January 1988. The sample
ends in December 2017, when the sample of foreign exchange portfolios
ends.

Table \ref{tab:Summary-Statistics} reports summary statistics of
the test assets and $h\left(f_{t}\lambda_{i}\right)$ estimated from
the sieve-based estimator. We report the mean and standard deviations
of the excess returns and $h\left(f_{t}\lambda_{i}\right)$ for the
whole sample and for each asset class. Returns are in excess of the
U.S. risk-free rate and reported in percentage. The ``All'' sample
includes 171 assets, each with 360 monthly return observations. The
average excess return is 0.32\% per month (3.84\% annualized). The
mean standard deviation is 4.46\% (15.44\% annualized). The average
excess return predicted by the HFL model across all assets exactly
matches the sample counterpart, with a monthly volatility which is
lower by one order of magnitude. The lower volatility is explained
by the fact that realized returns for assets include an idiosyncratic,
and volatile, component, which in our model is captured by the volatility
of the residuals. Equities and commodities, on average, have higher
and more volatile returns than other assets. Bonds, both U.S. and
international, have on average lower, and less volatile, returns.
Foreign exchange portfolios, in our sample, have average returns not
statistically different from zero. For the individual asset classes,
as for the whole sample of assets, the mean excess return predicted
by the HFL model is very close to the average asset excess return.

{\footnotesize{}}
\begin{table}[H]
{\footnotesize{}\caption{Summary Statistics of Test Assets\label{tab:Summary-Statistics}}
}{\footnotesize\par}

{\small{}\medskip{}
}{\small\par}

{\small{}This table reports summary statistics of the tests assets
and for $h\left(f_{t}\lambda_{i}\right)$. For the mean and the standard
deviation we report averages across assets. Returns are monthly and
in excess of the US risk-free rate, which we take from Ken French's
data library. The sample period is January 1988 to December 2017.}{\small\par}

{\footnotesize{}\bigskip{}
}{\footnotesize\par}
\centering{}{\small{}}%
\begin{tabular}{lcccccc}
\hline 
 &
All &
Equities &
US Bonds &
Intl Bonds &
Commodities &
FX\tabularnewline
\hline 
 &
 &
 &
 &
 &
 &
\tabularnewline
Mean ($R_{i,t}$, \%) &
0.320 &
0.796 &
0.156 &
0.217 &
0.312 &
-0.097\tabularnewline
Std ($R_{i,t}$, \%) &
4.467 &
5.210 &
0.819 &
1.137 &
8.810 &
2.397\tabularnewline
Mean ($h\left(f_{t}\lambda_{i}\right)$, \%) &
0.320 &
0.860 &
0.121 &
0.172 &
0.282 &
-0.110\tabularnewline
Std ($h\left(f_{t}\lambda_{i}\right)$, \%) &
0.464 &
0.347 &
0.029 &
0.084 &
0.398 &
0.186\tabularnewline
Assets &
171 &
53 &
11 &
40 &
21 &
46\tabularnewline
Months &
360 &
360 &
360 &
360 &
360 &
360\tabularnewline
 &
 &
 &
 &
 &
 &
\tabularnewline
\hline 
\end{tabular}{\small\par}
\end{table}
{\footnotesize\par}

\section{Cross-Sectional Asset Pricing Tests\label{sec:Cross-Section}}

\subsection{Baseline Results}

In this section, we consider cross-sectional asset pricing tests.
These tests assess whether the HFL model can account for cross-sectional
differences in average excess returns. The HFL model is

\begin{equation}
r_{it}=h(f_{t}\lambda_{i})+\varepsilon_{it}\quad i=1,\ldots,N\quad1,\ldots,T,\label{eq:KA_on_returns}
\end{equation}
where $r_{it}$ is the time $t$ excess return on asset $i$ and $\varepsilon_{it}$
is an idiosyncratic mean zero disturbance. The term $h(f_{t}\lambda_{i})$
includes the single factor $f_{t}$, common to all assets, and the
asset $i$ factor loading $\lambda_{i}$ which are combined by the
non-linear link function $h$ by Lemma 2.1. For each asset $i$, Equation
(\ref{eq:KA_on_returns}) implies $E[r_{it}]=E[h(f_{t}\lambda_{i})]$.
We empirically test this prediction by estimating the following linear
model

\begin{equation}
E[r_{it}]=\alpha+\beta E[h(f_{t}\lambda_{i})]+\epsilon_{i},\label{eq:Ka_on_returns_cross_section}
\end{equation}
where the expectation operator is replaced by the sample mean (i.e.,
$E[r_{it}]=1/T\sum_{t=1}^{T}r_{it}$), $h(f_{t}\lambda_{i})$'s are
replaced by the corresponding estimators $\widehat{h}(\widehat{f}_{t}\widehat{\lambda}_{i})$,
$\alpha$ and $\beta$ are the intercept and slope coefficients, and
$\epsilon$ is a mean zero residual term. Let $E[r_{it}]$ and $E[h(f_{t}\lambda_{i})]$
denote vectors containing, respectively, the averages of the excess
returns and predictions from the model.

In the empirical analysis we test several implications of our one
factor model, such as that the intercept is zero and that the slope
coefficient is equal to 1. Furthermore, we evaluate the fit of the
model with regression R-squared and mean absolute pricing error (MAPE).
The latter measures the mean absolute residuals in the cross-sectional
regression (\ref{eq:Ka_on_returns_cross_section}). Our empirical
strategy follows the classic two-step \citet{fama1973risk}'s methodology,
where the first step (i.e., the asset level time-series regressions
to retrieve the asset beta) is implicit in the algorithm to estimate
the HFL model. The second step, that is the cross-sectional regression
of mean asset returns on their betas, is replaced by the cross-sectional
regression (\ref{eq:Ka_on_returns_cross_section}).

\noindent \textbf{Remark.} In the Appendix, we demonstrate that unless
the HFL model is correct, the probability limits $\alpha_{0}$ and
$\beta_{0}$ of the OLS estimates $\widehat{\alpha}$ and $\widehat{\beta}$
of Equation (\ref{eq:KA_on_returns}) generally satisfy the restrictions
$\alpha_{0}=0$ and $\beta_{0}=1$ only if the factor $f_{t}$ is
constant over time. Since the latter does not hold in practice, it
follows that the OLS-based tests considered in this section are indeed
consistent.\qed

Column (1) of Table (\ref{tab:Cross-Sectional-Asset-Pricing}) reports
estimates of (\ref{eq:Ka_on_returns_cross_section}) using all assets
for the period of January 1988 to December 2017. We consider estimates
of the HFL model based on four degrees of the polynomial function
used to approximate the function $h(.)$.\footnote{We find that using higher order of polynomials generate qualitatively
similar results.} The table shows that the slope coefficient ($h\left(f_{t}\lambda_{i}\right)$)
is significantly different from zero at conventional confidence levels.
The pricing error, captured be the estimate of the intercept $\alpha$,
is not significantly different from zero. The model provides a good
fit with the adjusted $R^{2}$ of 89\% and the MAPE is small at 0.101\%. 

{\footnotesize{}}
\begin{table}[H]
{\footnotesize{}\caption{Cross-Sectional Asset Pricing Tests\label{tab:Cross-Sectional-Asset-Pricing}}
}{\footnotesize\par}

{\footnotesize{}\medskip{}
}{\footnotesize\par}

{\small{}This table presents the results of cross-sectional asset
pricing tests using all assets for the baseline HFL model, as well
as models augmented by, respectively, the US market excess return
(CAPM), the \citet{fama1993common}'s three factors (FF3), the \citet{fama2015five}'s
five factors (FF5), and the five factors augmented with the momentum
factor of \citet{jegadeesh1993returns} (FF5+MOM). We estimate the
cross-sectional linear model $E[r_{it}]=\alpha+\beta E[h(f_{t}\lambda_{i})]+\epsilon_{i}$
and report the estimates for the slope ($h\left(f_{t}\lambda_{i}\right)$)
and intercept ($\alpha$). Standard errors adjusted with the Fama-MacBeth
procedure are in parenthesis. The table additionally reports the regression
adjusted R-squared, and the mean absolute pricing error (MAPE) in
percentage terms. {*}, {*}{*}, and {*}{*}{*} represent significance
at the 1\%, 5\%, and 10\% level. We further report the number of assets
in the cross-sectional regression and the number of monthly observations
for each asset used in the estimation of the averages of the excess
returns and predictions from the model.}{\small\par}

{\footnotesize{}\bigskip{}
}{\footnotesize\par}
\centering{}{\small{}}%
\begin{tabular}{lccccc}
\hline 
 &
(1) &
(2) &
(3) &
(4) &
(5)\tabularnewline
\hline 
 &
Baseline &
CAPM &
FF3 &
FF5 &
FF5+MOM\tabularnewline
\hline 
 &
 &
 &
 &
 &
\tabularnewline
$h\left(f_{t}\lambda_{i}\right)$ &
0.816{*}{*}{*} &
0.715{*}{*}{*} &
0.811{*}{*}{*} &
0.823{*}{*}{*} &
0.821{*}{*}{*}\tabularnewline
 &
(0.177) &
(0.135) &
(0.154) &
(0.153) &
(0.169)\tabularnewline
MKTRF &
 &
0.147 &
-0.095 &
-0.106 &
-0.104\tabularnewline
 &
 &
(0.260) &
(0.259) &
(0.261) &
(0.281)\tabularnewline
SMB &
 &
 &
0.331{*} &
0.328{*} &
0.327{*}\tabularnewline
 &
 &
 &
(0.170) &
(0.168) &
(0.170)\tabularnewline
HML &
 &
 &
-0.105 &
-0.087 &
-0.085\tabularnewline
 &
 &
 &
(0.181) &
(0.177) &
(0.175)\tabularnewline
CMA &
 &
 &
 &
-0.163 &
-0.162\tabularnewline
 &
 &
 &
 &
(0.199) &
(0.203)\tabularnewline
RMW &
 &
 &
 &
-0.000 &
-0.001\tabularnewline
 &
 &
 &
 &
(0.217) &
(0.218)\tabularnewline
MOM &
 &
 &
 &
 &
0.111\tabularnewline
 &
 &
 &
 &
 &
(1.109)\tabularnewline
 &
 &
 &
 &
 &
\tabularnewline
$\alpha$ &
0.001 &
0.000 &
0.001 &
0.001 &
0.001\tabularnewline
 &
(0.001) &
(0.001) &
(0.001) &
(0.001) &
(0.001)\tabularnewline
Adj $R^{2}$ &
0.885 &
0.900 &
0.948 &
0.948 &
0.948\tabularnewline
MAPE, \% &
0.101 &
0.098 &
0.072 &
0.071 &
0.071\tabularnewline
Assets &
171 &
171 &
171 &
171 &
171\tabularnewline
Months &
360 &
360 &
360 &
360 &
360\tabularnewline
 &
 &
 &
 &
 &
\tabularnewline
\hline 
\end{tabular}{\small\par}
\end{table}
{\footnotesize\par}

The good fit of the model is visually confirmed by Figure \ref{fig:Actual-vs-Predicted_ALL},
which plots the average asset excess returns versus the predicted
excess returns using the HFL model. All assets line up along the 45
degree line. Figure \ref{fig:Actual-vs-Predicted_no_intercept} in
the Appendix reveals that this result is robust to the exclusion of
the intercept from the model used to predict average excess returns.

We compare the HFL model with the standard capital asset pricing model
(CAPM) of \citet{sharpe1964capital} and \citet{lintner1965security},
the \citet{fama1993common}'s three factor model, the \citet{fama2015five}'s
five factors model, and the five factors model augmented with the
momentum factor of \citet{jegadeesh1993returns}. 

Columns (2) to (5) of Table (\ref{tab:Cross-Sectional-Asset-Pricing})
present the results of cross-sectional asset pricing tests using all
assets. In all models, we include the HFL component ($h\left(f_{t}\lambda_{i}\right)$),
and evaluate the effect of including additional factors. First, the
slope estimate associated with the HFL component is always significantly
different from zero at the 1\% confidence level. Second, including
the additional factors increases the adjusted R-squareds only marginally,
from 89\% to a maximum of 95\%. In fact, with the exception of the
slope coefficient associated with the size factor, which is significant
at the 10\% level, the slope coefficients associated with all the
other factors are never statistically different from zero. 

\subsection{Comparison with PCA factor models}

In this section, we compare the HFL model with factor models based
on principal component analysis (see, e.g., \citet{connor1986performance,connor1988risk}).
Table \ref{tab:Comparison-with-PCA} presents the results of cross-sectional
asset pricing tests using all assets and three estimators of the latent
factors related to PCA. The first estimator is based on the standard
PCA of the covariance matrix of asset returns. The second estimator,
called RP-PCA, is a generalization of PCA, proposed by \citet{lettau2020factors},
which includes a penalty term to account for the pricing errors in
the cross-sectional regressions based on the RP-PCA factors. RP-PCA
is motivated by the poor performance of standard PCA in identifying
factors which are relevant to explain the cross-section of average
excess returns (see, e.g., \citet{onatski2012asymptotics} and \citet*{kozak2020shrinking}).
The third estimator is the kernel-PCA, proposed by \citet*{scholkopf1998nonlinear}.
Kernel-PCA (K-PCA) is a nonlinear form of PCA which allows for the
separability of nonlinear data by projecting it onto a higher dimensional
space where it becomes linearly separable using kernels.
\begin{center}
\begin{figure}[H]
\caption{Actual vs Predicted Expected Returns\label{fig:Actual-vs-Predicted_ALL}}

{\footnotesize{}\medskip{}
}{\footnotesize\par}

{\small{}The figure plots actual average excess returns on all tested
assets ($E[r_{it}]$) versus predicted excess returns using the cross-sectional
model based on the HFL model: $E[r_{it}]=\alpha+\beta E[h(\lambda_{i}f_{t})]+\epsilon_{i}$.
The HFL model is based on a degree of the polynomial used to approximate
the function $h(.)$ equal to 4.}{\small\par}

{\footnotesize{}\medskip{}
}{\footnotesize\par}
\centering{}\includegraphics[width=15cm]{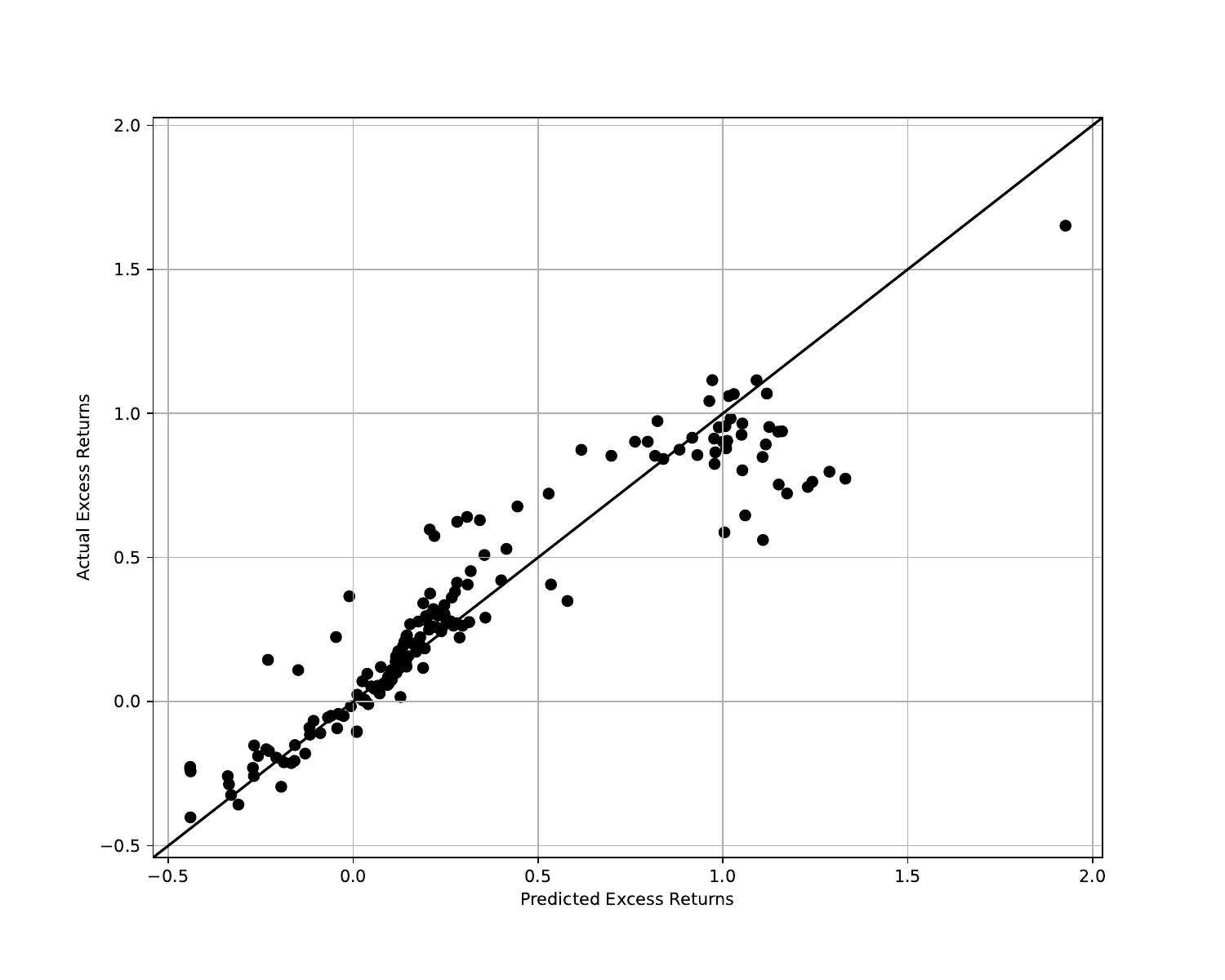}
\end{figure}
\par\end{center}

In all models, we include the HFL component ($h\left(f_{t}\lambda_{i}\right)$),
and evaluate the effect of including, respectively, the first principal
component, the first two principal components, and the first three
principal components, separately for standard PCA, RP-PCA and K-PCA.
The estimates from the cross-sectional asset pricing regressions reveal
that the slope coefficient associated with the HFL component is statistically
significant at conventional levels in all models, while the slopes
associated with the principal components, for all three estimators
we consider, are never statically different from zero. Furthermore,
including the principal components increases only marginally the cross-sectional
adjusted R-squared, from 89\% to a maximum of 91\%, while the pricing
error $\alpha$ is never statistically significant at conventional
condidence levels.

Table \ref{tab:Comparison-with-PCA-higher} in the Appendix reveals
similar conclusions for asset pricing estimates of PCA models which
include up to 6 factors. These estimates confirm the robustness of
the results based on PCA models including the first three components.

\subsection{Comparison with higher order factor models}

In this section, we compare the HFL model with higher order factor
models, related to the non-linear version of the CAPM of \citet{kraus1976skewness}.
\citet{harvey2000conditional} develop a three-moment conditional
CAPM where skewness is priced. \citet*{ang2006cross} develop a model
where aggregate market volatility is priced. Table \ref{tab:Comparison-with-Higher-Orders}
presents the results of cross-sectional asset pricing tests using
all assets. In all models, we include the HFL component ($h\left(f_{t}\lambda_{i}\right)$),
and evaluate the effect of including higher order factors. Specifically,
we consider both a model specification which additionally includes
the U.S. equity market excess returns, the squared U.S. equity market
excess returns, and the cubed U.S. equity market excess returns, and
the second model specification which additionally includes the first
principal component, the squared first principal component, and the
cubed first principal component. For the principal components, we
consider three estimators: the standard PCA based on the covariance
matrix of asset returns; the risk premium PCA (RP-PCA) of \citet{lettau2020factors},
and the kernel PCA (K-PCA) of \citet*{scholkopf1998nonlinear}. In
all modes, the slope coefficient associated to the HFL component is
statistically significant at conventional confidence levels, while
the slopes associated with the higher order factors are never statistically
different from zero. Furthermore, including the principal components
increases only marginally the cross-sectional R-squared, from 89\%
to a maximum of 92\%. Moreover, the pricing error $\alpha$ is not
statistically different from zero in all models, with the exception
of the CAPM model augmented with the equity market factor squared
and cubed (CAPM$^{3}$) for which the estimate for the intercept is
statistically significant at the 10\% level.

{\footnotesize{}}
\begin{sidewaystable}[H]
{\footnotesize{}\caption{Comparison with PCA Factor Models\label{tab:Comparison-with-PCA}}
}{\footnotesize\par}

{\footnotesize{}\medskip{}
}{\footnotesize\par}

{\small{}This table presents the results of cross-sectional asset
pricing tests using all assets. We estimate the cross-sectional linear
model $E[r_{it}]=\alpha+\beta E[h(f_{t}\lambda_{i})]+\epsilon_{i}$
augmented by, respectively, the first principal component (PC1), the
first two principal components (PC1+PC2) and the first three principal
components extracted by the test asset returns. For the principal
components, we consider three estimators: the standard PCA based on
the covariance matrix of asset returns; the risk premium PCA (RP-PCA)
of \citet{lettau2020factors}, and the kernel PCA (K-PCA) of \citet*{scholkopf1998nonlinear}.
We report the estimates for the slopes and intercept ($\alpha$),
and standard errors adjusted with the Fama-MacBeth procedure in parenthesis.
The table additionally reports the regression adjusted R-squared,
and the mean absolute pricing error (MAPE) in percentage terms. {*},
{*}{*}, and {*}{*}{*} represent significance at the 1\%, 5\%, and
10\% level. We further report the number of assets in the cross-sectional
regression and the number of monthly observations for each asset used
in the estimation of the averages of the excess returns and predictions
from the model. The HFL model is based on a degree of the polynomial
used to approximate the function $h(.)$ equal to 4.}{\small\par}

{\footnotesize{}\bigskip{}
}{\footnotesize\par}
\centering{}{\small{}}%
\begin{tabular}{lccccccccccc}
\hline 
 &
\multicolumn{3}{c}{Standard PCA} &
 &
\multicolumn{3}{c}{RP-PCA} &
 &
\multicolumn{3}{c}{K-PCA}\tabularnewline
\hline 
 &
PC1 &
PC1 + PC2 &
PC1--PC3 &
 &
PC1 &
PC1 + PC2 &
PC1--PC3 &
 &
PC1 &
PC1 + PC2 &
PC1--PC3\tabularnewline
$h\left(f_{t}\lambda_{i}\right)$ &
0.705{*}{*}{*} &
0.696{*}{*}{*} &
0.654{*}{*}{*} &
 &
0.692{*}{*}{*} &
0.656{*}{*}{*} &
0.635{*}{*}{*} &
 &
0.807{*}{*}{*} &
0.804{*}{*}{*} &
0.777{*}{*}{*}\tabularnewline
 &
(0.129) &
(0.120) &
(0.109) &
 &
(0.129) &
(0.117) &
(0.113) &
 &
(0.174) &
(0.176) &
(0.169)\tabularnewline
PC1 &
0.014 &
0.014 &
0.015 &
 &
0.014 &
0.016 &
0.016 &
 &
0.012 &
0.010 &
0.009\tabularnewline
 &
(0.020) &
(0.019) &
(0.019) &
 &
(0.020) &
(0.019) &
(0.019) &
 &
(0.018) &
(0.017) &
(0.017)\tabularnewline
PC2 &
 &
0.001 &
0.001 &
 &
 &
0.001 &
0.002 &
 &
 &
0.004 &
0.007\tabularnewline
 &
 &
(0.010) &
(0.010) &
 &
 &
(0.010) &
(0.010) &
 &
 &
(0.013) &
(0.014)\tabularnewline
PC3 &
 &
 &
-0.005 &
 &
 &
 &
0.003 &
 &
 &
 &
-0.008\tabularnewline
 &
 &
 &
(0.010) &
 &
 &
 &
(0.010) &
 &
 &
 &
(0.014)\tabularnewline
$\alpha$ &
0.000 &
0.000 &
0.000 &
 &
0.000 &
0.000 &
0.001 &
 &
0.001 &
0.001 &
0.001\tabularnewline
 &
(0.001) &
(0.001) &
(0.000) &
 &
(0.001) &
(0.001) &
(0.000) &
 &
(0.001) &
(0.001) &
(0.001)\tabularnewline
Adj $R^{2}$ &
0.907 &
0.907 &
0.912 &
 &
0.908 &
0.911 &
0.914 &
 &
0.888 &
0.888 &
0.895\tabularnewline
MAPE, \% &
0.096 &
0.097 &
0.090 &
 &
0.095 &
0.094 &
0.089 &
 &
0.103 &
0.103 &
0.104\tabularnewline
Assets &
171 &
171 &
171 &
 &
171 &
171 &
171 &
 &
171 &
171 &
171\tabularnewline
Months &
360 &
360 &
360 &
 &
360 &
360 &
360 &
 &
360 &
360 &
360\tabularnewline
 &
 &
 &
 &
 &
 &
 &
 &
 &
 &
 &
\tabularnewline
\hline 
\end{tabular}{\small\par}
\end{sidewaystable}
{\footnotesize\par}

{\scriptsize{}}
\begin{table}[H]
{\scriptsize{}\caption{Comparison with Higher Order Factor Models\label{tab:Comparison-with-Higher-Orders}}
}{\scriptsize\par}

{\scriptsize{}\medskip{}
}{\scriptsize\par}

{\footnotesize{}This table presents the results of cross-sectional
asset pricing tests using all assets. We estimate the cross-sectional
linear model $E_{t}[r_{it}]=\alpha+\beta E_{i}[h(f_{t}\lambda_{i})]+\epsilon_{i}$
augmented by the US market excess return and, respectively, by the
squared market excess return (CAPM$^{2}$) and the cubed market excess
return (CAPM$^{3}$); or augmented by the first principal component
and, respectively, the squared first principal component (PC1$^{2}$)
and the cubed first principal component (PC1$^{3}$). For the principal
components, we consider three estimators: the standard PCA based on
the covariance matrix of asset returns; the risk premium PCA (RP-PCA)
of \citet{lettau2020factors}, and the kernel PCA (K-PCA) of \citet*{scholkopf1998nonlinear}.
We report the estimates for the slopes and intercept ($\alpha$),
and standard errors adjusted with the Fama-MacBeth procedure in parenthesis.
The table additionally reports the regression adjusted R-squared,
and the mean absolute pricing error (MAPE) in percentage terms. {*},
{*}{*}, and {*}{*}{*} represent significance at the 1\%, 5\%, and
10\% level. We further report the number of assets in the cross-sectional
regression and the number of monthly observations for each asset used
in the estimation of the averages of the excess returns and predictions
from the model. The HFL model is based on a degree of the polynomial
used to approximate the function $h(.)$ equal to 4.}{\footnotesize\par}

{\scriptsize{}\medskip{}
}{\scriptsize\par}
\centering{}{\scriptsize{}}%
\begin{tabular}{lcccccccc}
\hline 
 &
{\scriptsize{}CAPM$^{2}$} &
{\scriptsize{}CAPM$^{3}$} &
{\scriptsize{}PC1$^{2}$} &
{\scriptsize{}PC1$^{3}$} &
{\scriptsize{}RP-PC1$^{2}$} &
{\scriptsize{}RP-PC1$^{3}$} &
{\scriptsize{}K-PCA1$^{2}$} &
{\scriptsize{}K-PCA1$^{3}$}\tabularnewline
\hline 
{\scriptsize{}$h\left(f_{t}\lambda_{i}\right)$} &
{\scriptsize{}0.715{*}{*}{*}} &
{\scriptsize{}0.711{*}{*}{*}} &
{\scriptsize{}0.744{*}{*}{*}} &
{\scriptsize{}0.749{*}{*}{*}} &
{\scriptsize{}0.725{*}{*}{*}} &
{\scriptsize{}0.726{*}{*}{*}} &
{\scriptsize{}0.793{*}{*}{*}} &
{\scriptsize{}0.672{*}{*}{*}}\tabularnewline
 &
{\scriptsize{}(0.135)} &
{\scriptsize{}(0.135)} &
{\scriptsize{}(0.141)} &
{\scriptsize{}(0.131)} &
{\scriptsize{}(0.139)} &
{\scriptsize{}(0.132)} &
{\scriptsize{}(0.167)} &
{\scriptsize{}(0.139)}\tabularnewline
{\scriptsize{}MKTRF} &
{\scriptsize{}0.126} &
{\scriptsize{}0.084} &
 &
 &
 &
 &
 &
\tabularnewline
 &
{\scriptsize{}(0.262)} &
{\scriptsize{}(0.257)} &
 &
 &
 &
 &
 &
\tabularnewline
{\scriptsize{}MKTRF$^{2}$} &
{\scriptsize{}-5.220} &
{\scriptsize{}-9.051} &
 &
 &
 &
 &
 &
\tabularnewline
 &
{\scriptsize{}(7.854)} &
{\scriptsize{}(7.146)} &
 &
 &
 &
 &
 &
\tabularnewline
{\scriptsize{}MKTRF$^{3}$} &
 &
{\scriptsize{}8.949} &
 &
 &
 &
 &
 &
\tabularnewline
 &
 &
{\scriptsize{}(62.955)} &
 &
 &
 &
 &
 &
\tabularnewline
{\scriptsize{}PC1} &
 &
 &
{\scriptsize{}0.010} &
{\scriptsize{}0.010} &
 &
 &
 &
\tabularnewline
 &
 &
 &
{\scriptsize{}(0.020)} &
{\scriptsize{}(0.020)} &
 &
 &
 &
\tabularnewline
{\scriptsize{}PC1$^{2}$} &
 &
 &
{\scriptsize{}0.046} &
{\scriptsize{}0.050} &
 &
 &
 &
\tabularnewline
 &
 &
 &
{\scriptsize{}(0.060)} &
{\scriptsize{}(0.046)} &
 &
 &
 &
\tabularnewline
{\scriptsize{}PC1$^{3}$} &
 &
 &
 &
{\scriptsize{}-0.020} &
 &
 &
 &
\tabularnewline
 &
 &
 &
 &
{\scriptsize{}(0.045)} &
 &
 &
 &
\tabularnewline
{\scriptsize{}RP-PC1} &
 &
 &
 &
 &
{\scriptsize{}0.011} &
{\scriptsize{}0.011} &
 &
\tabularnewline
 &
 &
 &
 &
 &
{\scriptsize{}(0.020)} &
{\scriptsize{}(0.019)} &
 &
\tabularnewline
{\scriptsize{}RP-PC1$^{2}$} &
 &
 &
 &
 &
{\scriptsize{}0.045} &
{\scriptsize{}0.045} &
 &
\tabularnewline
 &
 &
 &
 &
 &
{\scriptsize{}(0.056)} &
{\scriptsize{}(0.044)} &
 &
\tabularnewline
{\scriptsize{}RP-PC1$^{3}$} &
 &
 &
 &
 &
 &
{\scriptsize{}-0.020} &
 &
\tabularnewline
 &
 &
 &
 &
 &
 &
{\scriptsize{}(0.041)} &
 &
\tabularnewline
{\scriptsize{}KPC1} &
 &
 &
 &
 &
 &
 &
{\scriptsize{}0.003} &
{\scriptsize{}0.001}\tabularnewline
 &
 &
 &
 &
 &
 &
 &
{\scriptsize{}(0.021)} &
{\scriptsize{}(0.021)}\tabularnewline
{\scriptsize{}KPC1$^{2}$} &
 &
 &
 &
 &
 &
 &
{\scriptsize{}0.006} &
{\scriptsize{}0.007}\tabularnewline
 &
 &
 &
 &
 &
 &
 &
{\scriptsize{}(0.011)} &
{\scriptsize{}(0.011)}\tabularnewline
{\scriptsize{}KPC1$^{3}$} &
 &
 &
 &
 &
 &
 &
 &
{\scriptsize{}-0.005}\tabularnewline
 &
 &
 &
 &
 &
 &
 &
 &
{\scriptsize{}(0.007)}\tabularnewline
{\scriptsize{}$\alpha$} &
{\scriptsize{}0.000} &
{\scriptsize{}0.001{*}} &
{\scriptsize{}0.000} &
{\scriptsize{}0.000} &
{\scriptsize{}0.000} &
{\scriptsize{}0.000} &
{\scriptsize{}0.001} &
{\scriptsize{}0.001}\tabularnewline
 &
{\scriptsize{}(0.001)} &
{\scriptsize{}(0.000)} &
{\scriptsize{}(0.001)} &
{\scriptsize{}(0.000)} &
{\scriptsize{}(0.001)} &
{\scriptsize{}(0.000)} &
{\scriptsize{}(0.001)} &
{\scriptsize{}(0.001)}\tabularnewline
{\scriptsize{}Adj $R^{2}$} &
{\scriptsize{}0.905} &
{\scriptsize{}0.921} &
{\scriptsize{}0.915} &
{\scriptsize{}0.915} &
{\scriptsize{}0.917} &
{\scriptsize{}0.917} &
{\scriptsize{}0.897} &
{\scriptsize{}0.923}\tabularnewline
{\scriptsize{}MAPE, \%} &
{\scriptsize{}0.098} &
{\scriptsize{}0.086} &
{\scriptsize{}0.088} &
{\scriptsize{}0.089} &
{\scriptsize{}0.103} &
{\scriptsize{}0.104} &
{\scriptsize{}0.096} &
{\scriptsize{}0.086}\tabularnewline
{\scriptsize{}Assets} &
{\scriptsize{}171} &
{\scriptsize{}171} &
{\scriptsize{}171} &
{\scriptsize{}171} &
{\scriptsize{}171} &
{\scriptsize{}171} &
{\scriptsize{}171} &
{\scriptsize{}171}\tabularnewline
{\scriptsize{}Months} &
{\scriptsize{}360} &
{\scriptsize{}360} &
{\scriptsize{}360} &
{\scriptsize{}360} &
{\scriptsize{}360} &
{\scriptsize{}360} &
{\scriptsize{}360} &
{\scriptsize{}360}\tabularnewline
\hline 
\end{tabular}{\scriptsize\par}
\end{table}
{\scriptsize\par}

\subsection{Comparison with recent macro-factor models}

In this section, we compare the HFL model with factors from recent
recent macro-factor models. Specifically, we consider \citet*{he2017intermediary}'s
intermediary capital risk factor (HKM), the \citet*{lettau2014conditional}'s
downside equity market risk factor (LMW), and \citet{pastor2003liquidity}'s
liquidity risk factor (PS). Both the HKM and LMW factors are developed
to investigate cross-sectional differences in average returns across
large cross-section of assets from different asset classes. The PS
factor is developed to investigate cross-sectional differences in
equity average returns. In all cases, we compare the HFL model with
single-index models based on the macro factors.\footnote{For the HKM and PS factors, we rely on the data provided by Assaf
Manela and Lubos Pastor, respectively. We construct the LMW factor,
for the period of analysis, following \citet*{lettau2014conditional}.} 

Table \ref{tab:Comparison-with-Recent-Macro-Factor-Models} presents
the results of cross-sectional asset pricing tests using all assets.
In all models, we include the HFL component ($h\left(f_{t}\lambda_{i}\right)$),
and evaluate the effect of including, respectively, the HKM, LMW and
PS factors. In all modes, the slope coefficient associated to the
HFL component is statistically significant at conventional levels,
while the slopes associated with the macro-factor models are never
statistically different from zero. Furthermore, including the macro-factors
increases only marginally the cross-sectional adjusted R-squared,
from 89\% to a maximum of 91\%. The pricing error $\alpha$ is not
statistically different from zero in all specifications.

{\footnotesize{}}
\begin{table}[H]
{\footnotesize{}\caption{Comparison with Recent Macro Factor Models\label{tab:Comparison-with-Recent-Macro-Factor-Models}}
}{\footnotesize\par}

{\footnotesize{}\medskip{}
}{\footnotesize\par}

{\small{}This table presents the results of cross-sectional asset
pricing tests using all assets. We estimate the cross-sectional linear
model $E[r_{it}]=\alpha+\beta E[h(f_{t}\lambda_{i})]+\epsilon_{i}$
augmented by \citet*{he2017intermediary}'s intermediary capital risk
factor (HKM), \citet*{lettau2014conditional}'s downside equity market
risk factor (LMW), and \citet{pastor2003liquidity}'s liquidity risk
factor (PS). We report the estimates for the slopes and intercept
($\alpha$), and standard errors adjusted with the Fama-MacBeth procedure
in parenthesis. The table additionally reports the regression adjusted
R-squared, and the mean absolute pricing error (MAPE) in percentage
terms. {*}, {*}{*}, and {*}{*}{*} represent significance at the 1\%,
5\%, and 10\% level. We further report the number of assets in the
cross-sectional regression and the number of monthly observations
for each asset used in the estimation of the averages of the excess
returns and predictions from the model. The HFL model is based on
a degree of the polynomial used to approximate the function $h(.)$
equal to 4.}{\small\par}

{\footnotesize{}\bigskip{}
}{\footnotesize\par}
\centering{}{\small{}}%
\begin{tabular}{llccccc}
\hline 
 &
 &
HKM &
 &
LMW &
 &
PS\tabularnewline
\hline 
 &
 &
 &
 &
 &
 &
\tabularnewline
$h\left(f_{t}\lambda_{i}\right)$ &
 &
0.789{*}{*}{*} &
 &
0.714{*}{*}{*} &
 &
0.715{*}{*}{*}\tabularnewline
 &
 &
(0.152) &
 &
(0.135) &
 &
(0.141)\tabularnewline
MKTRF &
 &
0.098 &
 &
0.132 &
 &
0.147\tabularnewline
 &
 &
(0.265) &
 &
(0.357) &
 &
(0.261)\tabularnewline
HKM &
 &
-0.006 &
 &
 &
 &
\tabularnewline
 &
 &
(0.007) &
 &
 &
 &
\tabularnewline
LMW &
 &
 &
 &
0.016 &
 &
\tabularnewline
 &
 &
 &
 &
(0.232) &
 &
\tabularnewline
PS &
 &
 &
 &
 &
 &
0.000\tabularnewline
 &
 &
 &
 &
 &
 &
(0.015)\tabularnewline
$\alpha$ &
 &
0.000 &
 &
0.000 &
 &
0.000\tabularnewline
 &
 &
(0.001) &
 &
(0.001) &
 &
(0.001)\tabularnewline
Adj $R^{2}$ &
 &
0.918 &
 &
0.899 &
 &
0.899\tabularnewline
MAPE, \% &
 &
0.089 &
 &
0.099 &
 &
0.098\tabularnewline
Assets &
 &
171 &
 &
171 &
 &
171\tabularnewline
Months &
 &
360 &
 &
360 &
 &
360\tabularnewline
 &
 &
 &
 &
 &
 &
\tabularnewline
\hline 
\end{tabular}{\small\par}
\end{table}
{\footnotesize\par}

\begin{sidewaystable}[H]
\caption{Comparison with macro factors of \citet{ludvigson2009macro}\label{tab:Comparison-with-macro-LN}}

{\small{}\medskip{}
}{\small\par}

\begin{singlespace}
{\small{}This table presents the results of cross-sectional asset
pricing tests using all assets. We estimate the cross-sectional linear
model $E_{i}[r_{it}]=\alpha+\beta E_{i}[h(f_{t}\lambda_{i})]+\epsilon_{i}$
augmented by the 9 macro factors of \citet{ludvigson2009macro}. These
factors, denoted by $F_{LN}$, are eight macro factors extracted from
a sample of US government bonds plus the first factor cubed. We report
the estimates for the slopes and intercept ($\alpha$), and standard
errors adjusted with the Fama-MacBeth procedure in parenthesis. The
table additionally reports the regression adjusted R-squared. {*},
{*}{*}, and {*}{*}{*} represent significance at the 1\%, 5\%, and
10\% level. We further report the number of assets in the cross-sectional
regression and the number of monthly observations for each asset used
in the estimation of the averages of the excess returns and predictions
from the model. The HFL model is based on a degree of the polynomial
used to approximate the function $h(.)$ equal to 4.}{\small\par}
\end{singlespace}
\centering{}{\scriptsize{}\medskip{}
}{\tiny{}}%
\begin{tabular}{lcccccccccccccccccc}
\hline 
 &
\multicolumn{9}{c}{{\tiny{}Without HFL}} &
\multicolumn{9}{c}{{\tiny{}With HFL}}\tabularnewline
\hline 
 &
{\tiny{}$F_{LN,1}$} &
{\tiny{}$F_{LN,1}^{3}$} &
{\tiny{}$F_{LN,2}$} &
{\tiny{}$F_{LN,3}$} &
{\tiny{}$F_{LN,4}$} &
{\tiny{}$F_{LN,5}$} &
{\tiny{}$F_{LN6}$} &
{\tiny{}$F_{LN7}$} &
{\tiny{}$F_{LN8}$} &
{\tiny{}$F_{LN,1}$} &
{\tiny{}$F_{LN,1}^{3}$} &
{\tiny{}$F_{LN,2}$} &
{\tiny{}$F_{LN,3}$} &
{\tiny{}$F_{LN,4}$} &
{\tiny{}$F_{LN,5}$} &
{\tiny{}$F_{LN6}$} &
{\tiny{}$F_{LN7}$} &
{\tiny{}$F_{LN8}$}\tabularnewline
\hline 
{\tiny{}$F_{LN,1}$} &
{\tiny{}-0.125{*}} &
 &
 &
 &
 &
 &
 &
 &
 &
{\tiny{}-0.007} &
 &
 &
 &
 &
 &
 &
 &
\tabularnewline
 &
{\tiny{}(-1.830)} &
 &
 &
 &
 &
 &
 &
 &
 &
{\tiny{}(-0.118)} &
 &
 &
 &
 &
 &
 &
 &
\tabularnewline
{\tiny{}$F_{LN,1}^{3}$} &
 &
{\tiny{}-0.289{*}{*}} &
 &
 &
 &
 &
 &
 &
 &
 &
{\tiny{}-0.011} &
 &
 &
 &
 &
 &
 &
\tabularnewline
 &
 &
{\tiny{}(-2.469)} &
 &
 &
 &
 &
 &
 &
 &
 &
{\tiny{}(-0.113)} &
 &
 &
 &
 &
 &
 &
\tabularnewline
{\tiny{}$F_{LN,2}$} &
 &
 &
{\tiny{}0.049} &
 &
 &
 &
 &
 &
 &
 &
 &
{\tiny{}0.023} &
 &
 &
 &
 &
 &
\tabularnewline
 &
 &
 &
{\tiny{}(1.078)} &
 &
 &
 &
 &
 &
 &
 &
 &
{\tiny{}(0.536)} &
 &
 &
 &
 &
 &
\tabularnewline
{\tiny{}$F_{LN,3}$} &
 &
 &
 &
{\tiny{}0.062} &
 &
 &
 &
 &
 &
 &
 &
 &
{\tiny{}-0.002} &
 &
 &
 &
 &
\tabularnewline
 &
 &
 &
 &
{\tiny{}(1.621)} &
 &
 &
 &
 &
 &
 &
 &
 &
{\tiny{}(-0.042)} &
 &
 &
 &
 &
\tabularnewline
{\tiny{}$F_{LN,4}$} &
 &
 &
 &
 &
{\tiny{}-0.103{*}{*}} &
 &
 &
 &
 &
 &
 &
 &
 &
{\tiny{}-0.008} &
 &
 &
 &
\tabularnewline
 &
 &
 &
 &
 &
{\tiny{}(-2.128)} &
 &
 &
 &
 &
 &
 &
 &
 &
{\tiny{}(-0.200)} &
 &
 &
 &
\tabularnewline
{\tiny{}$F_{LN,5}$} &
 &
 &
 &
 &
 &
{\tiny{}0.037} &
 &
 &
 &
 &
 &
 &
 &
 &
{\tiny{}-0.013} &
 &
 &
\tabularnewline
 &
 &
 &
 &
 &
 &
{\tiny{}(0.989)} &
 &
 &
 &
 &
 &
 &
 &
 &
{\tiny{}(-0.324)} &
 &
 &
\tabularnewline
{\tiny{}$F_{LN6}$} &
 &
 &
 &
 &
 &
 &
{\tiny{}0.166{*}{*}{*}} &
 &
 &
 &
 &
 &
 &
 &
 &
{\tiny{}0.056} &
 &
\tabularnewline
 &
 &
 &
 &
 &
 &
 &
{\tiny{}(3.560)} &
 &
 &
 &
 &
 &
 &
 &
 &
{\tiny{}(1.524)} &
 &
\tabularnewline
{\tiny{}$F_{LN7}$} &
 &
 &
 &
 &
 &
 &
 &
{\tiny{}0.046{*}{*}{*}} &
 &
 &
 &
 &
 &
 &
 &
 &
{\tiny{}0.012} &
\tabularnewline
 &
 &
 &
 &
 &
 &
 &
 &
{\tiny{}(2.727)} &
 &
 &
 &
 &
 &
 &
 &
 &
{\tiny{}(0.648)} &
\tabularnewline
{\tiny{}$F_{LN8}$} &
 &
 &
 &
 &
 &
 &
 &
 &
{\tiny{}-0.047{*}{*}{*}} &
 &
 &
 &
 &
 &
 &
 &
 &
{\tiny{}-0.006}\tabularnewline
 &
 &
 &
 &
 &
 &
 &
 &
 &
{\tiny{}(-2.939)} &
 &
 &
 &
 &
 &
 &
 &
 &
{\tiny{}(-0.389)}\tabularnewline
{\tiny{}$h\left(f_{t}\lambda_{i}\right)$} &
 &
 &
 &
 &
 &
 &
 &
 &
 &
{\tiny{}0.810{*}{*}{*}} &
{\tiny{}0.808{*}{*}{*}} &
{\tiny{}0.806{*}{*}{*}} &
{\tiny{}0.817{*}{*}{*}} &
{\tiny{}0.801{*}{*}{*}} &
{\tiny{}0.825{*}{*}{*}} &
{\tiny{}0.774{*}{*}{*}} &
{\tiny{}0.701{*}{*}{*}} &
{\tiny{}0.794{*}{*}{*}}\tabularnewline
 &
 &
 &
 &
 &
 &
 &
 &
 &
 &
{\tiny{}(4.890)} &
{\tiny{}(5.008)} &
{\tiny{}(4.732)} &
{\tiny{}(4.635)} &
{\tiny{}(5.503)} &
{\tiny{}(4.440)} &
{\tiny{}(4.505)} &
{\tiny{}(5.230)} &
{\tiny{}(4.170)}\tabularnewline
{\tiny{}$\alpha$} &
{\tiny{}0.003{*}{*}{*}} &
{\tiny{}0.002{*}{*}{*}} &
{\tiny{}0.002{*}{*}{*}} &
{\tiny{}0.003{*}{*}{*}} &
{\tiny{}0.001{*}} &
{\tiny{}0.004{*}{*}{*}} &
{\tiny{}0.005{*}{*}{*}} &
{\tiny{}0.001{*}{*}} &
{\tiny{}0.005{*}{*}{*}} &
{\tiny{}0.001} &
{\tiny{}0.001} &
{\tiny{}0.000} &
{\tiny{}0.001} &
{\tiny{}0.000} &
{\tiny{}0.000} &
{\tiny{}0.001} &
{\tiny{}0.000} &
{\tiny{}0.001{*}}\tabularnewline
 &
{\tiny{}(3.473)} &
{\tiny{}(3.246)} &
{\tiny{}(4.026)} &
{\tiny{}(3.388)} &
{\tiny{}(1.775)} &
{\tiny{}(3.991)} &
{\tiny{}(3.865)} &
{\tiny{}(2.032)} &
{\tiny{}(4.555)} &
{\tiny{}(0.854)} &
{\tiny{}(0.859)} &
{\tiny{}(0.364)} &
{\tiny{}(0.886)} &
{\tiny{}(0.686)} &
{\tiny{}(0.851)} &
{\tiny{}(1.562)} &
{\tiny{}(0.700)} &
{\tiny{}(1.770)}\tabularnewline
{\tiny{}Adj $R^{2}$} &
{\tiny{}0.107} &
{\tiny{}0.213} &
{\tiny{}0.036} &
{\tiny{}0.058} &
{\tiny{}0.213} &
{\tiny{}0.016} &
{\tiny{}0.189} &
{\tiny{}0.559} &
{\tiny{}0.192} &
{\tiny{}0.885} &
{\tiny{}0.885} &
{\tiny{}0.894} &
{\tiny{}0.885} &
{\tiny{}0.886} &
{\tiny{}0.887} &
{\tiny{}0.904} &
{\tiny{}0.904} &
{\tiny{}0.888}\tabularnewline
{\tiny{}Months} &
{\tiny{}360} &
{\tiny{}360} &
{\tiny{}360} &
{\tiny{}360} &
{\tiny{}360} &
{\tiny{}360} &
{\tiny{}360} &
{\tiny{}360} &
{\tiny{}360} &
{\tiny{}360} &
{\tiny{}360} &
{\tiny{}360} &
{\tiny{}360} &
{\tiny{}360} &
{\tiny{}360} &
{\tiny{}360} &
{\tiny{}360} &
{\tiny{}360}\tabularnewline
\hline 
\end{tabular}{\tiny\par}
\end{sidewaystable}

\subsection{Comparison with macro factors of \citet{ludvigson2009macro}}

In this section, we compare the HFL model with nine macro factors
proposed by \citet{ludvigson2009macro} to study bond risk premia.\footnote{The macro factors of \citet{ludvigson2009macro} are from Sydney Ludvigson's
website.} These factors are eight factors extracted using dynamic factor analysis
and a panel of US government bonds, and the first factor cubed. For
each of these factors (denoted with $F_{LN}$), separately, we first
estimate the cross-sectional regression $E[r_{it}]=\alpha+\beta E[F_{LN}]+\epsilon_{i}$.
Next, we repeat the estimation additionally including the predicted
values obtained with the HFL model (i.e., $E[h(f_{t}\lambda_{i})])$. 

Table \ref{tab:Comparison-with-macro-LN} summarizes our results.
When taken individually, each of the macro factors is significant,
except for the second, third and fifth factors. When we additionally
include in the regression specification the HFL component, all of
the nine macro factors are not statistically different from zero.
In contrast, the HFL component is statistically significant in all
models.

\subsection{Comparison with the factor zoo\label{subsec:Comparison-with-the-factor-zoo}}

In this section, we compare the HFL model with a large set of factors
proposed in the last decades to account for the cross-section of asset
returns (see, e.g., \citet*{feng2020taming}). In particular, we consider
factors for 153 characteristics in 13 themes, using data from 93 countries
and four regions, constructed by \citet*{jensen2023there}.\footnote{The data for the factor zoo is available through the Global Factor
Data website. We use capped value weighted data for all factors and
all countries.} For each of these factors (denoted with $f^{zoo}$), separately,
we first estimate the cross-sectional regression $E[r_{it}]=\alpha+\beta^{zoo}E[f_{t}^{zoo}]+\epsilon_{i}$.
Next, we repeat the estimation additionally including the predicted
values obtained with the HFL model (i.e., $E[h(f_{t}\lambda_{i})])$.
Figure \ref{fig:Comparison-with-the-factor-zoo} summarizes our results.
It plots the average of the absolute values of the $t$-statistics
in a test of the intercepts and slope coefficients in the cross-sectional
regressions, respectively, equal to zero. In the left panel of the
figure, we consider the regressions which only include the factor
asset betas. In the right panel of the figure, we consider regressions
which additionally include the average predicted values from the HFL
model. The figure reveals that, for the models which do not include
the HFL component, the estimates of the slope coefficients are on
average significantly different from zero (average absolute $t$-statistics
equal to 2.06). For these models, though, the pricing errors, captured
by the estimates of the intercept, are on average statistically significant
(average absolute $t$-statistics equal to 2.91). Furthermore, the
figure reveals that, in the models which additionally include the
average predicted values from the HFL model, the slope coefficients
associated with the factor zoo become not statistically different
from zero (average absolute $t$-statistics equal to 0.85), while
those associated with the HFL component are highly significant (average
absolute t-statistics equal to 5.29). Moreover, for the models which
includes the HFL component, the estimates of the pricing error also
become not statistically different from zero (average $t$-statistics
equal to 0.21).
\begin{center}
\begin{figure}[H]
\caption{\textbf{Comparison with the Factor Zoo\label{fig:Comparison-with-the-factor-zoo}}}

{\footnotesize{}\medskip{}
}{\footnotesize\par}

{\small{}This figure shows the average of the absolute values of the
$t$-statistics in a test of the intercepts and slope coefficients
in the cross-sectional regressions, respectively, equal to zero. Standard
errors are adjusted with the Fama-MacBeth procedure. In the left panel,
we consider the cross-sectional regressions $E[r_{it}]=\alpha+\beta^{zoo}E[f_{t}^{zoo}]+\epsilon_{i}$
for each factor from the factor zoo. In the right panel, we consider
cross-sectional regressions which additionally include the average
predicted values based on the HFL component (i.e., $E[h(f_{t}\lambda_{i})])$.
The vertical bars denote one standard deviation around the mean. The
horizontal black dashed-line denote significance at the 10\% confidence
level. The data for 153 factors from the factor zoo are from \citet*{jensen2023there}.
The HFL model is based on a degree of the polynomial used to approximate
the function $h(.)$ equal to 4.}{\small\par}

{\footnotesize{}\medskip{}
}{\footnotesize\par}
\centering{}{\small{}}%
\subfloat{\centering{}{\small{}\includegraphics[width=8cm]{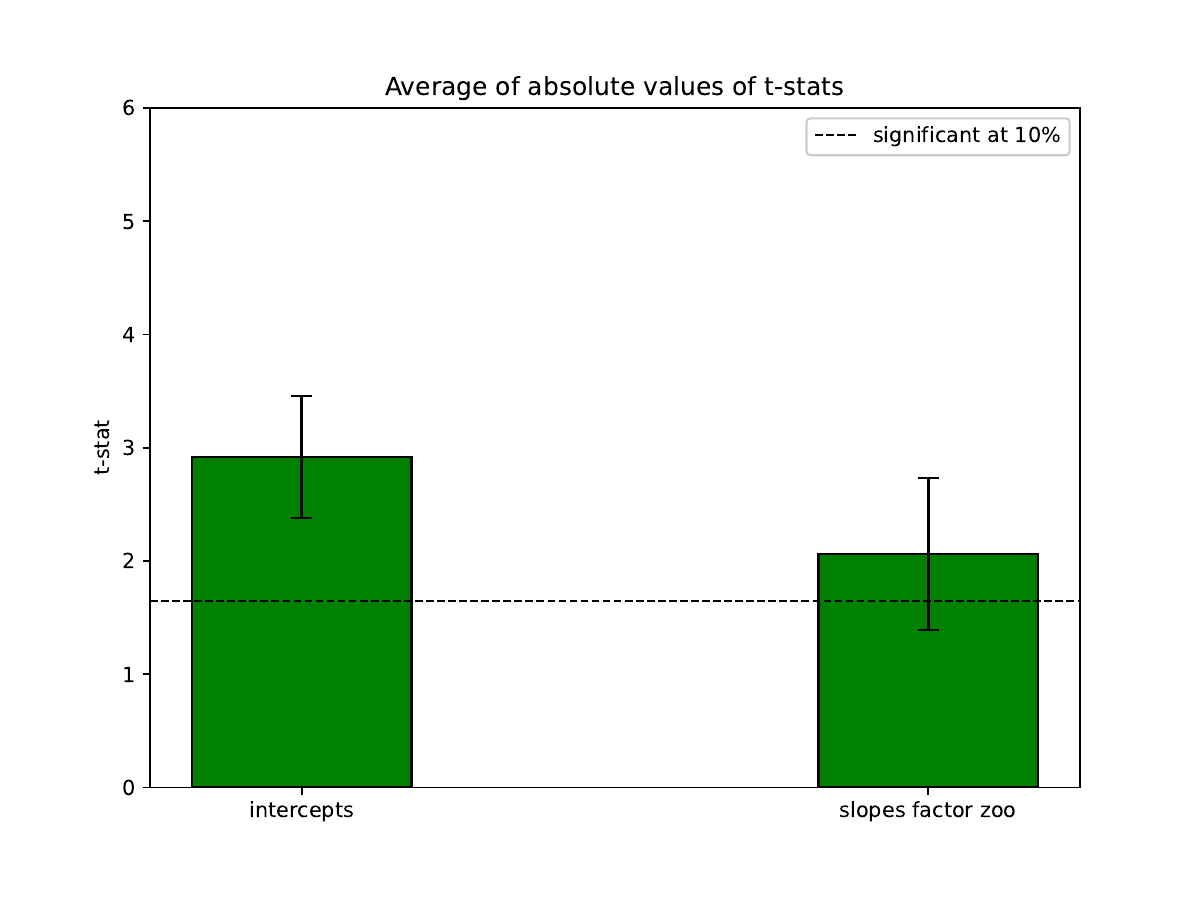}}}{\small{}}%
\subfloat{\centering{}{\small{}\includegraphics[width=8cm]{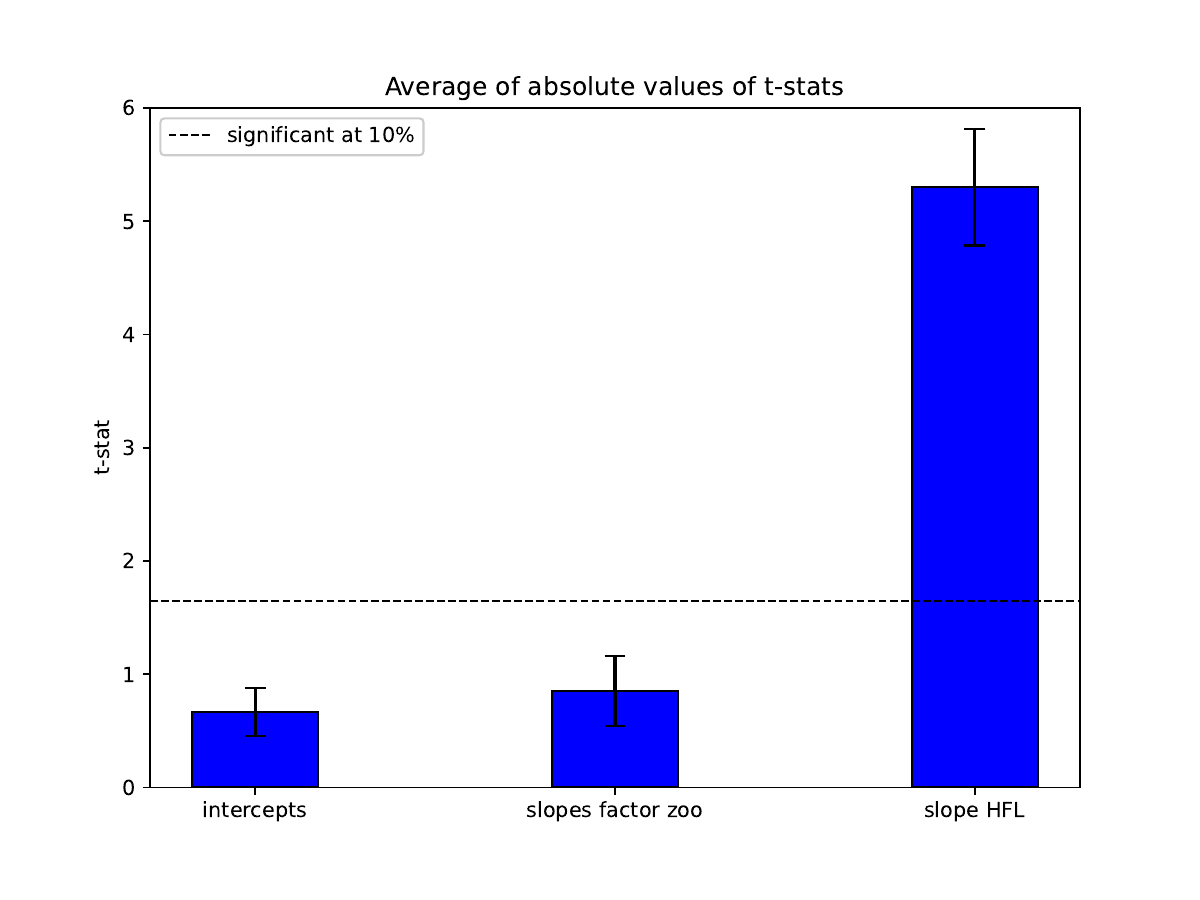}}}{\small\par}
\end{figure}
\par\end{center}

Table \ref{tab:Compairsaon-with-the-factor-zoo} provides further
details about the comparison with the factor zoo. If the HFL component
is not included (Panel A), the fraction of models with a slope coefficient
on the factor zoo which is significant at the 10\% confidence level
is approximately 78\%. When we further include the HFL component,
this fraction drops to approximately 2\%. Moreover, if the HFL component
is included (Panel B), all models have a slope on the HFL component
which is significant at the 1\% confidence level. In these models,
the intercept is never statistically different from zero at a confidence
level of 10\% or lower. Finally, we find that only three of the 153
factors from the factor zoo are significant with a confidence level
of 10\%. These factors are the capital turnover of \citet{haugen1996commonality};
the dollar trading volume of \citet*{brennan1998alternative}; and
the share turnover of \citet*{datar1998liquidity}.

The double-selection Lasso of \citet*{belloni2014inference} and \citet*{chernozhukov2018double}
allows estimating regressions where the number of right-hand side
variables could be large, even larger than the sample size. In practice,
double-selection Lasso estimates only one coefficient on the right-hand
side at a time by means of a two-step selection method. In the first,
factors with a low contribution to the cross-sectional pricing are
excluded from the set of controls. In the second step, factors whose
covariances with returns are highly correlated in the cross-section
with the covariance between returns and a given factor are added to
the set of controls. Therefore, the double-selection Lasso achieves
dimensionality reduction by eliminating useless and redundant factors.
\citet*{feng2020taming} propose double-selection Lasso as model selection
method to evaluate the contribution of factors in cross-sectional
asset pricing regressions. 

The double-selection Lasso is based on two Lasso regressions which
contain a regularization parameter. In setting these parameters, we
follow a cross-validation, or CV, procedure (see, e.g., \citet*{hastie2009elements}).
We first set a broad range of values for each regularization parameter.
Next, we split the data into 5 folds, and train the model on 4 folds
with one fold held out as validation set, cycling through all folds.
This process is repeated for each value in the range of values for
the regularization parameters and the model performance is evaluated
on the validation set using the mean squared error as the metric.
The parameter value with the lowest average mean squared error across
all folds is selected as the chosen regularization parameter.

The HFL component using the double-selection Lasso method is statistically
significant at the 1\% level ($t$-statistics equal to 7.34). In this
case, the intercept is not statistically different from zero ($t$-statistics
equal to 1.37). Furthermore, Panel C of table \ref{tab:Compairsaon-with-the-factor-zoo}
reports summary statistics on the statistical significance of slope
coefficients on the factor zoo, after controlling for the HFL component,
as well as for the remaining factors in the factor zoo. The table
reveals that the fraction of significant factors in the factor zoo,
with a confidence of 10\% or below, is approximately 10\%, or 14 factors
(see Table \ref{tab:Significant-Factors-from-DS-LASSO} in the Appendix
for the list of significant factors).

\subsection{Predictability\label{subsec:Predictability}}

In this section, we study whether the HFL model can predict returns
by investigating the properties of portfolios constructed on the basis
of the predictions of the HFL model. We start by estimating the model
on all 171 assets and the first 120 months of data.\footnote{The 171 assets contain three equity factors, and the results are qualitatively
similar when we exclude them in the predictability exercise.} Next, we use the model predicted returns to sort assets in five portfolios,
and compute the equally-weighted average portfolio return in the following
six months. Finally, we repeat the procedure expanding the estimation
window by 6 months each time until the end of the sample (i.e., we
repeat the procedure 40 times in order to reach the end of the sample
in December, 2017). In each period, the first four portfolios contain
34 assets each, and the last portfolio contains the remaining 35 assets.
This empirical methodology is designed to capture the strategy of
an investor who uses the model to predict future returns rebalancing
at a semi-annual frequency. In each period $t$, we sort assets only
using information available up to period $t$ and compute excess returns
between $t$ and $t+1,$where each period contains 6 months. Portfolio
1 groups the assets with, each period, the lowest predicted returns
by the HFL model. Portfolio 5 groups the assets with, each period,
the highest predicted returns by the HFL model. 

Table \ref{tab:HFL-Portfolios} describes the properties of the five
portfolios, as well as of a long-short strategy long in portfolio
5 and short in portfolio 1, and their risk-adjusted performance. The
analysis of Panel A reveals the predictive ability of the HFL model
in forecasting future returns. Across the portfolios, there is a sizable
cross-section in average monthly returns, ranging from 0.08\% for
the first portfolio to 0.77\% for the last one. Notably, a zero-cost
long-short strategy, wherein assets with the highest predicted returns
are bought while those with the lowest predicted returns are sold
short, yields a monthly average excess return of 0.69\% (8.28\% annualized),
with a monthly Sharpe ratio of 0.15 (0.51 annualized). Despite the
relatively short time-series, the cross-sectional return spread is
statistically significant. For instance, the standard error for the
long-short strategy is 30 basis points. Consequently, the average
excess return is more than two standard errors from zero. Similarly,
each individual portfolio, except the first two portfolios, offers
average returns which are statistically different from zero. Panel
B highlights the significant risk-adjusted performance of the strategy
based on the HFL model. For each portfolio, we report the estimates
of the intercept in a regression of the portfolio return on the \citet*{fama1993common}'s
three factors ($\alpha_{FF3})$, on the \citet*{fama2015five}'s five
factors ($\alpha_{FF5}$), and on the five factors augmented by the
momentum factor of \citet*{jegadeesh1993returns} ($\alpha_{FF5+MOM})$,
respectively. These estimates of the intercept are a measure of risk-adjusted
performance. First, the table reveals that the $\alpha$ estimates
increase from the first to the last portfolio in all three specifications.
Second, we find that the alphas for portfolio 3 to 5, as well as for
the long-short portfolio, are statistically different from zero at
conventional levels. For the long-short portfolio, the monthly alphas
are equal, respectively, to 0.60\%, 0.78\% and 0.73\%, and are statically
significant at the 5\% level.

\begin{table}[H]
\caption{Comparison with the Factor Zoo\label{tab:Compairsaon-with-the-factor-zoo}}
{\footnotesize{}\medskip{}
}{\footnotesize\par}

{\small{}This table provides a comparison of the HFL model and the
factor zoo. Panel A refers to cross-sectional regressions $E[r_{it}]=\alpha+\beta^{zoo}E[f_{t}^{zoo}]+\epsilon_{i}$
for each factor from the factor zoo. Standard errors are adjusted
with the Fama-MacBeth procedure. Panel B refers to cross-sectional
regressions which additionally include the average predicted values
based on the HFL component (i.e., $E[h(\lambda_{i}f_{t})])$. The
two panels report the fraction (in percentage) of intercepts and slope
coefficients with $p$-values less than 1\%, between 1 and 5\%, between
5 and 10\% and larger than 10\%. The $p$-values correspond to tests
for the coefficient equal to zero. Panel C refers to the model which
includes the HFL component and estimates based on the double-selection
Lasso method, and reports the fraction of slope coefficients on the
factor zoo with $p$-values less than 1\%, between 1 and 5\%, between
5 and 10\% and larger than 10\%. The HFL component using the double-selection
Lasso method is statistically significant at the 1\% level. The data
for 153 factors from the factor zoo are from \citet*{jensen2023there}.
The HFL model is based on a degree of the polynomial used to approximate
the function $h(.)$ equal to 4.}{\footnotesize{}\bigskip{}
}{\footnotesize\par}
\centering{}%
\begin{tabular}{ccccc}
\hline 
$p(coeff=0)$ &
$p\leq1\%$ &
$1\%<p\leq5\%$ &
$5\%<p\leq10\%$ &
$p>10\%$\tabularnewline
\hline 
 &
 &
 &
 &
\tabularnewline
 &
\multicolumn{4}{c}{Panel A: Model without HFL}\tabularnewline
\cline{2-5} \cline{3-5} \cline{4-5} \cline{5-5} 
 &
 &
 &
 &
\tabularnewline
$\alpha$ &
79.08 &
13.72 &
2.61 &
4.57\tabularnewline
$E[f_{t}^{zoo}]$ &
15.68 &
54.24 &
7.84 &
22.22\tabularnewline
 &
 &
 &
 &
\tabularnewline
 &
\multicolumn{4}{c}{Panel B: Model with HFL}\tabularnewline
\cline{2-5} \cline{3-5} \cline{4-5} \cline{5-5} 
 &
 &
 &
 &
\tabularnewline
$\alpha$ &
0 &
0 &
0 &
1\tabularnewline
$E[f_{t}^{zoo}]$ &
0 &
0 &
1.96 &
98.03\tabularnewline
$h\left(f_{t}\lambda_{i}\right)$ &
1 &
0 &
0 &
0\tabularnewline
 &
 &
 &
 &
\tabularnewline
 &
\multicolumn{4}{c}{Panel C: Model with HFL and double-selection Lasso}\tabularnewline
\cline{2-5} \cline{3-5} \cline{4-5} \cline{5-5} 
 &
 &
 &
 &
\tabularnewline
$\alpha$ &
98.03 &
1.96 &
0 &
0\tabularnewline
$E[f_{t}^{zoo}]$ &
1.30 &
4.57 &
3.26 &
90.84\tabularnewline
 &
 &
 &
 &
\tabularnewline
\hline 
\end{tabular}
\end{table}

\begin{table}[H]
\caption{HFL Portfolios\label{tab:HFL-Portfolios}}

{\footnotesize{}\medskip{}
}{\footnotesize\par}

{\small{}This table describes the properties of five portfolios formed
by sorting assets by the predicted values of the HFL model and their
risk-adjusted performance. Starting in January 1998, we sort all assets
by the predicted values from the HFL model estimated using the previous
120 months and compute the equally-weighted average portfolio return
in the following 6 months. Finally, we repeat the procedure expanding
the estimation window by 6 months each time until the end of the sample
(i.e., we repeat the procedure 40 times in order to reach the end
of the sample in December, 2017). For each portfolio, and for the
long-short portfolio long in the last portfolio and short in the first
portfolio, Panel A reports the mean (Mean), standard deviation (SD),
and standard error (SE), along the Sharpe ratio (SR) measured as the
ratio between mean and standard deviation. The last row reports the
number of assets in each portfolio in each period. Panel B reports
the estimates of the intercept in a regression of the portfolio return
on the \citet*{fama1993common}'s three factors ($\alpha_{FF3})$,
on the \citet*{fama2015five}'s five factors ($\alpha_{FF5}$), and
on the five factors augmented by the momentum factor of \citet*{jegadeesh1993returns}
($\alpha_{FF5+MOM})$, respectively. \citet{newey1987simple}'s standard
errors with a lag of 3 in parenthesis. {*}, {*}{*}, and {*}{*}{*}
represent significance at the 1\%, 5\%, and 10\% level. The HFL model
is based on a degree of the polynomial used to approximate the function
$h(.)$ equal to 4.}{\small\par}

{\footnotesize{}\bigskip{}
}{\footnotesize\par}
\begin{centering}
\begin{tabular}{lcccccc}
\hline 
\textit{Portfolio} &
P1 &
P2 &
P3 &
P4 &
P5 &
P5-P1\tabularnewline
\hline 
 &
 &
 &
 &
 &
 &
\tabularnewline
 &
\multicolumn{6}{c}{Panel A: Portfolio properties}\tabularnewline
\cline{2-7} \cline{3-7} \cline{4-7} \cline{5-7} \cline{6-7} \cline{7-7} 
Mean (\%) &
0.079 &
0.084 &
0.29 &
0.612 &
0.769 &
0.689\tabularnewline
 &
 &
 &
 &
 &
 &
\tabularnewline
SD (\%) &
3.443 &
2.29 &
1.704 &
2.522 &
4.123 &
4.703\tabularnewline
 &
 &
 &
 &
 &
 &
\tabularnewline
SE (\%) &
0.222 &
0.148 &
0.11 &
0.163 &
0.266 &
0.304\tabularnewline
 &
 &
 &
 &
 &
 &
\tabularnewline
SR &
0.023 &
0.037 &
0.17 &
0.243 &
0.186 &
0.147\tabularnewline
 &
 &
 &
 &
 &
 &
\tabularnewline
Assets &
34 &
34 &
34 &
34 &
35 &
69\tabularnewline
 &
 &
 &
 &
 &
 &
\tabularnewline
 &
\multicolumn{6}{c}{Panel B: Risk-adjusted performance}\tabularnewline
\cline{2-7} \cline{3-7} \cline{4-7} \cline{5-7} \cline{6-7} \cline{7-7} 
$\alpha_{FF3}$ &
-0.098 &
0.031 &
0.348$^{***}$ &
0.532$^{***}$ &
0.505$^{***}$ &
0.604$^{**}$\tabularnewline
 &
(0.181) &
(0.124) &
(0.11) &
(0.108) &
(0.176) &
(0.281)\tabularnewline
 &
 &
 &
 &
 &
 &
\tabularnewline
$\alpha_{FF5}$ &
-0.198 &
-0.079 &
0.286$^{***}$ &
0.565$^{***}$ &
0.583$^{***}$ &
0.781$^{***}$\tabularnewline
 &
(0.179) &
(0.113) &
(0.108) &
(0.113) &
(0.189) &
(0.285)\tabularnewline
 &
 &
 &
 &
 &
 &
\tabularnewline
$\alpha_{FF5+MOM}$ &
-0.169 &
-0.063 &
0.283$^{**}$ &
0.577$^{***}$ &
0.56$^{***}$ &
0.729$^{**}$\tabularnewline
 &
(0.181) &
(0.112) &
(0.112) &
(0.117) &
(0.192) &
(0.291)\tabularnewline
 &
 &
 &
 &
 &
 &
\tabularnewline
\hline 
\end{tabular}
\par\end{centering}
\end{table}

\section{Additional Results\label{sec:Additional-results}}

In this section, we describe four sets of additional results related
to our main findings. First, we report the results of asset pricing
tests for individual asset classes. Second, we report further asset
pricing tests using the first and second half of our sample. Third,
we investigate the time-variation in the asset pricing test using
an estimation based on an expanding window. Fourth, we consider a
larger set of test assets, which further includes the equity momentum
portfolios.

\subsection{Asset pricing tests for individual asset classes}

In this section, we present the result of cross-sectional asset pricing
tests for individual asset classes. Table \ref{tab:Cross-Sectional-Asset-Pricing_Individual_Asset_Classes}
summarizes these results. The slope associated with the HFL component
is statistically significant for each asset class. Specifically, it
is significant at the 1\% level for each asset classes except for
the case of commodity, for which the slope is significant at the 10\%
level. The estimates for the intercept are small in magnitude, and
statistically different from zero, at the 5\% level, only for US bonds
and international bonds. Furthermore, the adjusted R-squareds are
above 85\% for each asset class, with the exception of equities (47\%).
The MAPE is the highest for equities (0.13\%) and the smallest for
US bonds (0.006\%). Section \ref{subsec:Cross-sectional-asset-pricing-asset-class-appendix}
in the Appendix contains scatter plots of average excess returns against
predicted excess returns for each individual asset class.

{\footnotesize{}}
\begin{table}[H]
{\footnotesize{}\caption{Cross-Sectional Asset Pricing Tests: Individual Asset Classes\label{tab:Cross-Sectional-Asset-Pricing_Individual_Asset_Classes}}
}{\footnotesize\par}

{\footnotesize{}\medskip{}
}{\footnotesize\par}

{\small{}This table presents the results of cross-sectional asset
pricing tests for each asset class separately. We estimate the cross-sectional
linear model $E[r_{it}]=\alpha+\beta E[h(f_{t}\lambda_{i})]+\epsilon_{i}$
and report the estimates for the slope ($h\left(f_{t}\lambda_{i}\right)$)
and intercept ($\alpha$), and standard errors adjusted with the Fama-MacBeth
procedure in parenthesis. The table columns report estimates for the
different asset classes. The table additionally reports the regression
adjusted R-squared, the mean absolute pricing error (MAPE) in percentage
terms. {*}, {*}{*}, and {*}{*}{*} represent significance at the 1\%,
5\%, and 10\% level. We further report the number of assets in the
cross-sectional regression and the number of monthly observations
for each asset used in the estimation of the averages of the excess
returns and predictions from the model. The KA factor model is based
on a degree of the polynomial used to approximate the function $h(.)$
equal to 4.}{\small\par}

{\footnotesize{}\medskip{}
}{\footnotesize\par}
\centering{}{\small{}}%
\begin{tabular}{lcccccc}
\hline 
 &
All &
Equities &
US Bonds &
Intl Bonds &
Commodities &
FX\tabularnewline
\hline 
 &
 &
 &
 &
 &
 &
\tabularnewline
$h\left(f_{t}\lambda_{i}\right)$ &
0.816{*}{*}{*} &
0.441{*}{*}{*} &
2.536{*}{*}{*} &
1.617{*}{*}{*} &
0.813{*} &
0.787{*}{*}{*}\tabularnewline
 &
(0.177) &
(0.152) &
(0.767) &
(0.275) &
(0.490) &
(0.101)\tabularnewline
$\alpha$ &
0.001 &
0.004 &
-0.001{*}{*} &
-0.001{*}{*} &
0.001 &
-0.000\tabularnewline
 &
(0.001) &
(0.003) &
(0.001) &
(0.000) &
(0.002) &
(0.001)\tabularnewline
Adj $R^{2}$ &
0.885 &
0.472 &
0.987 &
0.913 &
0.898 &
0.888\tabularnewline
MAPE, \% &
0.101 &
0.131 &
0.006 &
0.032 &
0.077 &
0.037\tabularnewline
Assets &
171 &
53 &
11 &
40 &
21 &
46\tabularnewline
Months &
360 &
360 &
360 &
360 &
360 &
360\tabularnewline
 &
 &
 &
 &
 &
 &
\tabularnewline
\hline 
\end{tabular}{\small\par}
\end{table}
{\footnotesize\par}

\subsection{Asset-pricing using alternative samples\label{subsec:Asset-pricing-using-alternative}}

In this section, we present results of cross-sectional asset pricing
tests by asset class using two alternative samples. The first sample
starts in January 1988 and ends in December 2003 (first half). The
second sample starts in January 2004 and ends in December 2017 (second
half). We report the estimates for these two sub-samples, each including
180 monthly dates, in Table \ref{tab:Cross-Sectional-Asset-Pricing-Alt-Samples}.

Panel A refers to the estimates using the first half of the sample.
In the model with all assets, the slope coefficient associated with
$h\left(f_{t}\lambda_{i}\right)$ is statistically different from
zero. Furthermore, the estimate for the $\alpha$, the model pricing
error, is not statistically different from zero and the cross-sectional
regression adjusted R-squared is equal to 86\%. For the individual
asset classes, we reject the null of slope associated with the HFL
component equal to zero for all assets, except for commodities (for
equities the $p$-value is equal to 9.6\%). The cross-sectional regression
adjusted R-squared is highest for US bonds (96\%) and lowest for equities
(60\%). 

Panel B refers to the, more recent, second half of the sample. Similarly
to the results for the first half of the sample, in the model with
all assets, the slope coefficient associated with $h\left(f_{t}\lambda_{i}\right)$
is statistically different from zero. The slope point estimate is
equal to 0.82, whereas it is equal to 0.85 in the first half of the
sample. Furthermore, the estimate for the $\alpha$, the model pricing
error, is not statistically different from zero. However, the model
fit is lower than in the first half, with a regression adjusted R-squared
of 77\% and a MAPE of 0.14\%. For the individual asset classes, similarly
to the estimates for the first half of the sample, we reject the null
of slope equal to zero for the HFL component for all asset classes
with the exception of commodities. The cross-sectional regression
adjusted R-squared is highest for US bonds (98\%) and lowest for equities
(22\%). 

Table \ref{tab:Cross-Sectional-Asset-Pricing-Alt-Samples-Fincrisis}
in the Appendix presents asset pricing estimates for two samples before
and after the global financial crisis of 2007-08, which further confirm
the robustness of our results for alternative samples.

\subsection{Time variation in asset-pricing tests}

\begin{figure}[H]
\caption{Time-Varying Coefficient Estimates\label{fig:10-Year-Rolling-Beta}}

{\footnotesize{}\medskip{}
}{\footnotesize\par}

{\small{}The figure plots the estimates of the slope coefficient in
the cross-sectional model based on the HFL model: $E[r_{it}]=\alpha+\beta E[h(f_{t}\lambda_{i})]+\epsilon_{i}$
. The estimation is based on an expanding window, starting with a
minimum window size of 10 years of data at the monthly frequency following
\citet{eugene1992cross}. The HFL model is estimated using all assets.
The colored region denotes a two standard error confidence band. Standard
errors are adjusted with the Fama-MacBeth procedure. The HFL model
is based on a degree of the polynomial used to approximate the function
$h(.)$ equal to 4.}{\small\par}

{\footnotesize{}\medskip{}
}{\footnotesize\par}
\centering{}%
\subfloat{\centering{}\includegraphics[width=12cm]{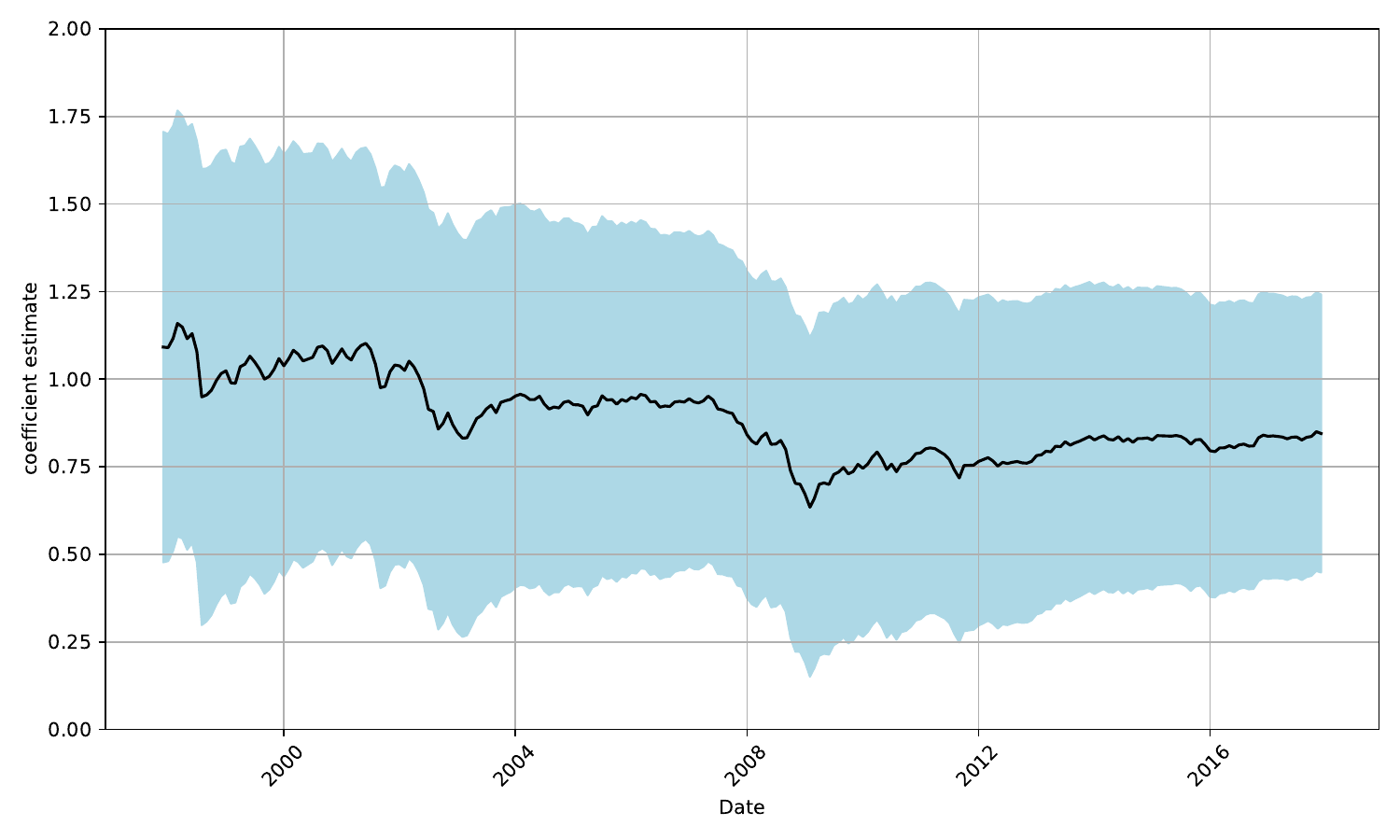}}
\end{figure}

In this section, we further investigate time-variation in the cross-sectional
asset pricing tests. In section \ref{subsec:Asset-pricing-using-alternative},
we showed that the estimates for the slope associated with the HFL
component, on all assets, are similar and statistically significant
in both the first and second half of the sample. In this section,
we provide further robustness evidence by plotting, in Figure \ref{fig:10-Year-Rolling-Beta},
time-varying monthly coefficient estimates. The time-varying estimates
are based on an expanding window and starting with a minimum window
size of 10 years of data at the monthly frequency following \citet{eugene1992cross}.
The figure reveals that the slope associated with $E[h(f_{t}\lambda_{i})]$
is statistically different from zero for all months, as highlighted
by the shaded region which denotes a two standard error confidence
band. Furthermore, we can never reject the hypothesis that the slope
is equal to 1, which is one of the cross-sectional predictions of
our model.

{\footnotesize{}}
\begin{table}[H]
{\footnotesize{}\caption{Cross-Sectional Asset Pricing by Asset Class: Alternative Samples\label{tab:Cross-Sectional-Asset-Pricing-Alt-Samples}}
}{\footnotesize\par}

{\footnotesize{}\medskip{}
}{\footnotesize\par}

{\small{}This table presents the results of cross-sectional asset
pricing tests for alternative samples. The first sample starts in
January 1988 and ends in December 2003 (first half). The second sample
starts in January 2004 and ends in December 2017 (second half). We
estimate the cross-sectional linear model $E[r_{it}]=\alpha+\beta E[h(f_{t}\lambda_{i})]+\epsilon_{i}$
and report the estimates for the slope ($h\left(f_{t}\lambda_{i}\right)$)
and intercept ($\alpha$), and standard errors adjusted with the Fama-MacBeth
procedure in parenthesis. Panel A reports estimates for the first
half of the sample. Panel B reports estimates for the second half
of the sample. The table columns report estimates for the different
asset classes. The table additionally reports the regression adjusted
R-squared, the mean absolute pricing error (MAPE) in percentage terms.
{*}, {*}{*}, and {*}{*}{*} represent significance at the 1\%, 5\%,
and 10\% level. We further report the number of assets in the cross-sectional
regression and the number of monthly observations for each asset used
in the estimation of the averages of the excess returns and predictions
from the model. The HFL model is based on a degree of the polynomial
used to approximate the function $h(.)$ equal to 4.}{\small\par}

{\footnotesize{}\bigskip{}
}{\footnotesize\par}
\centering{}{\small{}}%
\begin{tabular}{lcccccc}
\hline 
Panel A: First Half &
All &
Equities &
US Bonds &
Intl Bonds &
Commodities &
FX\tabularnewline
\hline 
 &
 &
 &
 &
 &
 &
\tabularnewline
$h\left(f_{t}\lambda_{i}\right)$ &
0.849{*}{*}{*} &
0.546{*} &
2.232{*}{*} &
1.411{*}{*}{*} &
0.782 &
1.003{*}{*}{*}\tabularnewline
 &
(0.233) &
(0.282) &
(0.913) &
(0.340) &
(0.576) &
(0.172)\tabularnewline
$\alpha$ &
-0.000 &
0.002 &
-0.001 &
-0.000 &
-0.001 &
-0.000\tabularnewline
 &
(0.001) &
(0.004) &
(0.001) &
(0.000) &
(0.002) &
(0.002)\tabularnewline
Adj $R^{2}$ &
0.859 &
0.606 &
0.961 &
0.782 &
0.654 &
0.789\tabularnewline
MAPE, \% &
0.133 &
0.127 &
0.012 &
0.058 &
0.218 &
0.072\tabularnewline
Assets &
171 &
53 &
11 &
40 &
21 &
46\tabularnewline
Months &
180 &
180 &
180 &
180 &
180 &
180\tabularnewline
 &
 &
 &
 &
 &
 &
\tabularnewline
\hline 
Panel B: Second Half &
All &
Equities &
US Bonds &
Intl Bonds &
Commodities &
FX\tabularnewline
\hline 
 &
 &
 &
 &
 &
 &
\tabularnewline
$h\left(f_{t}\lambda_{i}\right)$ &
0.819{*}{*}{*} &
0.370{*}{*}{*} &
2.984{*}{*} &
1.916{*}{*}{*} &
0.849 &
0.607{*}{*}{*}\tabularnewline
 &
(0.277) &
(0.126) &
(1.342) &
(0.462) &
(0.810) &
(0.117)\tabularnewline
$\alpha$ &
0.001 &
0.006{*} &
-0.003{*}{*} &
-0.001{*}{*}{*} &
0.003 &
0.001\tabularnewline
 &
(0.001) &
(0.003) &
(0.001) &
(0.000) &
(0.005) &
(0.002)\tabularnewline
Adj $R^{2}$ &
0.765 &
0.222 &
0.974 &
0.883 &
0.495 &
0.733\tabularnewline
MAPE, \% &
0.140 &
0.164 &
0.010 &
0.038 &
0.210 &
0.059\tabularnewline
Assets &
171 &
53 &
11 &
40 &
21 &
46\tabularnewline
Months &
180 &
180 &
180 &
180 &
180 &
180\tabularnewline
 &
 &
 &
 &
 &
 &
\tabularnewline
\hline 
\end{tabular}{\small\par}
\end{table}
{\footnotesize\par}

\subsection{Asset-pricing with momentum portfolios}

Table \ref{tab:Cross-Sectional-Asset-Pricing-Momentum} presents estimates
of cross-sectional asset pricing tests using a set of test assets
which additionally includes 10 momentum portfolios and the momentum
factor. Both the momentum portfolios and the momentum factor are from
Ken French's website and are based on U.S. stocks monthly returns
sorted by the 12-2 past return. The table reveals that the slope coefficient
associated with the averages of the predicted values from the HFL
component is significantly different from zero for all assets as well
as for each individual asset class. The pricing error, captured be
the estimate of the intercept, is significantly different from zero
only for US bonds (significant at the 5\%). The model provides the
closest fit for US bonds (adjusted-$R^{2}$ of 99\%), although the
model fit is similar across the specifications with the exception
of equities with a lower value for the adjusted R-squared of 54\%.
The MAPE ranges from 0.005\% (US bonds) to 0.122\% (equities). 

{\footnotesize{}}
\begin{table}[H]
{\footnotesize{}\caption{Cross-Sectional Asset Pricing Tests -- With Momentum\label{tab:Cross-Sectional-Asset-Pricing-Momentum}}
}{\footnotesize\par}

{\footnotesize{}\medskip{}
}{\footnotesize\par}

{\small{}This table presents the results of cross-sectional asset
pricing tests using all assets and, additionally, 10 momentum portfolios
and a momentum factor. We estimate the cross-sectional linear model
$E[r_{it}]=\alpha+\beta E[h(f_{t}\lambda_{i})]+\epsilon_{i}$ and
report the estimates for the slope ($h\left(f_{t}\lambda_{i}\right)$)
and intercept ($\alpha$), and standard errors adjusted with the Fama-MacBeth
procedure in parenthesis. The table additionally reports the regression
R-squared, the mean absolute pricing error (MAPE) in percentage terms.
{*}, {*}{*}, and {*}{*}{*} represent significance at the 1\%, 5\%,
and 10\% level. We further report the number of assets in the cross-sectional
regression and the number of monthly observations for each asset used
in the estimation of the averages of the excess returns and predictions
from the model. The HFL model is based on a degree of the polynomial
used to approximate the function $h(.)$ equal to 4.}{\small\par}

{\footnotesize{}\bigskip{}
}{\footnotesize\par}
\centering{}{\small{}}%
\begin{tabular}{lcccccc}
\hline 
 &
All &
Equities &
US Bonds &
Intl Bonds &
Commodities &
FX\tabularnewline
\hline 
 &
 &
 &
 &
 &
 &
\tabularnewline
$h\left(f_{t}\lambda_{i}\right)$ &
0.795{*}{*}{*} &
0.500{*}{*}{*} &
2.316{*}{*}{*} &
1.529{*}{*}{*} &
0.783{*} &
0.731{*}{*}{*}\tabularnewline
 &
(0.164) &
(0.148) &
(0.700) &
(0.261) &
(0.472) &
(0.093)\tabularnewline
$\alpha$ &
0.001 &
0.004 &
-0.001{*}{*} &
-0.000 &
0.001 &
-0.000\tabularnewline
 &
(0.001) &
(0.003) &
(0.000) &
(0.000) &
(0.002) &
(0.001)\tabularnewline
Adj $R^{2}$ &
0.877 &
0.544 &
0.993 &
0.918 &
0.900 &
0.879\tabularnewline
MAPE, \% &
0.109 &
0.122 &
0.005 &
0.031 &
0.078 &
0.036\tabularnewline
Assets &
182 &
64 &
11 &
40 &
21 &
46\tabularnewline
Months &
180 &
180 &
180 &
180 &
180 &
180\tabularnewline
 &
 &
 &
 &
 &
 &
\tabularnewline
\hline 
\end{tabular}{\small\par}
\end{table}
{\footnotesize\par}

\section{Conclusion}

\noindent We propose a parsimonious non-linear single factor asset
pricing model $r_{it}=h(f_{t}\lambda_{i})+\epsilon_{it}$. Despite
its parsimony, the HFL model represents exactly any non-linear model
with an arbitrary number of factors and factor loadings as a consequence
of the Kolmogorov-Arnold representation theorem. Empirically, we jointly
estimate the link function, the time factor, and the loading with
the sieve-based least square estimator.

Using 171 test assets across major asset classes across U.S. equity,
U.S. and international bonds, commodities, and currencies, we show
that this one factor model delivers superior cross-sectional empirical
asset pricing performance with a low-dimensional approximation of
the link function. Moreover, controlling for the HFL model, the majority
of the established factors for the cross-section of asset returns
becomes insignificant. Additionally, we use the HFL model to construct
a tradable strategy that generates statistically significant and sizable
risk-adjusted long-short excess returns

There are two other broader points we want to make. First, nonlinearity
and the higher-order interactions are the key feature of the HFL model.
Either as a stand alone tool or incorporating it in other asset pricing
settings thus allows a parsimonious, yet general, way to capture the
importance of these effects compared to linear asset pricing models.
Second, there is significant recent interest in using neural nets
and artificial intelligence models in asset-pricing. It is interesting
to speculate why these models may deliver superior performance to
the existing asset-pricing models. One answer is the Kolmogorov-Arnold
representation theorem. This theorem shows that any nonlinear continuous
function of many variables can be represented as a continuous outer
layer and an inner layer consisting of the sum of functions of single
variables. \citet*{hecht1987kolmogorov} argues that this is exactly
the structure of a general neural network. Our HFL model thus offers
a parsimonious method to capture the performance of neural network
models and to evaluate their performance for asset pricing.

\newpage{}

\section*{References}

\begin{btSect}[econ-aea]{4_Users_nicolaborri_Dropbox_Research_INDEX-EVERYTHING_lyx_bib_ka}
\btPrintCited
\end{btSect}

\newpage{}

\section*{Online Appendix}

\setcounter{section}{0} \renewcommand{\thesection}{\Alph{section}} \setcounter{table}{0} \renewcommand{\thetable}{A.\arabic{table}} \setcounter{figure}{0} \renewcommand{\thefigure}{A.\arabic{figure}}
\setcounter{page}{1}

\section{Econometrics Appendix}

\subsection{Proof of Theorem \ref{thm: consistency}\label{sec:proof rate of convergence}}

It follows from the definition of the least-squares sieve estimator
and Assumption \ref{as: approximation error} that
\[
\frac{1}{NT}\sum_{i=1}^{N}\sum_{t=1}^{T}(r_{it}-\widehat{h}(\widehat{f}_{t}\widehat{\lambda}_{i}))^{2}\leq\frac{1}{NT}\sum_{i=1}^{N}\sum_{t=1}^{T}(r_{it}-h_{K}(f_{t}\lambda_{i}))^{2}.
\]
Substituting here $r_{it}=h(f_{t}\lambda_{i})+\varepsilon_{it}$,
we obtain
\begin{align*}
 & \frac{1}{NT}\sum_{i=1}^{N}\sum_{t=1}^{T}(\widehat{h}(\widehat{f}_{t}\widehat{\lambda}_{i})-h(f_{t}\lambda_{i}))^{2}\\
 & \qquad\leq\frac{1}{NT}\sum_{i=1}^{N}\sum_{t=1}^{T}(h_{K}(f_{t}\lambda_{i})-h(f_{t}\lambda_{i}))^{2}+\frac{2}{NT}\sum_{i=1}^{N}\sum_{t=1}^{T}\varepsilon_{it}(\widehat{h}(\widehat{f}_{t}\widehat{\lambda}_{i})-h_{K}(f_{t}\lambda_{i})).
\end{align*}
In addition, by the triangle inequality,
\begin{align*}
 & \frac{1}{NT}\sum_{i=1}^{N}\sum_{t=1}^{T}(\widehat{h}(\widehat{f}_{t}\widehat{\lambda}_{i})-h_{K}(f_{t}\lambda_{i}))^{2}\\
 & \qquad\leq\frac{2}{NT}\sum_{i=1}^{N}\sum_{t=1}^{T}(\widehat{h}(\widehat{f}_{t}\widehat{\lambda}_{i})-h(f_{t}\lambda_{i}))^{2}+\frac{2}{NT}\sum_{i=1}^{N}\sum_{t=1}^{T}(h_{K}(f_{t}\lambda_{i})-h(f_{t}\lambda_{i}))^{2}.
\end{align*}
Combining the last two displayed equations, we have
\begin{align}
 & \frac{1}{NT}\sum_{i=1}^{N}\sum_{t=1}^{T}(\widehat{h}(\widehat{f}_{t}\widehat{\lambda}_{i})-h_{K}(f_{t}\lambda_{i}))^{2}\nonumber \\
 & \qquad\leq\frac{4}{NT}\sum_{i=1}^{N}\sum_{t=1}^{T}(h_{K}(f_{t}\lambda_{i})-h(f_{t}\lambda_{i}))^{2}+\frac{4}{NT}\sum_{i=1}^{N}\sum_{t=1}^{T}\varepsilon_{it}(\widehat{h}(\widehat{f}_{t}\widehat{\lambda}_{i})-h_{K}(f_{t}\lambda_{i})).\label{eq: basic inequality}
\end{align}
The first term on the right-hand side here can be bounded using Assumption
\ref{as: approximation error}. To bound the second term, we use the
empirical process theory. Denote $\mathcal{E}=[-L,L]^{K+1}\times[0,1]^{T}\times[0,1]^{N}$
and let $d$ be the metric on $\mathcal{E}$ defined so that for each
$\xi_{1}=(\{c_{j1}\}_{j=0}^{K},\{\phi_{t1}\}_{t=1}^{T},\{l_{i1}\}_{i=1}^{N})$
and $\xi_{2}=(\{c_{j2}\}_{j=0}^{K},\{\phi_{t2}\}_{t=1}^{T},\{l_{i2}\}_{i=1}^{N})$
in $\mathcal{E}$, we have
\[
d(\xi_{1},\xi_{2})^{2}=\frac{1}{NT}\sum_{i=1}^{N}\sum_{t=1}^{T}\left(\sum_{j=0}^{K}c_{j1}(\phi_{t1}l_{i1})^{j}-\sum_{j=0}^{K}c_{j2}(\phi_{t2}l_{i2})^{j}\right)^{2}.
\]
It is straightforward to show that covering numbers $N(\epsilon,\mathcal{E},d)$
of the metric space $(\mathcal{E},d)$ satisfy
\begin{equation}
N(\epsilon,\mathcal{E},d)\leq\begin{cases}
(1+2L(K+1)/\epsilon)^{T+N+K+1}, & \text{if }\epsilon\leq2L(K+1),\\
1 & \text{if }\epsilon>2L(K+1).
\end{cases}\label{eq: covering number}
\end{equation}
It is also well--known that the corresponding packing numbers $D(\epsilon,\mathcal{E},d)$
satisfy $D(\epsilon,\mathcal{E},d)\leq N(\epsilon/2,\mathcal{E},d)$
for all $\epsilon>0$; see page 98 of \citet{VW96} for the definitions
of covering and packing numbers and relations between them. 

Further, for each $\xi=(\{c_{j}\}_{j=0}^{K},\{\phi_{t}\}_{t=1}^{T},\{l_{i}\}_{i=1}^{N})$
in $\mathcal{E}$, denote
\[
X_{\xi}=\frac{1}{\sqrt{NT}}\sum_{i=1}^{N}\sum_{t=1}^{T}\varepsilon_{it}\left(\sum_{j=0}^{K}c_{j}(\phi_{t}l_{i})^{j}\right).
\]
Then there exists a constant $C>0$ such that
\begin{equation}
\|X_{\xi_{1}}-X_{\xi_{2}}\|_{\psi_{2}}\leq Cd(\xi_{1},\xi_{2}),\quad\text{for all }\xi_{1},\xi_{2}\in\mathcal{E},\label{eq: psi-norm bound}
\end{equation}
by Assumption \ref{as: idiosyncratic noise}(ii) and Proposition 2.6.1
in \citet{V18}. Thus, denoting $\delta=1/\sqrt{NT}$, it follows
from Corollary 2.2.8 in \citet{VW96}, applied to $X_{\xi}/C$ instead
of $X_{\xi}$, that if
\begin{equation}
\frac{1}{NT}\sum_{i=1}^{N}\sum_{t=1}^{T}(\widehat{h}(\widehat{f}_{t}\widehat{\lambda}_{i})-h_{K}(f_{t}\lambda_{i}))^{2}\leq\delta^{2},\label{eq: low radius}
\end{equation}
then
\[
\left|\frac{1}{\sqrt{NT}}\sum_{i=1}^{N}\sum_{t=1}^{T}\varepsilon_{it}(\widehat{h}(\widehat{f}_{t}\widehat{\lambda}_{i})-h_{K}(f_{t}\lambda_{i}))\right|\leq\sup_{\underset{d(\xi_{1},\xi_{2})\leq\delta}{\xi_{1},\xi_{2}\in\mathcal{E}\colon}}|X_{\xi_{1}}-X_{\xi_{2}}|\lesssim_{P}\int_{0}^{\delta}\sqrt{\log D(\epsilon,\mathcal{E},d)}d\epsilon
\]
Here, using (\ref{eq: covering number}), $D(\epsilon,\mathcal{E},d)\leq N(\epsilon/2,\mathcal{E},d)$,
and $\delta=1/\sqrt{NT}$, we have
\begin{align*}
\int_{0}^{\delta}\sqrt{\log D(\epsilon,\mathcal{E},d)}d\epsilon & \lesssim\int_{0}^{\delta}\sqrt{(T+N+K+1)\log\left(\frac{8L(K+1)}{\epsilon}\right)}d\epsilon\\
 & \lesssim\delta\sqrt{(T+N+K+1)\log\left(\frac{8L(K+1)}{\delta}\right)}\lesssim\sqrt{\frac{(T+N+K)\log(TNK)}{NT}}
\end{align*}
by standard integral calculations.

It thus remains to consider the case when (\ref{eq: low radius})
does not hold. To cover this case, denote $\xi_{0}=(\{h_{j}^{K}\}_{j=0}^{K},\{f_{t}\}_{t=1}^{T},\{\lambda_{i}\}_{i=1}^{N})$.
Also, for all $\xi\in\mathcal{E}$, denote $Y_{\xi}=(X_{\xi}-X_{\xi_{0}})/d(\xi,\xi_{0})$.
Then $\|Y_{\xi}\|_{\psi_{2}}\leq C$ by (\ref{eq: psi-norm bound}).
Also, for all $\xi_{1},\xi_{2}\in\mathcal{E}$ such $d(\xi_{1},\xi_{0})>\delta$,
\begin{align*}
\|Y_{\xi_{1}}-Y_{\xi_{2}}\|_{\psi_{2}} & =\left\Vert \frac{X_{\xi_{1}}-X_{\xi_{0}}}{d(\xi_{1},\xi_{0})}-\frac{X_{\xi_{2}}-X_{\xi_{0}}}{d(\xi_{2},\xi_{0})}\right\Vert _{\psi_{2}}\leq\frac{\|X_{\xi_{1}}-X_{\xi_{2}}\|_{\psi_{2}}}{d(\xi_{1},\xi_{0})}+\|X_{\xi_{2}}-X_{\xi_{0}}\|_{\psi_{2}}\left|\frac{1}{d(\xi_{1},\xi_{0})}-\frac{1}{d(\xi_{2},\xi_{0})}\right|\\
 & \leq\frac{Cd(\xi_{1},\xi_{2})}{\delta}+\frac{C|d(\xi_{1},\xi_{0})-d(\xi_{2},\xi_{0})|}{\delta}\leq\frac{2Cd(\xi_{1},\xi_{2})}{\delta},
\end{align*}
where the first and the third inequalities follow from the triangle
inequality, and the second from (\ref{eq: psi-norm bound}). Hence,
denoting $\mathcal{E}_{1}=\{\xi\in\mathcal{E}\colon d(\xi,\xi_{0})>\delta\}$
and letting $d_{1}$ be the metric $\mathcal{E}_{1}$ such that $d_{1}(\xi_{1},\xi_{2})=\|Y_{\xi_{1}}-Y_{\xi_{2}}\|_{\psi_{2}}$
for all $\xi_{1},\xi_{2}\in\mathcal{E}_{1}$, it follows that the
packing numbers $D(\epsilon,\mathcal{E}_{1},d_{1})$ and the covering
numbers $N(\epsilon,\mathcal{E}_{1},d_{1})$ of the metric space $(\mathcal{E}_{1},d_{1})$
satisfy
\[
D(\epsilon,\mathcal{E}_{1},d_{1})\leq N(\epsilon/2,\mathcal{E}_{1},d_{1})\leq N\left(\frac{\delta\epsilon}{4C},\mathcal{E},d\right),\quad\text{for all }\epsilon>0.
\]
Thus, it follows from Corollary 2.2.8 in \citet{VW96} that if (\ref{eq: low radius})
does not hold, then
\begin{align*}
\frac{\left|\frac{1}{\sqrt{NT}}\sum_{i=1}^{N}\sum_{t=1}^{T}\varepsilon_{it}(\widehat{h}(\widehat{f}_{t}\widehat{\lambda}_{i})-h_{K}(f_{t}\lambda_{i}))\right|}{\sqrt{\frac{1}{NT}\sum_{i=1}^{N}\sum_{t=1}^{T}(\widehat{h}(\widehat{f}_{t}\widehat{\lambda}_{i})-h_{K}(f_{t}\lambda_{i}))^{2}}} & \leq\sup_{\xi\in\mathcal{E}_{1}}|Y_{\xi}|\lesssim_{P}\int_{0}^{\infty}\sqrt{\log D(\epsilon,\mathcal{E}_{1},d_{1})}d\epsilon\\
 & =\int_{0}^{2C}\sqrt{\log D(\epsilon,\mathcal{E}_{1},d_{1})}d\epsilon\\
 & \leq\int_{0}^{2C}\sqrt{\log N\left(\frac{\delta\epsilon}{4C},\mathcal{E},d\right)}d\epsilon\\
 & =\frac{4C}{\delta}\int_{0}^{\delta/2}\sqrt{\log N(\epsilon,\mathcal{E},d)}d\epsilon\lesssim\sqrt{(T+N+K)\log(TNK)}
\end{align*}
by standard integral calculations. Therefore, we conclude that
\begin{align*}
 & \left|\frac{1}{NT}\sum_{i=1}^{N}\sum_{t=1}^{T}\varepsilon_{it}(\widehat{h}(\widehat{f}_{t}\widehat{\lambda}_{i})-h_{K}(f_{t}\lambda_{i}))\right|\\
 & \qquad\lesssim_{P}\sqrt{\frac{(T+N+K)\log(TNK)}{NT}}\left(\frac{1}{\sqrt{NT}}+\sqrt{\frac{1}{NT}\sum_{i=1}^{N}\sum_{t=1}^{T}(\widehat{h}(\widehat{f}_{t}\widehat{\lambda}_{i})-h_{K}(f_{t}\lambda_{i}))^{2}}\right).
\end{align*}
Combining this bound with (\ref{eq: basic inequality}) and using
Assumption \ref{as: approximation error}, it follows that
\[
\frac{1}{NT}\sum_{i=1}^{N}\sum_{t=1}^{T}(\widehat{h}(\widehat{f}_{t}\widehat{\lambda}_{i})-h_{K}(f_{t}\lambda_{i}))^{2}\lesssim_{P}\frac{(T+N+K)\log(TNK)}{NT}+K^{-2\alpha}.
\]
The asserted claim now follows by combining this bound with the triangle
inequality and Assumption \ref{as: approximation error}.

\noindent 

\subsection{Estimation procedure and cross-sectional asset pricing tests}

The HFL model

\begin{equation}
r_{it}=h(f_{t}\lambda_{i})+\varepsilon_{it},\quad i=1,\dots,N,\,t=1,\dots,T,\label{eq: model all over again}
\end{equation}
implies that the coefficients $\alpha$ and $\beta$ in the linear
regression
\[
E[r_{it}]=\alpha+\beta E[h(f_{t}\lambda_{i})]+\epsilon_{i},\quad i=1,\dots,N,
\]
should satisfy the following restrictions: $\alpha=0$ and $\beta=1$.
In Section \ref{sec:Cross-Section}, we tested these restrictions
by (i) estimating the HFL model to get $\widehat{h}$, $\{\widehat{f}_{t}\}_{t=1}^{T}$,
and $\{\widehat{\lambda}_{i}\}_{i=1}^{N}$ and (ii) running the OLS
of $T^{-1}\sum_{t=1}^{T}r_{it}$ on the constant one and $T^{-1}\sum_{t=1}^{T}\widehat{h}(\widehat{f}_{t}\widehat{\lambda}_{i})$
to get $\widehat{\alpha}$ and $\widehat{\beta}$ and comparing them
with $0$ and $1$, respecticely. Here, we show that this test indeed
has power in the sense that unless model (\ref{eq: model all over again})
is correct, the probability limits $\alpha_{0}$ and $\beta_{0}$
of the OLS estimates $\widehat{\alpha}$ and $\widehat{\beta}$ generally
satisfy the restrictions $\alpha_{0}=0$ and $\beta_{0}=1$ only if
the factor $f_{t}$ is constant over time, which is not the case in
practice.

Assume, for simplicity, that we approximate the function $h$ with
the following polynomial of the second order:
\[
h(x)=\gamma_{0}+\gamma_{1}x+\gamma_{2}x^{2}.
\]
In this case, the estimator for the HFL model that we propose is:
\[
(\hat{\gamma}_{0},\hat{\gamma}_{1},\hat{\gamma}_{2},\{\hat{f}_{t}\},\{\hat{\lambda}_{i}\})=\arg\min_{\gamma_{0},\gamma_{1},\gamma_{2},\{\phi_{t}\},\{l_{i}\}}\frac{1}{NT}\sum_{i=1}^{N}\sum_{t=1}^{T}\left(r_{it}-\gamma_{0}-\gamma_{1}\phi_{t}l_{i}-\gamma_{2}(\phi_{t}l_{i})^{2}\right)^{2}.
\]
The first order conditions of this optimization problem with respect
to $\gamma_{0},\gamma_{1},\gamma_{2}$ are 
\[
\frac{1}{NT}\sum_{i=1}^{N}\sum_{t=1}^{T}(r_{it}-\hat{\gamma}_{0}-\hat{\gamma}_{1}\hat{f}_{t}\hat{\lambda}_{i}-\hat{\gamma}_{2}(\hat{f}_{t}\hat{\lambda}_{i})^{2})=0,
\]
\[
\frac{1}{NT}\sum_{i=1}^{N}\sum_{t=1}^{T}(r_{it}-\hat{\gamma}_{0}-\hat{\gamma}_{1}\hat{f}_{t}\hat{\lambda}_{i}-\hat{\gamma}_{2}(\hat{f}_{t}\hat{\lambda}_{i})^{2})\hat{f}_{t}\hat{\gamma}_{i}=0,
\]
\[
\frac{1}{NT}\sum_{i=1}^{N}\sum_{t=1}^{T}(r_{it}-\hat{\gamma}_{0}-\hat{\gamma}_{1}\hat{f}_{t}\hat{\lambda}_{i}-\hat{\gamma}_{2}(\hat{f}_{t}\hat{\lambda}_{i})^{2})(\hat{f}_{t}\hat{\gamma}_{i})^{2}=0,
\]
respectively. Substituting
\[
\hat{h}(\hat{f}_{t}\hat{\lambda}_{i})=\hat{\gamma}_{0}+\hat{\gamma}_{1}\hat{f}_{t}\hat{\lambda}_{i}+\hat{\gamma}_{2}(\hat{f}_{t}\hat{\lambda}_{i})^{2}
\]
into these equations, we obtain
\begin{equation}
\frac{1}{NT}\sum_{i=1}^{T}\sum_{t=1}^{T}(r_{it}-\hat{h}(\hat{f}_{t}\hat{\lambda}_{i}))=0,\label{eq: 1}
\end{equation}
\[
\frac{1}{NT}\sum_{i=1}^{N}\sum_{t=1}^{T}(r_{it}-\hat{h}(\hat{f}_{t}\hat{\lambda}_{i}))\hat{f}_{t}\hat{\lambda}_{i}=0,
\]
\[
\frac{1}{NT}\sum_{i=1}^{N}\sum_{t=1}^{T}(r_{it}-\hat{h}(\hat{f}_{t}\hat{\lambda}_{i}))(\hat{f}_{t}\hat{\lambda}_{i})^{2}=0.
\]
Further, multiplying these equations by $\hat{\gamma}_{0}$, $\hat{\gamma}_{1}$,
and $\hat{\gamma}_{2}$, respectively, and taking the sum, we obtain
in particular that
\[
\frac{1}{NT}\sum_{i=1}^{N}\sum_{t=1}^{T}(r_{it}-\hat{h}(\hat{f}_{t}\hat{\lambda}_{i}))(\hat{\gamma}_{0}+\hat{\gamma}_{1}\hat{f}_{t}\hat{\lambda}_{i}+\hat{\gamma}_{2}(\hat{f}_{t}\hat{\lambda}_{i})^{2})=0,
\]
or, equivalently, that
\begin{equation}
\frac{1}{NT}\sum_{i=1}^{N}\sum_{t=1}^{T}(r_{it}-\hat{h}(\hat{f}_{t}\hat{\lambda}_{i}))\hat{h}(\hat{f}_{t}\hat{\lambda}_{i})=0.\label{eq: 2}
\end{equation}
Taking the probability limits on the left-hand sides of equations
(\ref{eq: 1}) and (\ref{eq: 2}) and using Theorem \ref{thm: consistency},
it thus follows that
\begin{equation}
\frac{1}{NT}\sum_{i=1}^{N}\sum_{t=1}^{T}E[r_{it}-h(f_{t}\lambda_{i})]=o(1),\label{eq: limit 1}
\end{equation}
\begin{equation}
\frac{1}{NT}\sum_{i=1}^{N}\sum_{t=1}^{T}E[(r_{it}-h(f_{t}\lambda_{i}))h(f_{t}\lambda_{i})]=o(1).\label{eq: limit 2}
\end{equation}
Note that these equations are true regardless of whether the HFL model
is correct or not. In case the HFL model is not correct, the quantities
$h(f_{t}\lambda_{i})$ should be understood as the probability limits
of the estimators $\widehat{h}(\widehat{f}_{t}\widehat{\lambda}_{i})$.
Intuitively, equations (\ref{eq: limit 1}) and (\ref{eq: limit 2})
imply that the mean, across time and assets, of the residuals should
be zero and that the covariance between the residuals and the predictor
should be zero as well. Note also that even though we used the second-order
polynomial approximation to the function $h$, equations (\ref{eq: limit 1})
and (\ref{eq: limit 2}) remain valid for polynomial approximation
of any order.

Next, to derive the probability limits $\alpha_{0}$ and $\beta_{0}$
of the OLS estimates $\widehat{\alpha}$ and $\widehat{\beta}$, observe
that
\[
(\hat{\alpha},\hat{\beta})=\arg\min_{\alpha,\beta}\frac{1}{N}\sum_{i=1}^{N}\left(\frac{1}{T}\sum_{t=1}^{T}r_{it}-\alpha-\beta\times\frac{1}{T}\sum_{t}\hat{h}(\hat{f}_{t}\hat{\lambda}_{i})\right)^{2}.
\]
The first-order condition with respect to $\alpha$ is
\[
\frac{1}{N}\sum_{i=1}^{N}\left(\frac{1}{T}\sum_{t=1}^{T}r_{it}-\hat{\alpha}-\hat{\beta}\times\frac{1}{T}\sum_{t=1}^{T}\hat{h}(\hat{f}_{t}\hat{\lambda}_{i})\right)=0.
\]
The first-order condition with respect to $\beta$ is
\[
\frac{1}{N}\sum_{i=1}^{N}\left(\frac{1}{T}\sum_{t=1}^{T}r_{it}-\hat{\alpha}-\hat{\beta}\times\frac{1}{T}\sum_{t=1}^{T}\hat{h}(\hat{f}_{t}\hat{\lambda}_{i})\right)\left(\frac{1}{T}\sum_{t=1}^{T}\hat{h}(\hat{f}_{t}\hat{\lambda}_{i})\right)=0.
\]
These two conditions can be trivially rewritten as
\begin{equation}
\frac{1}{NT}\sum_{i=1}^{N}\sum_{t=1}^{T}(r_{it}-\hat{\alpha}-\hat{\beta}\hat{h}(\hat{f}_{t}\hat{\lambda}_{i}))=0,\label{eq: 3}
\end{equation}
\begin{equation}
\frac{1}{NT}\sum_{i=1}^{T}\sum_{t=1}^{T}(r_{it}-\hat{\alpha}-\hat{\beta}\hat{h}(\hat{f}_{t}\hat{\lambda}_{i}))\widehat{H}_{i}=0,\label{eq: 4}
\end{equation}
where we denoted
\[
\widehat{H}_{i}=\frac{1}{T}\sum_{t}\hat{h}(\hat{f}_{t}\hat{\lambda}_{i}),\quad\text{for all }i=1,\dots,N.
\]
Taking the probability limits of these two equations and using Theorem
\ref{thm: consistency}, it thus follows that $\alpha_{0}$ and $\beta_{0}$
satisfy
\begin{equation}
\frac{1}{NT}\sum_{i=1}^{N}\sum_{t=1}^{T}E[r_{it}-\alpha_{0}-\beta_{0}h(f_{t}\lambda_{i})]=o(1),\label{eq: limit 3}
\end{equation}
\begin{equation}
\frac{1}{NT}\sum_{i=1}^{N}\sum_{t=1}^{T}E[(r_{it}-\alpha_{0}-\beta_{0}h(f_{t}\lambda_{i}))\bar{H}_{i}]=o(1),\label{eq: limit 4}
\end{equation}
where we denoted
\[
\bar{H}_{i}=\frac{1}{T}\sum_{t=1}^{T}E[h(f_{t}\lambda_{i})],\quad i=1,\dots,N.
\]
Comparing Equations (\ref{eq: limit 3})-(\ref{eq: limit 4}) with
(\ref{eq: limit 1})-(\ref{eq: limit 2}), it thus follows that unless
the HFL model is correct, we generally have $\alpha_{0}=0$ and $\beta_{0}=1$
only if $\overline{H}_{i}=h(f_{t}\lambda_{i})$ for all $i=1,\dots,N$
and $t=1,\dots,T$, which is equivalent to $h(f_{t}\lambda_{i})=h(f_{s}\lambda_{i})$
for all $i=1,\dots,N$ and $t,s=1,\dots,T$, meaning that there is
no time-series variation in the factor $f_{t}$.

\subsection{Related Econometrics Literature}

Our model (\ref{eq: model}) is related to but different from those
in the recent literature on non-linear factor and two-way fixed effect
models in econometrics. For example, \citet{FW16}, \citet{BL17},
\citet{C17}, \citet*{CFW21}, \citet{W22}, and \citet*{GLPY23}
considered the model of the form $r_{it}\mid x_{it}\sim h(\cdot\mid x_{it}^{\top}\beta+f_{t}^{\top}\lambda_{i})$
and its variants, where $x_{it}$ is a vector of covariates, $\beta$
is a vector of coefficients, and $h$ is some function. The key difference
from our model here is that their function $h$ is assumed to be known.
\citet{MW22} studied the model of the form $r_{it}\mid x_{it}\sim h(\cdot\mid x_{it}^{\top}+f_{t}+\gamma_{i})$,
where the function $v\mapsto h(\cdot\mid v)$ is unknown but strictly
increasing, and assumed the existence of the compensating variable,
which is a component of $x_{it}$ that can ``undo'' the variation
in $f_{t}+\lambda_{i}$. \citet*{CDG21} and \citet{AB20} analyzed
the quantile factor model, where the $\tau$th quantile of the distribution
of $y_{it}$ is equal to $f_{t}^{\top}\lambda_{i}$ or to $x_{it}^{\top}\beta+f_{t}^{\top}\lambda_{i}$,
respectively. \citet*{CFW20} studied the distribution model, where
$P(r_{it}\leq\cdot\mid x_{it})=\Lambda(x_{it}^{\top}\beta(\cdot)+f_{t}(\cdot)+\lambda_{i}(\cdot))$
for some known function $\Lambda$.

In addition, our model (\ref{eq: model}) is related to those in the
literature on network models but our analysis is different. For example,
\citet{Z20} worked with the model $r_{ij}=x_{ij}^{\top}\beta+h(\xi_{i},\xi_{j})+e_{ij}$,
where $i,j=1,\dots,N$, $x_{ij}$ is a pair-specific covariate, $\xi_{i}$
and $\xi_{j}$ are individual effects, and $h$ is an unknown function,
but his analysis is about identification and estimation of the vector
of parameters $\beta$, which relies on partialling out $h(\xi_{i},\xi_{j})$
rather than estimating it. \citet{G20} worked with the model of the
form $r_{ij}\mid(x_{i},x_{j})\sim h(\cdot\mid w(x_{i},x_{j})+\xi_{i}+\xi_{j})$,
where $w$ is an unknown function but $v\mapsto h(\cdot\mid v)$ is
\textit{strictly increasing}.

More generally, it appears that non-linear factor models were first
developed in the psychology literature; see \citet{B53} and \citet{M62}.
\citet{M79} developed an estimator of parametric non-linear factor
models. \citet{ZL99} studied a non-linear factor model of the form
$r_{it}=\lambda_{i}^{\top}h(f_{t})+\varepsilon_{it}$, where $h$
is a vector of known functions. Other relevant references to classic
results can be found in a review \citet{YA01}. \citet{W22} also
provided several references for the analysis of non-linear factor
models with small $N$.

\section{Empirical Appendix}

This section presents the results of additional cross-sectional asset
pricing tests. Section \ref{subsec:Cross-sectional-asset-pricing-nocost-appendix}
presents results of the cross-sectional tests with no intercept. Section
\ref{subsec:Cross-sectional-asset-pricing-asset-class-appendix} presents
results of the cross-sectional tests by asset class. Section \ref{subsec:Significant-factors-in-dslasso}
presents further results about the significant factors in the double-selection
Lasso estimation. Section \ref{subsec:Comparison-with-factor-zoo-chen-zimmermann}
presents a comparison with an alternative factor zoo. Section \ref{subsec:Comparison-with-PCA6}
presents a comparison with PCA models with up to six factors. Section
\ref{subsec:Before-and-after} presents a further robustness analysis
considering the sample before and after the 2008 financial crisis.
Section \ref{subsec:Correlation-with-benchmark} presents a correlation
analysis with benchmark equity factors.

\subsection{Cross-sectional asset pricing tests with no constant\label{subsec:Cross-sectional-asset-pricing-nocost-appendix}}

Figure \ref{fig:Actual-vs-Predicted_no_intercept} plots the average
asset excess returns versus the predicted excess returns using the
HFL model with the constraint that the intercept is equal to zero
(i.e., $E[r_{it}]=\beta E[h(\lambda_{i}f_{t})]+\epsilon_{i}$). Note
that the latter is one of the implications of our factor model, which
we test in the main analysis. The figure reveals that, as for the
specification based on the model with a constant, all assets line
up along the 45 degree line. 
\begin{center}
\begin{figure}[H]
\caption{Actual vs Predicted Expected Returns -- No Intercept\label{fig:Actual-vs-Predicted_no_intercept}}

{\footnotesize{}\medskip{}
}{\footnotesize\par}

{\small{}The figure plots actual average excess returns on all tested
assets ($E[r_{it}]$) versus predicted excess returns using the cross-sectional
model based on the HFL model with no intercept: $E[r_{it}]=\beta E[h(\lambda_{i}f_{t})]+\epsilon_{i}$.
The red line denotes the 45 degree line. The HFL model is based on
a degree of the polynomial used to approximate the function $h(.)$
equal to 4.}{\small\par}

{\footnotesize{}\medskip{}
}{\footnotesize\par}
\centering{}\includegraphics[width=15cm]{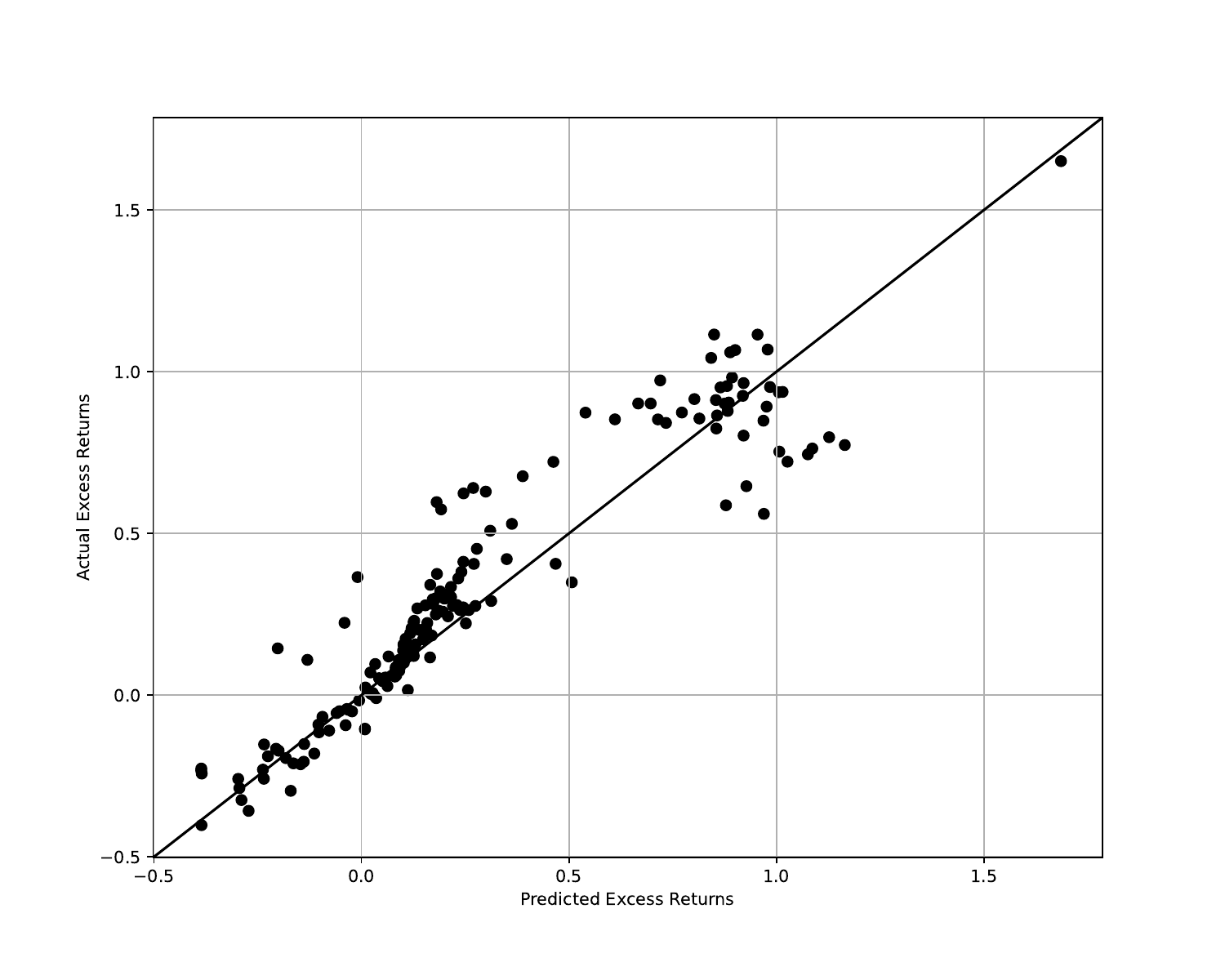}
\end{figure}
\par\end{center}

\subsection{Cross-sectional asset pricing tests by asset class\label{subsec:Cross-sectional-asset-pricing-asset-class-appendix}}

In this section we present visual evidence of the cross-sectional
performance of the HFL model in asset pricing tests by asset class.
Figure \ref{fig:Actual-vs-Predicted_subplots} plots, separetely for
all assets and for the different asset classes, actual average excess
returns ($E[r_{it}]$) versus predicted excess returns using the cross-sectional
model based on the HFL model: $E[r_{it}]=\alpha+\beta E[h(\lambda_{i}f_{t})]+\epsilon_{i}$.
The asset classes are: equities, US bonds, international bonds, commodities
and foreign exchange. The figures show that, also for the individual
asset classes, as for the model tested on all assets, individual assets
line up along the 45 degree line. 
\begin{center}
\begin{figure}[H]
\caption{Actual vs Predicted Expected Returns -- By Asset Class\label{fig:Actual-vs-Predicted_subplots}}

{\footnotesize{}\medskip{}
}{\footnotesize\par}

{\small{}The figure plots, separetely for all assets and for the different
asset classes, actual average excess returns ($E[r_{it}]$) versus
predicted excess returns using the cross-sectional model based on
the HFL model: $E[r_{it}]=\alpha+\beta E[h(\lambda_{i}f_{t})]+\epsilon_{i}$.
The asset classes are: equities, US bonds, international bonds, commodities
and foreign exchange. The HFL model is estimated using all assets.
The red line denotes the 45 degree line. The HFL model is based on
a degree of the polynomial used to approximate the function $h(.)$
equal to 4.}{\small\par}

{\footnotesize{}\bigskip{}
}{\footnotesize\par}
\begin{centering}
\subfloat[All assets]%
{\begin{centering}
\includegraphics[width=8cm]{0_Users_nicolaborri_Dropbox_Research_INDEX-EVERYTHING_lyx_Figure_Scatter_ALL.pdf}
\par\end{centering}
}%
\subfloat[Equities]%
{\begin{centering}
\includegraphics[width=8cm]{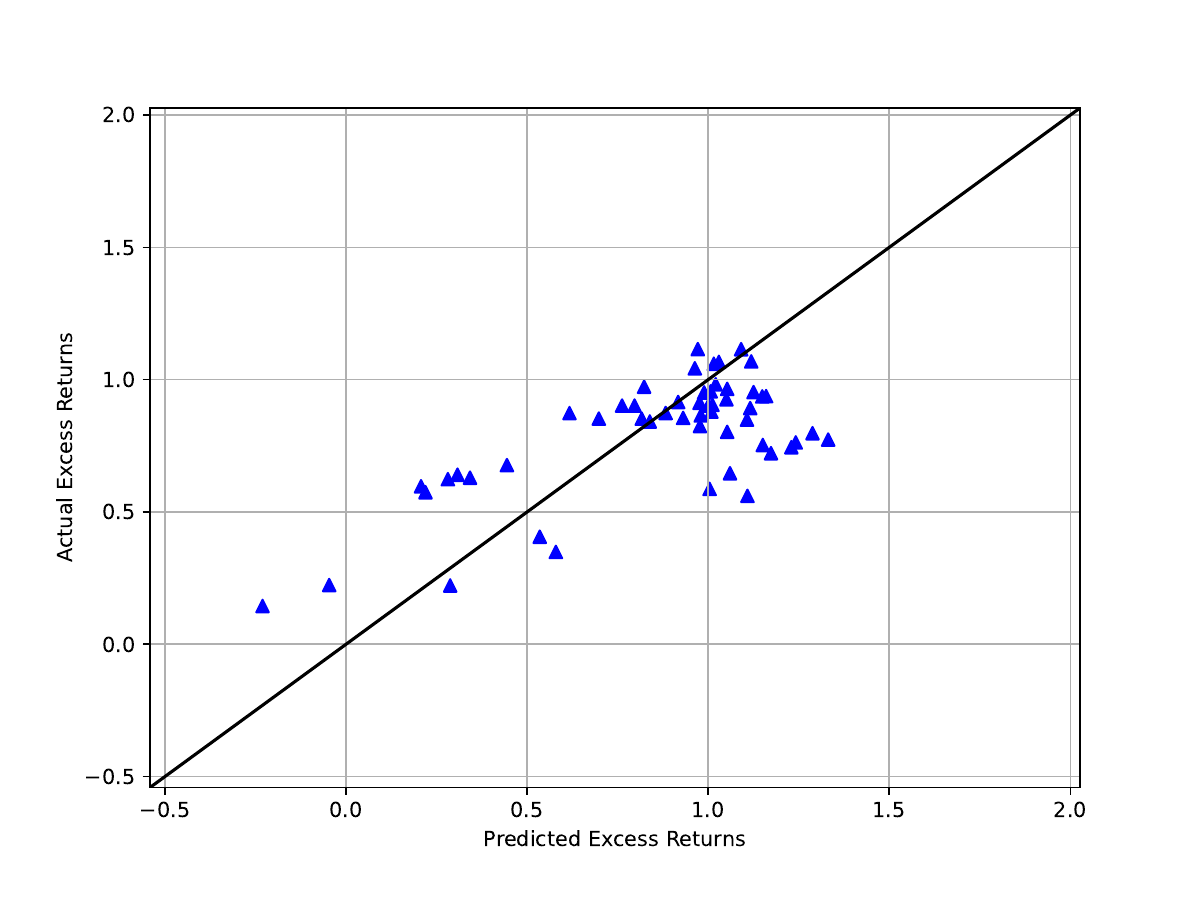}
\par\end{centering}
}
\par\end{centering}
\begin{centering}
\subfloat[US Bonds]%
{\begin{centering}
\includegraphics[width=8cm]{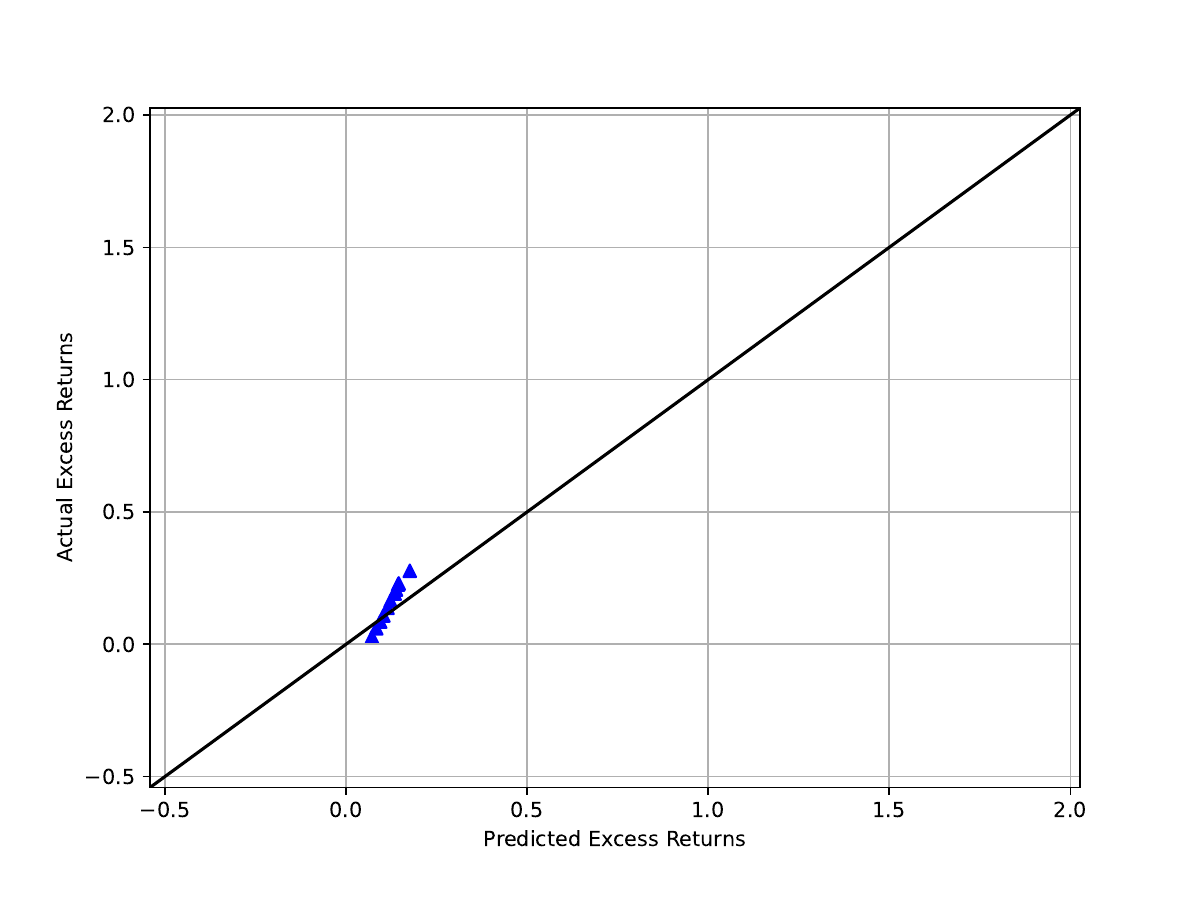}
\par\end{centering}
}%
\subfloat[International Bonds]%
{\begin{centering}
\includegraphics[width=8cm]{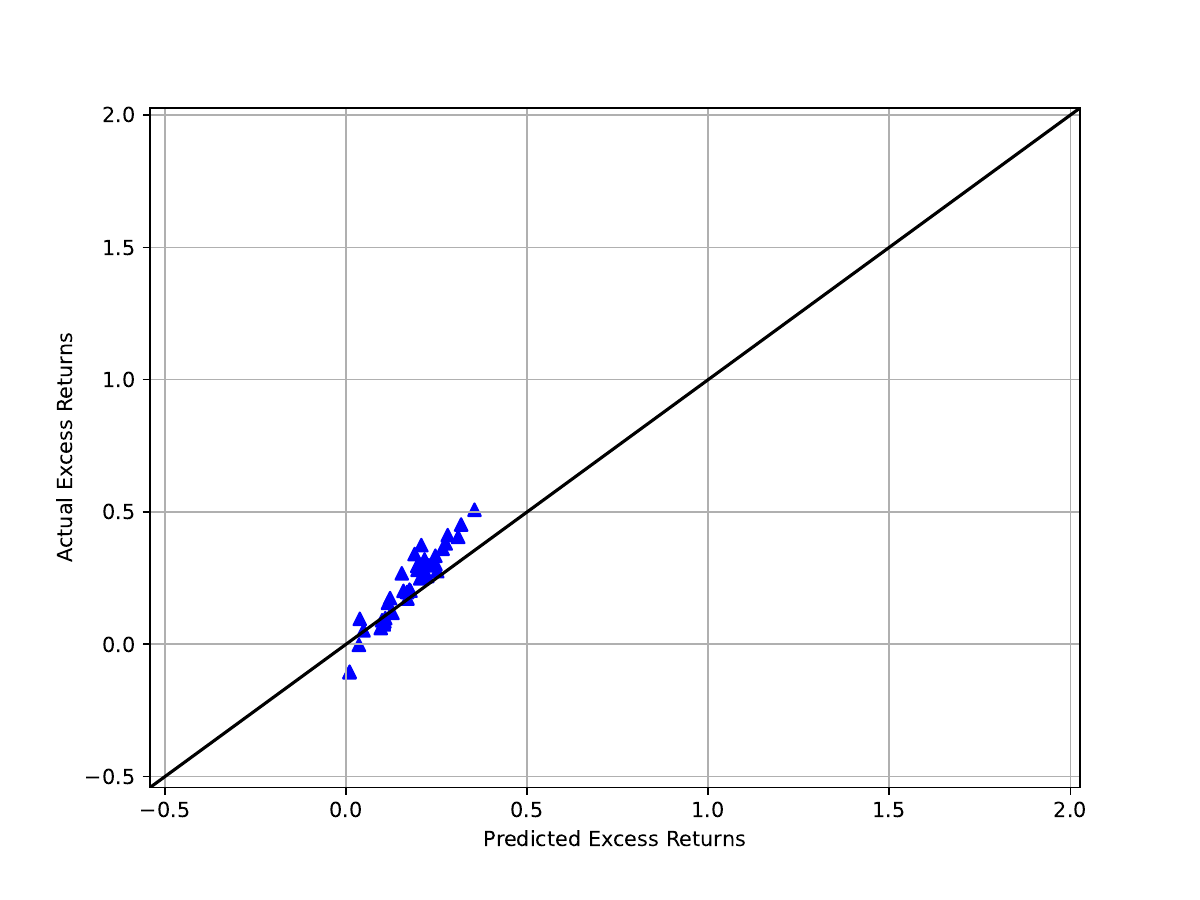}
\par\end{centering}
}
\par\end{centering}
\centering{}%
\subfloat[Commodities]%
{\begin{centering}
\includegraphics[width=8cm]{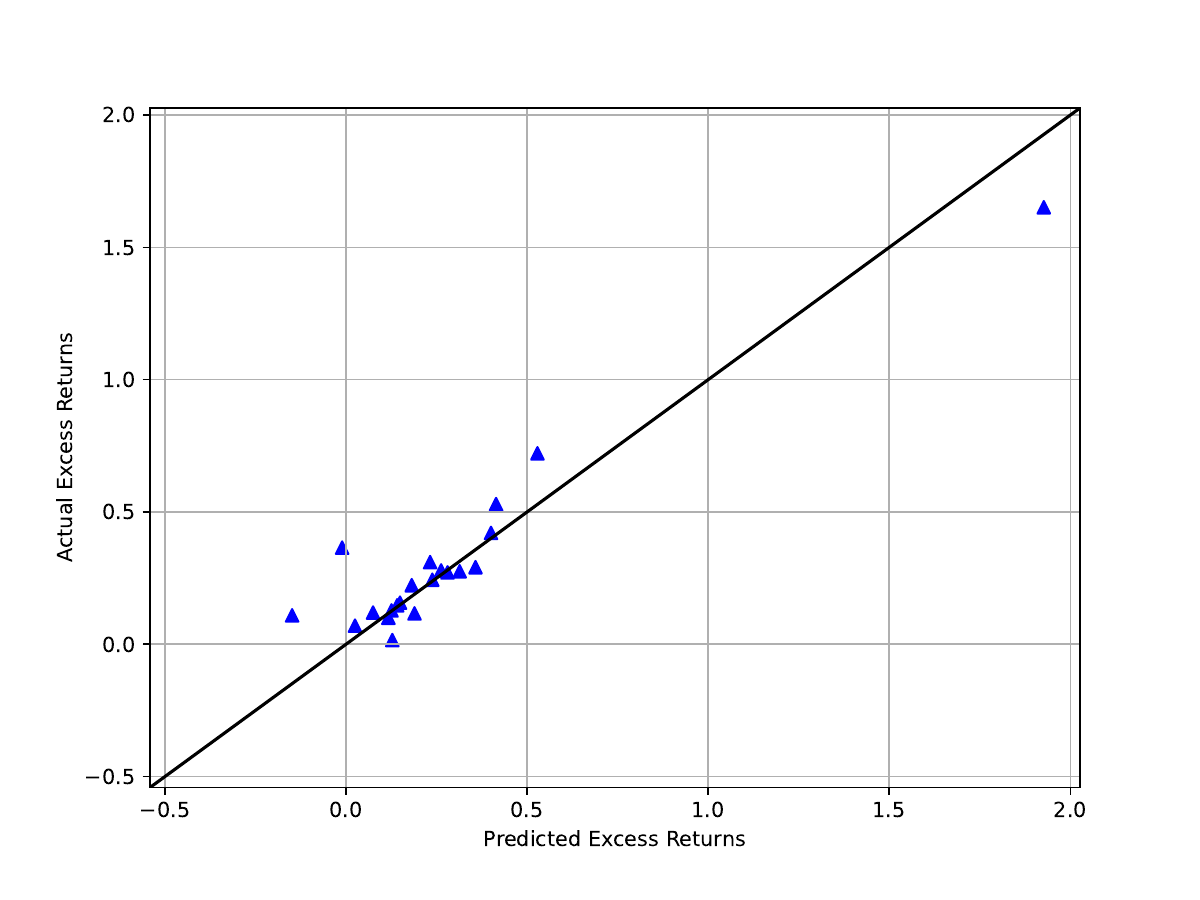}
\par\end{centering}
}%
\subfloat[Foreign exchange]%
{\begin{centering}
\includegraphics[width=8cm]{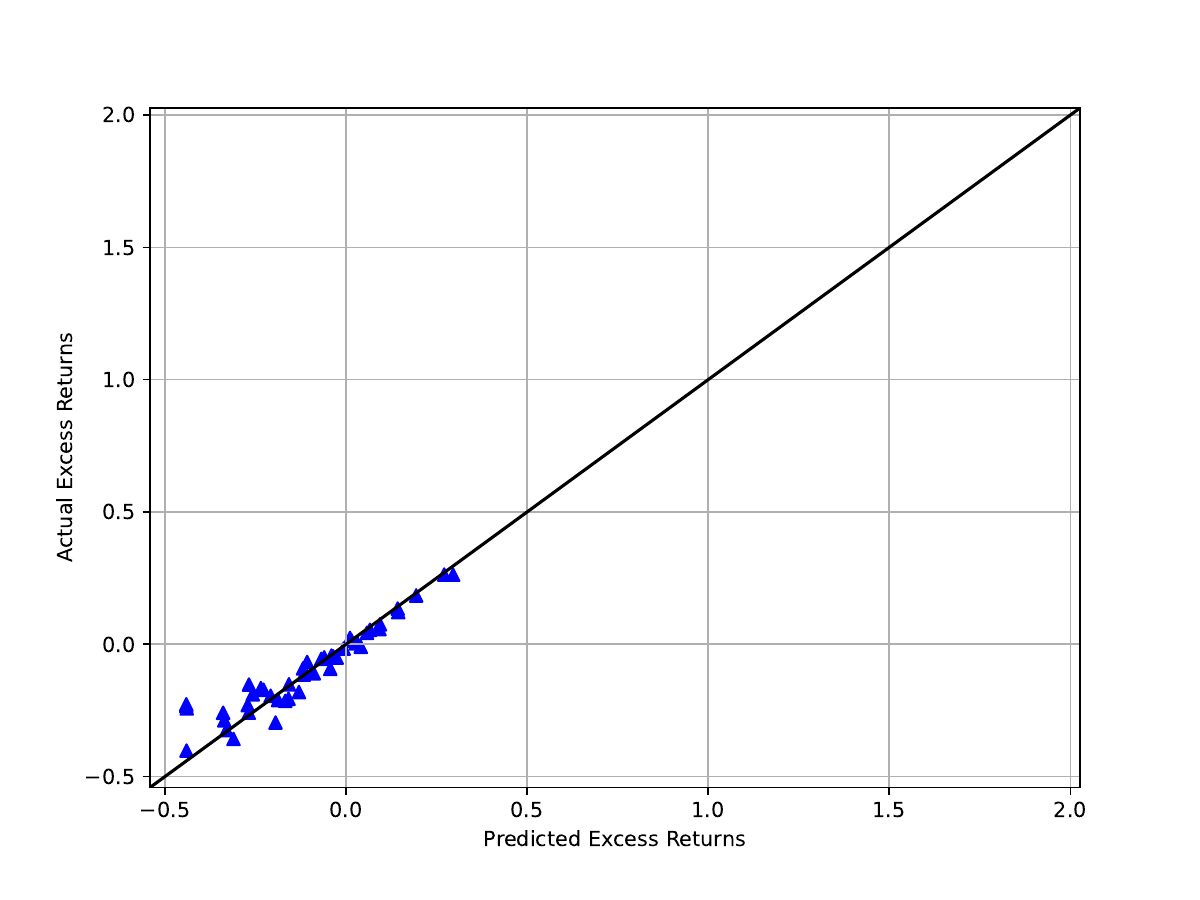}
\par\end{centering}
}
\end{figure}
\par\end{center}

\subsection{Significant factors in the factor zoo using double-selection Lasso\label{subsec:Significant-factors-in-dslasso}}

In this section, we report further results of the double-selection
Lasso estimation discussed in Section \ref{subsec:Comparison-with-the-factor-zoo}.
Specifically, for each factor in the factor zoo, we estimate the double-selection
Lasso model using all the remaining factors from the factor zoo and
the HFL component as controls. The factor zoo includes factors for
153 characteristics in 13 themes, using data from 93 countries and
four regions, constructed by \citet*{jensen2023there}. Table \ref{tab:Significant-Factors-from-DS-LASSO}
summarizises the additional results of the double-selection estimation,
listing all the factors from the factor zoo significant at least at
the 10\% confidence level, along the corresponding $p$-value and
$t$-statistics. The factor and factor names are from \citet*{jensen2023there}.

\begin{table}[H]
\caption{Significant Factors in the Factor Zoo Using Double-Selection Lasso\label{tab:Significant-Factors-from-DS-LASSO}}

\medskip{}

{\small{}This table reports further results from the double-selection
Lasso estimation for the factors in the factor zoo. For each factor
in the factor zoo, we estimation the double-selection Lasso model
using all the remaining factors from the factor zoo and the HFL component
as controls. The table lists all the factors from the factor zoo which
are significant at least at the 10\% confidence level, along the corresponding
$p$-value and $t$-statistics. Standard errors are adjusted with
the Fama-MacBeth procedure. The factor and the factor names are from
\citet*{jensen2023there}.}{\small\par}

\bigskip{}

\centering{}%
\begin{tabular}{lcc}
\hline 
{\small{}Factors in Factor Zoo} &
{\small{}$p$-value} &
{\small{}$t$-stat}\tabularnewline
\hline 
 &
 &
\tabularnewline
at\_turnover &
0 &
3.969\tabularnewline
beta\_60m &
0.037 &
2.093\tabularnewline
beta\_dimson\_21d &
0.078 &
1.765\tabularnewline
eq\_dur &
0.037 &
2.094\tabularnewline
inv\_gr1a &
0.094 &
1.682\tabularnewline
niq\_su &
0.023 &
-2.275\tabularnewline
nncoa\_gr1a &
0.021 &
-2.313\tabularnewline
qmj &
0.032 &
-2.158\tabularnewline
rd\_sale &
0.07 &
1.82\tabularnewline
resff3\_12\_1 &
0.004 &
2.894\tabularnewline
sale\_emp\_gr1 &
0.046 &
2.005\tabularnewline
seas\_2\_5an &
0.017 &
-2.402\tabularnewline
taccruals\_ni &
0.058 &
1.905\tabularnewline
zero\_trades\_252d &
0.054 &
-1.937\tabularnewline
 &
 &
\tabularnewline
\hline 
\end{tabular}
\end{table}

\subsection{Comparison with the factors from \citet{chen2021open}\label{subsec:Comparison-with-factor-zoo-chen-zimmermann}}

In section \ref{subsec:Comparison-with-the-factor-zoo} of the main
paper, we compared the HFL model with the factors from the factor
zoo presented in \citet*{jensen2023there} and showed that they become
insignificant after controlling for the HFL component. In this section,
we show that also the factors from the factor zoo presented in \citet{chen2021open}
become insignificant after controlling for the HFL component. 

In particular, we consider 212 factors based on stock return predictors.\footnote{The data for the factor zoo is available through the Open Source Asset
Pricing webpage.} For each of these factors (denoted with $f^{zoo}$), separately,
we first estimate the cross-sectional regression $E[r_{it}]=\alpha+\beta^{zoo}E[f_{t}^{zoo}]+\epsilon_{i}$.
Next, we repeat the estimation additionally including the predicted
values obtained with the HFL model (i.e., $E[h(\lambda_{i}f_{t})])$.
Figure \ref{fig:Comparison-with-the-factor-zoo-chen-zimmermann} summarizes
our results. It plots the average of the absolute values of the $t$-statistics
in a test of the intercepts and slope coefficients in the cross-sectional
regressions, respectively, equal to zero. In the left panel of the
figure, we consider the regressions which only include the average
predicted values based on the factor zoo. In the right panel of the
figure, we consider regressions which additionally include the average
predicted values from the HFL model. The figure reveals that, for
the models which do not include the HFL component, the estimates of
the slope coefficients are on average significantly different from
zero (average absolute $t$-statistics equal to 2.10). For these models,
though, also the pricing error, captured by the estimates of the intercept,
is on average statistically significant (average absolute $t$-statistics
equal to 2.83). Furthermore, the figure reveals that, in models which
additionally include the average predicted values from the HFL model,
the slope coefficients associated with the factor zoo become not statistically
different from zero (average absolute $t$-statistics equal to 0.80),
while those associated with the HFL component are highly significant
(average absolute t-statistics equal to 5.25). Moreover, for the models
which includes the HFL component, the estimates of the pricing error
also become not statistically different from zero (average $t$-statistics
equal to 0.70).
\begin{center}
\begin{figure}[H]
\caption{Comparison with the Factors from \citet{chen2021open}\label{fig:Comparison-with-the-factor-zoo-chen-zimmermann}}

{\footnotesize{}\medskip{}
}{\footnotesize\par}

{\small{}This figure shows the average of the absolute values of the
$t$-statistics in a test of the intercepts and slope coefficients
in the cross-sectional regressions, respectively, equal to zero. Standard
errors are adjusted with the Fama-MacBeth procedure. In the left panel,
we consider the cross-sectional regressions $E[r_{it}]=\alpha+\beta^{zoo}E[f_{t}^{zoo}]+\epsilon_{i}$
for each factor from the factor zoo. In the right panel, we consider
cross-sectional regressions which additionally include the average
predicted values based on the HFL component (i.e., $E[h(\lambda_{i}f_{t})])$.
The vertical bars denote one standard deviation around the mean. The
horizontal black dashed-line denote significance at the 10\% confidence
level. The data for 212 factors from the factor zoo are from }\citet{chen2021open}{\small{}.
The HFL model is based on a degree of the polynomial used to approximate
the function $h(.)$ equal to 4.}{\small\par}

{\footnotesize{}\medskip{}
}{\footnotesize\par}
\centering{}{\small{}}%
\subfloat{\centering{}{\small{}\includegraphics[width=8cm]{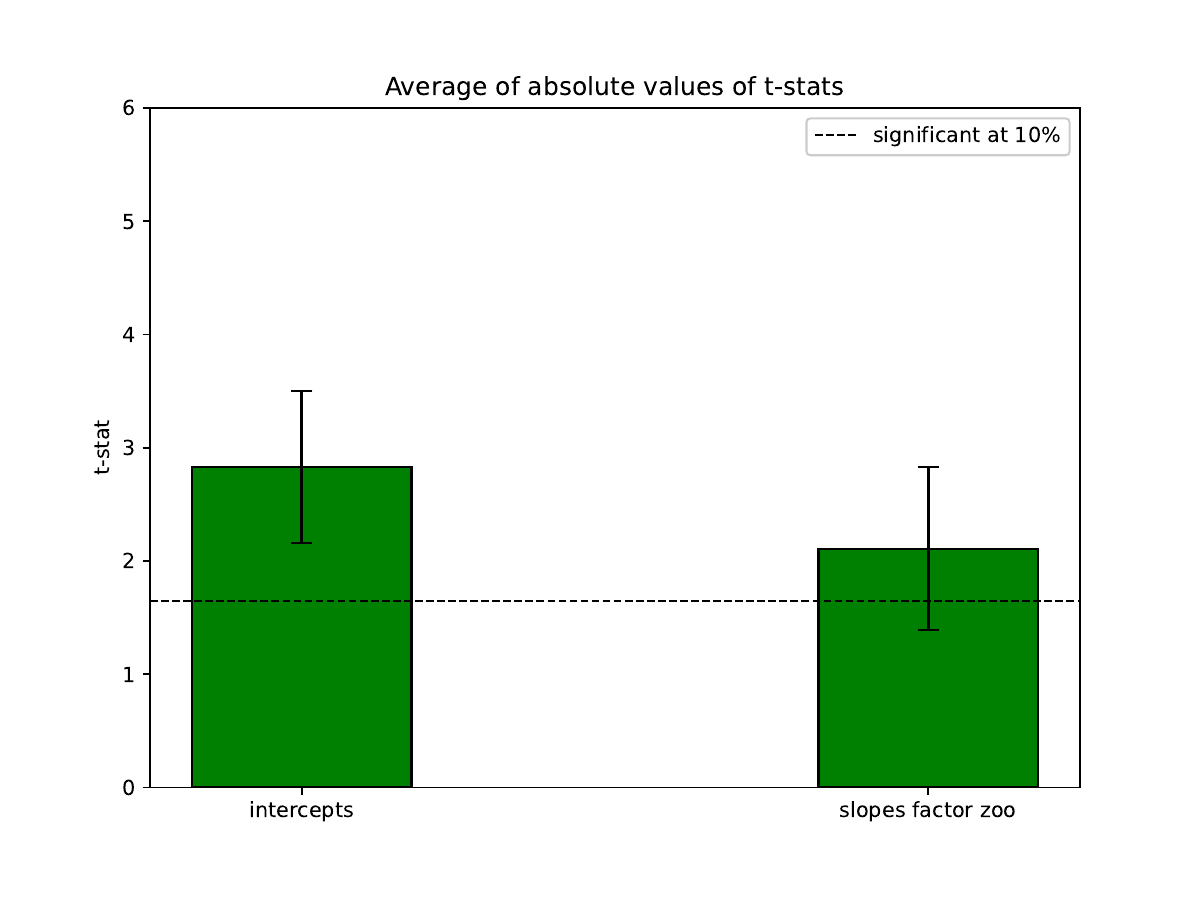}}}{\small{}}%
\subfloat{\centering{}{\small{}\includegraphics[width=8cm]{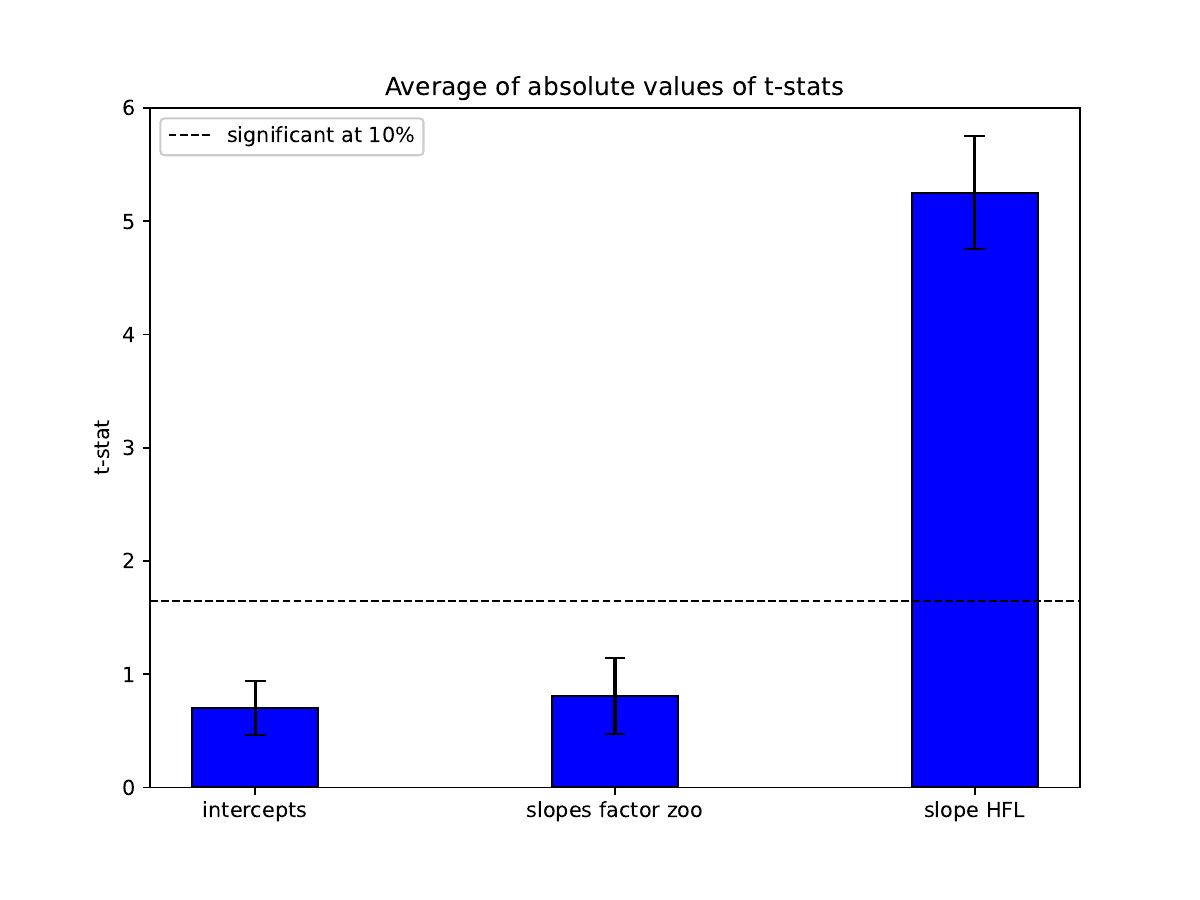}}}{\small\par}
\end{figure}
\par\end{center}

Table \ref{tab:Compairsaon-with-the-factor-zoo-chen-zimmerman} provides
further details about the comparison with the factor zoo. If the HFL
component is not included (Panel A), the fraction of models with a
slope coefficient on the factor zoo which is significant at the 10\%
confidence level is approximately 80\%. When we further include the
HFL component, this fraction drops to less than 1\%. Moreover, if
the HFL component is included (Panel B), the fraction of models with
a slope on the HFL component which is significant at the 1\% confidence
level is about 100\%. In these models, the intercept is statisticall
different from zero at a confidence level of 10\% or below in only
0.5\% of the cases. Finally, we find that only one of the 212 factors
from the factor zoo is significant with a confidence level of 10\%.
This factor is the industry return of big firms of \citet{hou2007industry}.

Furthermore, the table also reports the results of cross-sectional
asset pricing regressions based on the double-selection Lasso method
of \citet*{belloni2014inference} and \citet*{chernozhukov2018double}
(Panel C). Double-selection Lasso allows estimating regressions where
the number of right-hand side variables could be large, even larger
than the sample size. In practice, double-selection Lasso estimates
only one coefficient on the right-hand side at a time by means of
a two-step selection method. In the first, factors with a low contribution
to the cross-sectional pricing are excluded from the set of controls.
In the second step, factors whose covariances with returns are highly
correlated in the cross-section with the covariance between returns
and a given factor are added to the set of controls. \citet*{feng2020taming}
propose double-selection Lasso as model selection method to evaluate
the contribution of factors in cross-sectional asset pricing regressions.
The estimates based on the double-selection Lasso method confirm the
results based on pairwise comparisons of the factor zoo with the HFL
compoment. In particular, the HFL component using the double-selection
Lasso method is statistically significant at the 1\% level ($t$-statistics
of 7.67). The fraction of slope coefficients on the factor zoo with
$p$-values greater than 10\% is 94.33. 

\subsection{Comparison with PCA models with up to six factors\label{subsec:Comparison-with-PCA6}}

In this section, we complement the comparison with PCA models (see
Table \ref{tab:Comparison-with-PCA}). Specifically, we compare the
HFL model with PCA models with up to six factors. Including additional
principal components is important as some components might explain
a large fraction of the time-series variation, but very little of
the cross-sectional variation (e.g., the first component which is
often only a ``level'' factor for the cross-section of asset returns).
Additional components, on the contrary, might explain a small fraction
of the time-series variation, but a large fraction of the cross-sectional
variation in returns (see, e.g., \citet{lettau2020factors}, \citet*{kozak2020shrinking}
or \citet{giglio2021asset} and the review paper by \citet*{giglio2022factor}).

Table \ref{tab:Comparison-with-PCA-higher} presents the results of
cross-sectional asset pricing tests using all assets and three estimators
of the latent factors related to PCA. The first estimator is based
on the standard PCA of the covariance matrix of asset returns. The
second estimator, called RP-PCA, is a generalization of PCA, proposed
by \citet{lettau2020factors}, which includes a penalty term to account
for the pricing errors in the cross-sectional regressions based on
the RP-PCA factors. RP-PCA is motivated by the poor performance of
standard PCA in identifying factors which are relevant to explain
the cross-section of average excess returns (see, e.g., \citet{onatski2012asymptotics}
and \citet*{kozak2020shrinking}). The third estimator is the kernel-PCA,
proposed by \citet*{scholkopf1998nonlinear}. Kernel-PCA (K-PCA) is
a nonlinear form of PCA which allows for the separability of nonlinear
data by projecting it onto a higher dimensional space where it becomes
linearly separable using kernels.

In all models, we include the HFL component ($h\left(f_{t}\lambda_{i}\right)$),
and evaluate the effect of including, respectively, the first four
principal components, the first five principal components, and the
first six principal components, separately for standard PCA, RP-PCA
and K-PCA. The estimates from the cross-sectional asset pricing regressions
reveal that the slope coefficient associated with the HFL component
is statistically significant at conventional levels in all models,
while the slopes associated with the principal components, for all
three estimators we consider, are never statically different from
zero.

\begin{table}[H]
\caption{Comparison with the Factors from \citet{chen2021open}\label{tab:Compairsaon-with-the-factor-zoo-chen-zimmerman}}
{\footnotesize{}\medskip{}
}{\footnotesize\par}

{\small{}This table provides a comparison of the HFL model and the
factor zoo. Panel A refers to cross-sectional regressions $E[r_{it}]=\alpha+\beta^{zoo}E[f_{t}^{zoo}]+\epsilon_{i}$
for each factor from the factor zoo. Standard errors are adjusted
with the Fama-MacBeth procedure. Panel B refers to cross-sectional
regressions which additionally include the average predicted values
based on the HFL component (i.e., $E[h(\lambda_{i}f_{t})])$. The
two panels report the fraction (in percentage) of intercepts and slope
coefficients with $p$-values less than 1\%, between 1 and 5\%, between
5 and 10\% and larger than 10\%. The $p$-values correspond to tests
for the coefficient equal to zero. Panel C refers to the model which
includes the HFL component and estimates based on the double-selection
Lasso method, and reports the fraction of slope coefficients on the
factor zoo with $p$-values less than 1\%, between 1 and 5\%, between
5 and 10\% and larger than 10\%. The HFL component using the double-selection
Lasso method is statistically significant at the 1\% level. The data
for 212 factors from the factor zoo are from }\citet{chen2021open}{\small{}.
The HFL model is based on a degree of the polynomial used to approximate
the function $h(.)$ equal to 4.}{\footnotesize{}\bigskip{}
}{\footnotesize\par}
\centering{}%
\begin{tabular}{ccccc}
\hline 
$p(coeff=0)$ &
$p\leq1\%$ &
$1\%<p\leq5\%$ &
$5\%<p\leq10\%$ &
$p>10\%$\tabularnewline
\hline 
 &
 &
 &
 &
\tabularnewline
 &
\multicolumn{4}{c}{Panel A: Model without HFL}\tabularnewline
\cline{2-5} \cline{3-5} \cline{4-5} \cline{5-5} 
 &
 &
 &
 &
\tabularnewline
$\alpha$ &
70.28 &
17.92 &
3.30 &
8.49\tabularnewline
$E[f_{t}^{zoo}]$ &
24.05 &
45.75 &
8.49 &
21.69\tabularnewline
 &
 &
 &
 &
\tabularnewline
 &
\multicolumn{4}{c}{Panel B: Model with HFL}\tabularnewline
\cline{2-5} \cline{3-5} \cline{4-5} \cline{5-5} 
 &
 &
 &
 &
\tabularnewline
$\alpha$ &
0 &
0 &
0.48 &
99.52\tabularnewline
$E[f_{t}^{zoo}]$ &
0 &
0 &
0.48 &
99.52\tabularnewline
$h\left(f_{t}\lambda_{i}\right)$ &
1 &
0 &
0 &
0\tabularnewline
 &
 &
 &
 &
\tabularnewline
 &
\multicolumn{4}{c}{Panel C: Model with HFL and double-selection Lasso}\tabularnewline
\cline{2-5} \cline{3-5} \cline{4-5} \cline{5-5} 
 &
 &
 &
 &
\tabularnewline
$\alpha$ &
32.54 &
65.56 &
1.88 &
0\tabularnewline
$E[f_{t}^{zoo}]$ &
0.47 &
0.94 &
4.24 &
94.33\tabularnewline
 &
 &
 &
 &
\tabularnewline
\hline 
\end{tabular}
\end{table}

{\footnotesize{}}
\begin{sidewaystable}[H]
{\footnotesize{}\caption{Comparison with PCA Factor Models up to Degree Six\label{tab:Comparison-with-PCA-higher}}
}{\footnotesize\par}

{\footnotesize{}\medskip{}
}{\footnotesize\par}

{\small{}This table presents the results of cross-sectional asset
pricing tests using all assets. We estimate the cross-sectional linear
model $E[r_{it}]=\alpha+\beta E[h(\lambda_{i}f_{t})]+\epsilon_{i}$
augmented by, respectively, the first four principal components, the
first five principal components and the first six principal components,
extracted by the test asset returns. For the principal components,
we consider three estimators: the standard PCA based on the covariance
matrix of asset returns; the risk premium PCA (RP-PCA) of \citet{lettau2020factors},
and the kernel PCA (K-PCA) of \citet*{scholkopf1998nonlinear}. We
report the estimates for the slopes and intercept ($\alpha$), and
standard errors adjusted with the Fama-MacBeth procedure in parenthesis.
The table additionally reports the regression adjusted R-squared,
and the mean absolute pricing error (MAPE) in percentage terms. {*},
{*}{*}, and {*}{*}{*} represent significance at the 1\%, 5\%, and
10\% level. We further report the number of assets in the cross-sectional
regression and the number of monthly observations for each asset used
in the estimation of the averages of the excess returns and predictions
from the model. The HFL model is based on a degree of the polynomial
used to approximate the function $h(.)$ equal to 4.}{\small\par}

{\footnotesize{}\bigskip{}
}{\footnotesize\par}
\centering{}{\small{}}%
\begin{tabular}{lccccccccccc}
\hline 
 &
\multicolumn{3}{c}{Standard PCA} &
 &
\multicolumn{3}{c}{RP-PCA} &
 &
\multicolumn{3}{c}{K-PCA}\tabularnewline
\hline 
 &
PC1--PC4 &
PC1--PC5 &
PC1--PC6 &
 &
PC1--PC4 &
PC1--PC5 &
PC1--PC6 &
 &
PC1--PC4 &
PC1--PC5 &
PC1--PC6\tabularnewline
$h\left(f_{t}\lambda_{i}\right)$ &
0.657{*}{*}{*} &
0.613{*}{*}{*} &
0.595{*}{*}{*} &
 &
0.633{*}{*}{*} &
0.515{*}{*}{*} &
0.612{*}{*}{*} &
 &
0.734{*}{*}{*} &
0.746{*}{*}{*} &
0.757{*}{*}{*}\tabularnewline
 &
(0.109) &
(0.109) &
(0.109) &
 &
(0.113) &
(0.123) &
(0.106) &
 &
(0.127) &
(0.125) &
(0.126)\tabularnewline
PC1 &
0.014 &
0.013 &
0.012 &
 &
0.016 &
0.021 &
0.014 &
 &
0.007 &
0.008 &
0.006\tabularnewline
 &
(0.019) &
(0.019) &
(0.019) &
 &
(0.019) &
(0.019) &
(0.019) &
 &
(0.016) &
(0.016) &
(0.016)\tabularnewline
PC2 &
0.001 &
0.001 &
0.001 &
 &
0.002 &
0.005 &
0.003 &
 &
0.008 &
0.009 &
0.008\tabularnewline
 &
(0.010) &
(0.010) &
(0.010) &
 &
(0.010) &
(0.010) &
(0.010) &
 &
(0.013) &
(0.013) &
(0.013)\tabularnewline
PC3 &
-0.005 &
-0.007 &
-0.009 &
 &
0.003 &
0.005 &
0.005 &
 &
-0.008 &
-0.010 &
-0.005\tabularnewline
 &
(0.010) &
(0.010) &
(0.010) &
 &
(0.010) &
(0.010) &
(0.010) &
 &
(0.014) &
(0.013) &
(0.014)\tabularnewline
PC4 &
-0.003 &
-0.003 &
-0.004 &
 &
0.003 &
0.003 &
0.003 &
 &
-0.008 &
-0.006 &
-0.002\tabularnewline
 &
(0.007) &
(0.007) &
(0.007) &
 &
(0.007) &
(0.007) &
(0.007) &
 &
(0.016) &
(0.015) &
(0.015)\tabularnewline
PC5 &
 &
-0.008 &
-0.009 &
 &
 &
0.002 &
0.001 &
 &
 &
0.004 &
0.007\tabularnewline
 &
 &
(0.007) &
(0.007) &
 &
 &
(0.007) &
(0.007) &
 &
 &
(0.014) &
(0.013)\tabularnewline
PC6 &
 &
 &
-0.005 &
 &
 &
 &
-0.008 &
 &
 &
 &
-0.023\tabularnewline
 &
 &
 &
(0.006) &
 &
 &
 &
(0.006) &
 &
 &
 &
(0.019)\tabularnewline
$\alpha$ &
0.001 &
0.001{*}{*} &
0.001{*}{*}{*} &
 &
0.001 &
0.001{*}{*} &
0.001{*}{*} &
 &
0.001 &
0.000 &
0.000\tabularnewline
 &
(0.000) &
(0.000) &
(0.000) &
 &
(0.000) &
(0.000) &
(0.000) &
 &
(0.001) &
(0.001) &
(0.001)\tabularnewline
Adj $R^{2}$ &
0.916 &
0.936 &
0.943 &
 &
0.919 &
0.932 &
0.937 &
 &
0.902 &
0.903 &
0.917\tabularnewline
MAPE, \% &
0.091 &
0.074 &
0.071 &
 &
0.090 &
0.079 &
0.073 &
 &
0.103 &
0.102 &
0.092\tabularnewline
Assets &
171 &
171 &
171 &
 &
171 &
171 &
171 &
 &
171 &
171 &
171\tabularnewline
Months &
360 &
360 &
360 &
 &
360 &
360 &
360 &
 &
360 &
360 &
360\tabularnewline
 &
 &
 &
 &
 &
 &
 &
 &
 &
 &
 &
\tabularnewline
\hline 
\end{tabular}{\small\par}
\end{sidewaystable}
{\footnotesize\par}

\subsection{Before and after the 2008 financial crisis\label{subsec:Before-and-after}}

In this section, we present the results of cross-sectional asset pricing
tests in two alternative samples. The first is a pre-financial crisis
sample, which starts in January 1988 and ends in December 2008. The
second is a post-financial crisis sample, which starts in January
2009 and ends in December 2017. Table \ref{tab:Cross-Sectional-Asset-Pricing-Alt-Samples-Fincrisis}
presents the results for both all assets, and separately for each
asset class. When we include all assets, the slope coefficient associated
with the HKL component is statistically significant in both sample,
and quantitatively very similar. The slope is equal to 0.753 in the
pre-crisis sample, and 0.792 in the post-crisis sample. When we consider
the different asset classes separately, we find a significant slope
for each asset class, except for commodities, in both the pre-crisis
and post-crisis samples.

\subsection{Correlation with benchmark equity factors\label{subsec:Correlation-with-benchmark} }

In this section, we discuss the correlation matrix for the HFL component
and benchmark equity factors. Specifically, we consider the \citet{fama1993common}'s
three factors, the additional two factors from the \citet{fama2015five}'s
five factors model, and the momentum factor of \citet{jegadeesh1993returns}.
All equity factors are from Ken French's website. 

We first estimate $h(\lambda_{i}f_{t})$ for each asset $i$. Next,
we compute the sample correlation coefficient between the asset level
HFL components and the asset level factor betas. The latter are slope
coefficients in separate time-series regressions of asset excess returns
on each benchmark equity factor. 

Table \ref{tab:Correlation-Matrix} reports the correlation matrix.
The HFL component is highly positive correlated with assets' beta
exposures to the US equity market excess return (correlation coefficient
with MKTRF is 0.74), and positively related with assets' beta exposures
to the size factor (correlation coefficient with SMB is 0.43), the
value factor (correlation coefficient with HML is 0.48) and the operating
profitability factor (correlation coefficient with RMW 0.24). Moreover,
the HFL component is somewhat negatively related to the investment
factor (correlation coefficient with CMA is -0.10) and the momentum
factor (correlation coefficient with MOM is -0.12). Finally, the HFL
component is negatively correlated with assets' beta exposures to
$\Delta\log\left(VIX\right)$.

{\footnotesize{}}
\begin{table}[H]
{\footnotesize{}\caption{Cross-Sectional Asset Pricing by Asset Class: Before and After the
2008 Financial Crisis\label{tab:Cross-Sectional-Asset-Pricing-Alt-Samples-Fincrisis}}
}{\footnotesize\par}

{\footnotesize{}\medskip{}
}{\footnotesize\par}

{\small{}This table presents the results of cross-sectional asset
pricing tests for alternative samples: before and after the 2008 financial
crisis. The pre-financial crisis sample starts in January 1988 and
ends in December 2008. The post-financial crisis sample starts in
January 2009 and ends in December 2017. We estimate the cross-sectional
linear model $E[r_{it}]=\alpha+\beta E[h(\lambda_{i}f_{t})]+\epsilon_{i}$
and report the estimates for the slope ($h\left(f_{t}\lambda_{i}\right)$)
and intercept ($\alpha$), and standard errors adjusted with the Fama-MacBeth
procedure in parenthesis. Panel A reports estimates for the pre financial
crisis sample. Panel B reports estimates for the post financial crisis
sample. The table columns report estimates for the different asset
classes. The table additionally reports the regression adjusted R-squared,
the mean absolute pricing error (MAPE) in percentage terms. {*}, {*}{*},
and {*}{*}{*} represent significance at the 1\%, 5\%, and 10\% level.
We further report the number of assets in the cross-sectional regression
and the number of monthly observations for each asset used in the
estimation of the averages of the excess returns and predictions from
the model. The HFL model is based on a degree of the polynomial used
to approximate the function $h(.)$ equal to 4.}{\small\par}

{\footnotesize{}\bigskip{}
}{\footnotesize\par}
\centering{}{\small{}}%
\begin{tabular}{lcccccc}
\hline 
Panel A: pre-crisis &
All &
Equities &
US Bonds &
Intl Bonds &
Commodites &
FX\tabularnewline
\hline 
 &
 &
 &
 &
 &
 &
\tabularnewline
$h\left(f_{t}\lambda_{i}\right)$ &
0.753{*}{*}{*} &
0.438{*}{*} &
1.894{*}{*} &
1.271{*}{*}{*} &
0.549 &
1.079{*}{*}{*}\tabularnewline
 &
(0.188) &
(0.217) &
(0.783) &
(0.276) &
(0.476) &
(0.144)\tabularnewline
$\alpha$ &
0.001 &
0.004 &
-0.001 &
-0.000 &
0.002 &
0.002\tabularnewline
 &
(0.001) &
(0.004) &
(0.000) &
(0.000) &
(0.002) &
(0.001)\tabularnewline
Adj $R^{2}$ &
0.835 &
0.487 &
0.970 &
0.795 &
0.542 &
0.873\tabularnewline
MAPE, \% &
0.124 &
0.132 &
0.009 &
0.048 &
0.177 &
0.057\tabularnewline
Assets &
171 &
53 &
11 &
40 &
21 &
46\tabularnewline
Months &
240 &
240 &
240 &
240 &
240 &
240\tabularnewline
 &
 &
 &
 &
 &
 &
\tabularnewline
\hline 
Panel B: post-crisis &
All &
Equities &
US Bonds &
Intl Bonds &
Commodites &
FX\tabularnewline
\hline 
 &
 &
 &
 &
 &
 &
\tabularnewline
$h\left(f_{t}\lambda_{i}\right)$ &
0.792{*}{*} &
0.417{*}{*} &
3.829{*}{*} &
2.259{*}{*}{*} &
1.182 &
0.340{*}{*}\tabularnewline
 &
(0.374) &
(0.169) &
(1.673) &
(0.620) &
(1.080) &
(0.131)\tabularnewline
$\alpha$ &
0.001 &
0.005 &
-0.003{*}{*} &
-0.001{*} &
-0.002 &
-0.000\tabularnewline
 &
(0.001) &
(0.004) &
(0.002) &
(0.000) &
(0.006) &
(0.002)\tabularnewline
Adj $R^{2}$ &
0.697 &
0.278 &
0.979 &
0.827 &
0.507 &
0.342\tabularnewline
MAPE, \% &
0.169 &
0.158 &
0.011 &
0.054 &
0.275 &
0.075\tabularnewline
Assets &
171 &
53 &
11 &
40 &
21 &
46\tabularnewline
Months &
180 &
180 &
180 &
180 &
180 &
180\tabularnewline
 &
 &
 &
 &
 &
 &
\tabularnewline
\hline 
\end{tabular}{\small\par}
\end{table}
{\footnotesize\par}

{\footnotesize{}}
\begin{table}[H]
{\footnotesize{}\caption{Correlation Matrix\label{tab:Correlation-Matrix}}
}{\footnotesize\par}

{\footnotesize{}\medskip{}
}{\footnotesize\par}

{\small{}This table reports the correlation matrix for the HFL component,
the individual asset's beta exposures to the \citet{fama2015five}'s
five factors, and the momentum factor of \citet{jegadeesh1993returns},
and }$\Delta\log\left(VIX\right)${\small{}. For the HFL component,
we first estimate $h(\lambda_{i}f_{t})$ for each asset $i$, and
then construct the cross-sectional average. The HFL model is based
on a degree of the polynomial used to approximate the function $h(.)$
equal to 4.}{\small\par}

{\footnotesize{}\bigskip{}
}{\footnotesize\par}
\centering{}{\small{}}%
\begin{tabular}{lcccccccc}
\hline 
 &
HFL &
MKTRF &
SMB &
HML &
RMW &
CMA &
MOM &
$\Delta\log\left(VIX\right)$\tabularnewline
\hline 
HFL &
1.00 &
 &
 &
 &
 &
 &
 &
\tabularnewline
MKTRF &
0.74 &
1.00 &
 &
 &
 &
 &
 &
\tabularnewline
SMB &
0.43 &
0.64 &
1.00 &
 &
 &
 &
 &
\tabularnewline
HML &
0.48 &
0.28 &
0.14 &
1.00 &
 &
 &
 &
\tabularnewline
RMW &
0.24 &
0.05 &
0.04 &
0.13 &
1.00 &
 &
 &
\tabularnewline
CMA &
-0.10 &
0.14 &
-0.01 &
-0.24 &
0.12 &
1.00 &
 &
\tabularnewline
MOM &
-0.12 &
-0.30 &
-0.12 &
-0.15 &
-0.09 &
-0.10 &
1.00 &
\tabularnewline
$\Delta\log\left(VIX\right)$ &
-0.65 &
-0.98 &
-0.68 &
-0.20 &
0.07 &
-0.10 &
0.32 &
1.00\tabularnewline
\hline 
\end{tabular}{\small\par}
\end{table}
{\footnotesize\par}

\pagebreak{}

\section*{Appendix References}

\begin{btSect}[econ-aea]{13_Users_nicolaborri_Dropbox_Research_INDEX-EVERYTHING_lyx_ref}
\btPrintCited
\end{btSect}


@book{V18,
	author = {Vershynin, Roman},
	publisher = {Cambridge Series in Statistical and Probabilistic Mathematics},
	title = {High-Dimensional Probability},
	year = {2018}}

@book{VW96,
	author = {Van der Vaart, Aad and Wellner, Jon},
	publisher = {Springer Series in Statistics},
	title = {Weak Convergence and Empirical Processes},
	year = {1996}}

@article{SSM98,
	author = {Scholkopf, Bernard and Smola, Alexander and Muller, Klaus-Robert},
	journal = {Neural Computation},
	pages = {1299--1319},
	title = {Nonlinear component analysis as a kernel eigenvalue problem},
	volume = {10},
	year = {1998}}

@article{YA01,
	author = {Yalchin, Ilker and Amemiya, Yasuo},
	journal = {Statistical Science},
	pages = {275--294},
	title = {Nonlinear factor analysis as a statistical method},
	volume = {16},
	year = {2001}}

@article{B53,
	author = {Bartlett, M},
	journal = {Uppsala Symposium on Phychological Factor Analysis},
	pages = {23--24},
	title = {Factor analysis in psychology as a statistician sees it},
	year = {1953}}

@article{M62,
	author = {McDonald, R},
	journal = {Psychometrika},
	pages = {397--415},
	title = {A general approach to nonlinear factor analysis},
	volume = {27},
	year = {1962}}

@article{M79,
	author = {McDonald, R},
	journal = {British Journal of Mathematical and Statistical Psychology},
	pages = {212--228},
	title = {The simultaneous estimation of factor loadings and scores},
	volume = {32},
	year = {1979}}

@article{ZL99,
	author = {Zhu, Hong-Tu and Lee, Sik-Yum},
	journal = {British Journal of Mathematical and Statistical Psychology},
	pages = {225--242},
	title = {Statistical analysis of nonlinear factor analysis models},
	volume = {52},
	year = {1999}}

@article{FW16,
	author = {Fernandez-Val, Ivan and Weidner, Martin},
	journal = {Journal of Econometrics},
	pages = {291--312},
	title = {Individual and time effects in nonlinear panel models with large $N$, $T$},
	volume = {192},
	year = {2016}}

@article{CFW21,
	author = {Chen, Mingli and Fernandez-Val, Ivan and Weidner, Martin},
	journal = {Journal of Econometrics},
	pages = {296--324},
	title = {Nonlinear factor models for network and panel data},
	volume = {220},
	year = {2021}}

@article{W22,
	author = {Wang, Fa},
	journal = {Journal of Econometrics},
	pages = {180--200},
	title = {Maximum likelihood estimation and inference for high dimensional generalized factor models with application to factor-augmented regressions},
	volume = {229},
	year = {2022}}

@article{BL17,
	author = {Boneva, Lena and Linton, Oliver},
	journal = {Journal of Applied Econometrics},
	pages = {1226--1243},
	title = {A discrete-choice model for large heterogeneous panels with interactive fixed effects with an application to the determinants of corporate bond issuance},
	volume = {32},
	year = {2017}}

@article{C17,
	author = {Charbonneau, Karyne},
	journal = {Econometrics Journal},
	pages = {1--13},
	title = {Multiple fixed effects in binary response panel data models},
	volume = {20},
	year = {2017}}

@article{GLPY23,
	author = {Gao, Jiti and Liu, Fei and Peng, Bin and Yan, Yayi},
	journal = {Journal of Econometrics},
	pages = {1654--1679},
	title = {Binary response models for heterogeneous panel data with interactive fixed effects},
	volume = {235},
	year = {2023}}

@article{MW22,
	author = {Mugnier, Martin and Wang, Ao},
	journal = {Working Paper},
	pages = {1--64},
	title = {Identification and fast estimation of large nonlinear panel models with two-way fixed effects},
	year = {2022}}

@article{CDG21,
	author = {Chen, Liang and Dolado, Juan and Gonzalo, Jesus},
	journal = {Econometrica},
	pages = {875--910},
	title = {Quantile factor models},
	volume = {89},
	year = {2021}}

@article{AB20,
	author = {Ando, Tomohiro and Bai, Jushan},
	journal = {Journal of American Statistical Association},
	pages = {266--279},
	title = {Quantile co-movement in financial markets: a panel quantile model with unobserved heterogeneity},
	volume = {115},
	year = {2020}}

@article{CFW20,
	author = {Chernozhukov, Victor and Fernandez-Val, Ivan and Weidner, Martin},
	journal = {Journal of Econometrics},
	number = {2},
	title = {Network and panel quantile effects via distribution regression},
	volume = {240},
	year = {2024}}

@article{Z20,
	author = {Zeleneev, Andrei},
	journal = {Working paper},
	title = {Identification and estimation of network models with nonparametric unobserved heterogeneity},
	year = {2020}}

@article{G20,
	author = {Gao, Wayne},
	journal = {Journal of Econometrics},
	pages = {399--413},
	title = {Nonparametric identification in index models with link formation},
	volume = {215},
	year = {2020}}

@article{jensen2023there,
	author = {Jensen, Theis Ingerslev and Kelly, Bryan and Pedersen, Lasse Heje},
	journal = {Journal of Finance},
	number = {5},
	pages = {2465--2518},
	publisher = {Wiley Online Library},
	title = {Is there a replication crisis in finance?},
	volume = {78},
	year = {2023}}

@article{chen2021open,
	author = {Chen, Andrew Y and Zimmermann, Tom},
	journal = {Critical Finance Review},
	number = {2},
	title = {Open source cross-sectional asset pricing},
	volume = {11},
	year = {2022}}

@article{hou2007industry,
	author = {Hou, Kewei},
	journal = {Review of Financial Studies},
	number = {4},
	pages = {1113--1138},
	publisher = {Oxford University Press},
	title = {Industry information diffusion and the lead-lag effect in stock returns},
	volume = {20},
	year = {2007}}

@article{belloni2014inference,
	author = {Belloni, Alexandre and Chernozhukov, Victor and Hansen, Christian},
	journal = {Review of Economic Studies},
	number = {2},
	pages = {608--650},
	publisher = {Oxford University Press},
	title = {Inference on treatment effects after selection among high-dimensional controls},
	volume = {81},
	year = {2014}}

@article{feng2020taming,
	author = {Feng, Guanhao and Giglio, Stefano and Xiu, Dacheng},
	journal = {Journal of Finance},
	number = {3},
	pages = {1327--1370},
	publisher = {Wiley Online Library},
	title = {Taming the factor zoo: A test of new factors},
	volume = {75},
	year = {2020}}

@article{lettau2020factors,
	author = {Lettau, Martin and Pelger, Markus},
	journal = {Review of Financial Studies},
	number = {5},
	pages = {2274--2325},
	publisher = {Oxford University Press},
	title = {Factors that fit the time series and cross-section of stock returns},
	volume = {33},
	year = {2020}}

@article{kozak2020shrinking,
	author = {Kozak, Serhiy and Nagel, Stefan and Santosh, Shrihari},
	journal = {Journal of Financial Economics},
	number = {2},
	pages = {271--292},
	publisher = {Elsevier},
	title = {Shrinking the cross-section},
	volume = {135},
	year = {2020}}

@article{giglio2021asset,
	author = {Giglio, Stefano and Xiu, Dacheng},
	journal = {Journal of Political Economy},
	number = {7},
	pages = {1947--1990},
	publisher = {The University of Chicago Press Chicago, IL},
	title = {Asset pricing with omitted factors},
	volume = {129},
	year = {2021}}

@article{giglio2022factor,
	author = {Giglio, Stefano and Kelly, Bryan and Xiu, Dacheng},
	journal = {Annual Review of Financial Economics},
	pages = {337--368},
	publisher = {Annual Reviews},
	title = {Factor models, machine learning, and asset pricing},
	volume = {14},
	year = {2022}}

@article{onatski2012asymptotics,
	author = {Onatski, Alexei},
	journal = {Journal of Econometrics},
	number = {2},
	pages = {244--258},
	publisher = {Elsevier},
	title = {Asymptotics of the principal components estimator of large factor models with weakly influential factors},
	volume = {168},
	year = {2012}}

@article{fama1996multifactor,
	author = {Fama, Eugene F and French, Kenneth R},
	journal = {Journal of Finance},
	number = {1},
	pages = {55--84},
	publisher = {Wiley Online Library},
	title = {Multifactor explanations of asset pricing anomalies},
	volume = {51},
	year = {1996}}

@article{fama2015five,
	author = {Fama, Eugene F and French, Kenneth R},
	journal = {Journal of Financial Economics},
	number = {1},
	pages = {1--22},
	publisher = {Elsevier},
	title = {A five-factor asset pricing model},
	volume = {116},
	year = {2015}}

@article{jegadeesh1993returns,
	author = {Jegadeesh, Narasimhan and Titman, Sheridan},
	journal = {Journal of Finance},
	number = {1},
	pages = {65--91},
	publisher = {Wiley Online Library},
	title = {Returns to buying winners and selling losers: Implications for stock market efficiency},
	volume = {48},
	year = {1993}}

@article{fama1993common,
	author = {Fama, Eugene F. and French, Kenneth R.},
	journal = {Journal of Financial Economics},
	number = {1},
	pages = {3--56},
	publisher = {Elsevier},
	title = {Common risk factors in the returns on stocks and bonds},
	volume = {33},
	year = {1993}}

@inproceedings{hecht1987kolmogorov,
	author = {Hecht-Nielsen, Robert},
	booktitle = {Proceedings of the International Conference on Neural Networks},
	organization = {IEEE press New York, NY, USA},
	pages = {11--14},
	title = {Kolmogorov's mapping neural network existence theorem},
	volume = {3},
	year = {1987}}

@article{hutchinson1994nonparametric,
	author = {Hutchinson, James M and Lo, Andrew W and Poggio, Tomaso},
	journal = {Journal of Finance},
	number = {3},
	pages = {851--889},
	publisher = {Wiley Online Library},
	title = {A nonparametric approach to pricing and hedging derivative securities via learning networks},
	volume = {49},
	year = {1994}}

@article{newey1987simple,
	author = {Newey, Whitney K and West, Kenneth D},
	journal = {Econometrica},
	number = {3},
	pages = {703--708},
	title = {A simple, positive semi-definite, heteroskedasticity and autocorrelation consistent covariance matrix},
	volume = {55},
	year = {1987}}

@article{koijen2020investors,
	author = {Koijen, Ralph SJ and Richmond, Robert J and Yogo, Motohiro},
	journal = {Review of Economic Studies},
	title = {Which investors matter for equity valuations and expected returns?},
	year = {2024}}

@article{koijen2018carry,
	author = {Koijen, Ralph SJ and Moskowitz, Tobias J and Pedersen, Lasse Heje and Vrugt, Evert B},
	journal = {Journal of Financial Economics},
	number = {2},
	pages = {197--225},
	publisher = {Elsevier},
	title = {Carry},
	volume = {127},
	year = {2018}}

@article{koijen2019demand,
	author = {Koijen, Ralph SJ and Yogo, Motohiro},
	journal = {Journal of Political Economy},
	number = {4},
	pages = {1475--1515},
	publisher = {The University of Chicago Press Chicago, IL},
	title = {A demand system approach to asset pricing},
	volume = {127},
	year = {2019}}

@article{asness2013value,
	author = {Asness, Clifford S and Moskowitz, Tobias J and Pedersen, Lasse Heje},
	journal = {Journal of Finance},
	number = {3},
	pages = {929--985},
	publisher = {Wiley Online Library},
	title = {Value and momentum everywhere},
	volume = {68},
	year = {2013}}

@article{bollerslev2018risk,
	author = {Bollerslev, Tim and Hood, Benjamin and Huss, John and Pedersen, Lasse Heje},
	journal = {Review of Financial Studies},
	number = {7},
	pages = {2729--2773},
	publisher = {Oxford University Press},
	title = {Risk everywhere: Modeling and managing volatility},
	volume = {31},
	year = {2018}}

@article{chernozhukov2018double,
	author = {Chernozhukov, Victor and Chetverikov, Denis and Demirer, Mert and Duflo, Esther and Hansen, Christian and Newey, Whitney and Robins, James},
	journal = {Econometrics Journal},
	number = {1},
	publisher = {Oxford University Press},
	title = {Double/debiased machine learning for treatment and structural parameters: Double/debiased machine learning},
	volume = {21},
	year = {2018}}

@article{ludvigson2009macro,
	author = {Ludvigson, Sydney C and Ng, Serena},
	journal = {Review of Financial Studies},
	number = {12},
	pages = {5027--5067},
	publisher = {Oxford University Press},
	title = {Macro factors in bond risk premia},
	volume = {22},
	year = {2009}}

@article{nucera2024currency,
	author = {Nucera, Federico and Sarno, Lucio and Zinna, Gabriele},
	journal = {Review of Financial Studies},
	number = {2},
	pages = {356--408},
	publisher = {Oxford University Press},
	title = {Currency risk premiums redux},
	volume = {37},
	year = {2024}}

@article{fama1973risk,
	author = {Fama, Eugene F and MacBeth, James D},
	journal = {Journal of Political Economy},
	number = {3},
	pages = {607--636},
	publisher = {The University of Chicago Press},
	title = {Risk, return, and equilibrium: Empirical tests},
	volume = {81},
	year = {1973}}

@article{lettau2014conditional,
	author = {Lettau, Martin and Maggiori, Matteo and Weber, Michael},
	journal = {Journal of Financial Economics},
	number = {2},
	pages = {197--225},
	publisher = {Elsevier},
	title = {Conditional risk premia in currency markets and other asset classes},
	volume = {114},
	year = {2014}}

@article{he2017intermediary,
	author = {He, Zhiguo and Kelly, Bryan and Manela, Asaf},
	journal = {Journal of Financial Economics},
	number = {1},
	pages = {1--35},
	publisher = {Elsevier},
	title = {Intermediary asset pricing: New evidence from many asset classes},
	volume = {126},
	year = {2017}}

@book{hastie2009elements,
	author = {Hastie, Trevor and Tibshirani, Robert and Friedman, Jerome H and Friedman, Jerome H},
	publisher = {Springer},
	title = {The elements of statistical learning: data mining, inference, and prediction},
	volume = {2},
	year = {2009}}

@article{pastor2003liquidity,
	author = {P{\'a}stor, L'ubo{\v{s}} and Stambaugh, Robert F},
	journal = {Journal of Political Economy},
	number = {3},
	pages = {642--685},
	publisher = {The University of Chicago Press},
	title = {Liquidity risk and expected stock returns},
	volume = {111},
	year = {2003}}

@article{brennan1998alternative,
	author = {Brennan, Michael J and Chordia, Tarun and Subrahmanyam, Avanidhar},
	journal = {Journal of Financial Economics},
	number = {3},
	pages = {345--373},
	publisher = {Elsevier},
	title = {Alternative factor specifications, security characteristics, and the cross-section of expected stock returns},
	volume = {49},
	year = {1998}}

@article{datar1998liquidity,
	author = {Datar, Vinay T and Naik, Narayan Y and Radcliffe, Robert},
	journal = {Journal of Financial Markets},
	number = {2},
	pages = {203--219},
	publisher = {Elsevier},
	title = {Liquidity and stock returns: An alternative test},
	volume = {1},
	year = {1998}}

@article{haugen1996commonality,
	author = {Haugen, Robert A and Baker, Nardin L},
	journal = {Journal of Financial Economics},
	number = {3},
	pages = {401--439},
	publisher = {Elsevier},
	title = {Commonality in the determinants of expected stock returns},
	volume = {41},
	year = {1996}}

@article{adrian2014financial,
	author = {Adrian, Tobias and Etula, Erkko and Muir, Tyler},
	journal = {Journal of Finance},
	number = {6},
	pages = {2557--2596},
	publisher = {Wiley Online Library},
	title = {Financial intermediaries and the cross-section of asset returns},
	volume = {69},
	year = {2014}}

@article{connor1986performance,
	author = {Connor, Gregory and Korajczyk, Robert A},
	journal = {Journal of Financial Economics},
	number = {3},
	pages = {373--394},
	publisher = {Elsevier},
	title = {Performance measurement with the arbitrage pricing theory: A new framework for analysis},
	volume = {15},
	year = {1986}}

@article{eugene1992cross,
	author = {Fama, Eugene and French, Kenneth},
	journal = {Journal of Finance},
	number = {2},
	pages = {427--465},
	title = {The cross-section of expected stock returns},
	volume = {47},
	year = {1992}}

@article{connor1988risk,
	author = {Connor, Gregory and Korajczyk, Robert A},
	journal = {Journal of Financial Economics},
	number = {2},
	pages = {255--289},
	publisher = {Elsevier},
	title = {Risk and return in an equilibrium APT: Application of a new test methodology},
	volume = {21},
	year = {1988}}

@article{harvey2000conditional,
	author = {Harvey, Campbell R and Siddique, Akhtar},
	journal = {Journal of Finance},
	number = {3},
	pages = {1263--1295},
	publisher = {Wiley Online Library},
	title = {Conditional skewness in asset pricing tests},
	volume = {55},
	year = {2000}}

@article{ang2006cross,
	author = {Ang, Andrew and Hodrick, Robert J and Xing, Yuhang and Zhang, Xiaoyan},
	journal = {Journal of Finance},
	number = {1},
	pages = {259--299},
	publisher = {Wiley Online Library},
	title = {The cross-section of volatility and expected returns},
	volume = {61},
	year = {2006}}

@article{sharpe1964capital,
	author = {Sharpe, William F},
	journal = {Journal of Finance},
	number = {3},
	pages = {425--442},
	publisher = {Wiley Online Library},
	title = {Capital asset prices: A theory of market equilibrium under conditions of risk},
	volume = {19},
	year = {1964}}

@article{lintner1965security,
	author = {Lintner, John},
	journal = {Journal of Finance},
	number = {4},
	pages = {587--615},
	publisher = {JSTOR},
	title = {Security prices, risk, and maximal gains from diversification},
	volume = {20},
	year = {1965}}

@article{kraus1976skewness,
	author = {Kraus, Alan and Litzenberger, Robert H},
	journal = {Journal of Finance},
	number = {4},
	pages = {1085--1100},
	publisher = {JSTOR},
	title = {Skewness preference and the valuation of risk assets},
	volume = {31},
	year = {1976}}

@article{scholkopf1998nonlinear,
	author = {Sch{\"o}lkopf, Bernhard and Smola, Alexander and M{\"u}ller, Klaus-Robert},
	journal = {Neural Computation},
	number = {5},
	pages = {1299--1319},
	publisher = {MIT Press},
	title = {Nonlinear component analysis as a kernel eigenvalue problem},
	volume = {10},
	year = {1998}}

@article{Carhart1997,
	author = {Carhart, Mark M},
	journal = {Journal of Finance},
	number = {1},
	pages = {57--82},
	publisher = {Wiley Online Library},
	title = {On persistence in mutual fund performance},
	volume = {52},
	year = {1997}}

@article{lettau2001resurrecting,
	author = {Lettau, Martin and Ludvigson, Sydney},
	journal = {Journal of Political Economy},
	number = {6},
	pages = {1238--1287},
	publisher = {The University of Chicago Press},
	title = {Resurrecting the (C) CAPM: A cross-sectional test when risk premia are time-varying},
	volume = {109},
	year = {2001}}

@article{jagannathan1996conditional,
	author = {Jagannathan, Ravi and Wang, Zhenyu},
	journal = {Journal of Finance},
	number = {1},
	pages = {3--53},
	publisher = {Wiley Online Library},
	title = {The conditional CAPM and the cross-section of expected returns},
	volume = {51},
	year = {1996}}

@article{kelly2019characteristics,
	author = {Kelly, Bryan T and Pruitt, Seth and Su, Yinan},
	journal = {Journal of Financial Economics},
	number = {3},
	pages = {501--524},
	publisher = {Elsevier},
	title = {Characteristics are covariances: A unified model of risk and return},
	volume = {134},
	year = {2019}}

@article{bansal1993no,
	author = {Bansal, Ravi and Viswanathan, Salim},
	journal = {Journal of Finance},
	number = {4},
	pages = {1231--1262},
	publisher = {Wiley Online Library},
	title = {No arbitrage and arbitrage pricing: A new approach},
	volume = {48},
	year = {1993}}

@article{koijen2017cross,
	author = {Koijen, Ralph SJ and Lustig, Hanno and Van Nieuwerburgh, Stijn},
	journal = {Journal of Monetary Economics},
	pages = {50--69},
	publisher = {Elsevier},
	title = {The cross-section and time series of stock and bond returns},
	volume = {88},
	year = {2017}}

@article{kelly2024virtue,
	author = {Kelly, Bryan and Malamud, Semyon and Zhou, Kangying},
	journal = {Journal of Finance},
	number = {1},
	pages = {459--503},
	publisher = {Wiley Online Library},
	title = {The virtue of complexity in return prediction},
	volume = {79},
	year = {2024}}

@article{kelly2023principal,
	author = {Kelly, Bryan and Malamud, Semyon and Pedersen, Lasse Heje},
	journal = {Journal of Finance},
	number = {1},
	pages = {347--387},
	publisher = {Wiley Online Library},
	title = {Principal portfolios},
	volume = {78},
	year = {2023}}

@article{gu2020empirical,
	author = {Gu, Shihao and Kelly, Bryan and Xiu, Dacheng},
	journal = {Review of Financial Studies},
	number = {5},
	pages = {2223--2273},
	publisher = {Oxford University Press},
	title = {Empirical asset pricing via machine learning},
	volume = {33},
	year = {2020}}

@article{Kolmogorov1956on,
	author = {Kolmogorov, Andrey},
	journal = {Proceedings of the USSR Academy of Sciences},
	pages = {179-182},
	title = {On the representation of continuous functions of seversal variables by superpositions of continuous functions of a smaller number of variables},
	volume = {108},
	year = {1956}}

@article{Arnold1957on,
	author = {Arnold, Vladimir},
	journal = {Proceedings of the USSR Academy of Sciences},
	pages = {679--681},
	title = {On functions of three variables},
	volume = {114},
	year = {1957}}

@article{Borri2024factor,
	author = {Borri, Nicola and Chetverikov, Denis and Liu, Yukun and Tsyvinski, Aleh},
	journal = {Working Paper},
	title = {Factor model with nonparametric link function},
	year = {2024}}

@article{SH21,
	author = {Schmidt-Hieber, Johannes},
	journal = {Neural Networks},
	pages = {119--126},
	title = {The {K}olmogorov-{A}rnold representation theorem revisited},
	volume = {137},
	year = {2021}}

@article{K91,
	author = {Kurkova, Vera},
	journal = {Neural computation},
	pages = {617--622},
	title = {Kolmogorov's theorem is relevant},
	volume = {3},
	year = {1991}}

@article{MP99,
	author = {Maiorov, Vitaly and Pinkus, Allan},
	journal = {Neurocomputing},
	pages = {81--91},
	title = {Lower bounds for approximation by MLP neural networks},
	volume = {25},
	year = {1999}}

@article{C04,
	author = {Coppejans, Mark},
	journal = {Journal of Econometrics},
	pages = {1--31},
	title = {On Kolmogorov's representation of functions of several variables by functions of one variable},
	volume = {123},
	year = {2004}}

@article{gu2021autoencoder,
	author = {Gu, Shihao and Kelly, Bryan and Xiu, Dacheng},
	journal = {Journal of Econometrics},
	number = {1},
	pages = {429--450},
	publisher = {Elsevier},
	title = {Autoencoder asset pricing models},
	volume = {222},
	year = {2021}}
\end{document}